\documentclass[aps,preprint,showpacs,superscriptaddress,nofootinbib]{revtex4-1}

\usepackage{amsmath}
\usepackage[normalem]{ulem}
\usepackage{amssymb}
\usepackage{gensymb}
\usepackage{graphicx}
\usepackage{epstopdf}
\usepackage{soul}
\usepackage{setspace}
\usepackage{color}
\usepackage{siunitx}
\usepackage[version=3]{mhchem}
\usepackage[T1]{fontenc} 

\begin{document}
	
	\title{Proton distribution visualization in perovskite  nickelate devices utilizing nanofocused X-rays}
	
	\author{Ivan A. Zaluzhnyy}
	\email[Corresponding authors: ]{izaluzhnyy@physics.ucsd.edu, oshpyrko@physics.ucsd.edu, afrano@ucsd.edu}
	\affiliation{Department of Physics, University of California San Diego, La Jolla, CA 92093, USA}

	\author{Peter~O.~Sprau}
	\thanks{These two authors contributed equally: Ivan A. Zaluzhnyy and Peter O. Sprau}
	\affiliation{Department of Physics, University of California San Diego, La Jolla, CA 92093, USA}
	
	\author{Richard~Tran}
	\affiliation{Department of NanoEngineering, University of California San Diego, La Jolla, CA 92093, USA}
	
	\author{Qi~Wang}
	\affiliation{School of Materials Engineering, Purdue University, West Lafayette, IN 47907, USA}
	
	\author{Hai-Tian~Zhang}
	\affiliation{School of Materials Engineering, Purdue University, West Lafayette, IN 47907, USA}
	\affiliation{Lillian Gilbreth Fellowship Program, College of Engineering, Purdue University, West Lafayette, IN 47907, USA}
	
	\author{Zhen~Zhang}
	\affiliation{School of Materials Engineering, Purdue University, West Lafayette, IN 47907, USA}
	
	\author{Tae Joon Park}
	\affiliation{School of Materials Engineering, Purdue University, West Lafayette, IN 47907, USA}
	
	\author{Nelson Hua}
	\affiliation{Department of Physics, University of California San Diego, La Jolla, CA 92093, USA}	
	
	\author{Boyan Stoychev}
	\affiliation{Department of Physics, University of California San Diego, La Jolla, CA 92093, USA}	
	
	\author{Mathew~J.~Cherukara}
	\affiliation{Center for nanoscale materials, Argonne National Laboratory, Argonne, IL 60439, USA}
	
	\author{Martin~V.~Holt}
	\affiliation{Center for nanoscale materials, Argonne National Laboratory, Argonne, IL 60439, USA}
	
	\author{Evgeny~Nazarertski}
	\affiliation{National Synchrotron Light Source II, Brookhaven National Laboratory, Upton, NY 11973, USA}
	
	\author{Xiaojing~Huang}
	\affiliation{National Synchrotron Light Source II, Brookhaven National Laboratory, Upton, NY 11973, USA}
	
	\author{Hanfei~Yan}
	\affiliation{National Synchrotron Light Source II, Brookhaven National Laboratory, Upton, NY 11973, USA}
	
	\author{Ajith~Pattammattel}
	\affiliation{National Synchrotron Light Source II, Brookhaven National Laboratory, Upton, NY 11973, USA}
	
	\author{Yong~S.~Chu}
	\affiliation{National Synchrotron Light Source II, Brookhaven National Laboratory, Upton, NY 11973, USA}
	
	\author{Shyue~Ping~Ong}
	\affiliation{Department of NanoEngineering, University of California San Diego, La Jolla, CA 92093, USA}
	
	\author{Shriram~Ramanathan}
	\affiliation{School of Materials Engineering, Purdue University, West Lafayette, IN 47907, USA}
	
	\author{Oleg~G.~Shpyrko}
	\email[Corresponding authors: ]{izaluzhnyy@physics.ucsd.edu, oshpyrko@physics.ucsd.edu, afrano@ucsd.edu}
	\affiliation{Department of Physics, University of California San Diego, La Jolla, CA 92093, USA}	
	
	\author{Alex~Frano}
	\email[Corresponding authors: ]{izaluzhnyy@physics.ucsd.edu, oshpyrko@physics.ucsd.edu, afrano@ucsd.edu}
	\affiliation{Department of Physics, University of California San Diego, La Jolla, CA 92093, USA}


	\begin{abstract}

		We use a 30-nm x-ray beam to study the spatially resolved properties of a \ce{SmNiO3}-based nanodevice that is doped with protons.
		The x-ray absorption spectra supported by density-functional theory (DFT) simulations show partial reduction of nickel valence in the region with high proton concentration, which leads to the insulating behavior. Concurrently, x-ray diffraction reveals only a small lattice distortion in the doped regions. 
		Together, our results directly show that the knob which proton doping modifies is the electronic valency, and not the crystal lattice. 
		The studies are relevant to on-going efforts to disentangle structural and electronic effects across metal-insulator phase transitions in correlated oxides.
		
	\end{abstract}
	
	\maketitle

	\section{Introduction}

	Among possible replacements for density-limited silicon transistors on integrated circuits are quantum materials, such as transition metal oxides. These are promising because their electronic properties can be tuned efficiently and reversibly. 
	For example, they have recently gained attention as a platform for devices that could enable neuromorphic computing, which offers a new level of computational efficiency by creating artificial systems that can emulate the operation of animal brains \cite{Ramanathan2018, Roy2019}.
	This requires development of hardware elements whose electrical resistance changes under external stimuli (e.g., voltage or light pulse), emulating synaptic memory links between neurons \cite{Zhang2020, DelValle2018}. 
	Prospective materials for these new electrical elements must a) be electrically switched by a small external stimulus, b) have a wide range of electrical resistance, c) increase or decrease resistance with stimuli, and d) operate at room temperature. 
	
	Not many materials exist that can satisfy all these requirements. 
	Among them, rare-earth nickelate \ce{SmNiO3} (SNO) doped with protons H$^+$ (H-SNO) is an extremely promising candidate, especially due to controllability of the switching process \cite{Zhang2020a, Zhang2020, Shi2013, Zhu2020}. 
	Moreover, the design of the nanodevice with synaptic functionality is as simple as a proton-doped SNO film between two metallic electrodes \cite{Ramadoss2018}. 
	The synaptic behavior of such a memory device fundamentally relies on the motion and redistribution of protons influenced by electric field pulses. This raises two questions: a) how does proton doping change the structure and properties of SNO at the nanoscale, b) how protons are distributed in the SNO film? 
	Understanding these phenomena is important from a fundamental point of view.
	
	X-rays have been traditionally used to determine the properties of pristine nickelate heterostructures \cite{Staub2002, Lu2016}, but in small synaptic devices, the key challenge is to determine how the spatial distribution of protons affects the electrical properties of SNO. 
	Additionally, x-rays do not directly detect light ions easily, but instead they are a perfect tool to study the influence of doping on the electronic and crystal structure of SNO.
	To overcome these challenges, we uniquely probed the device with nanofocused x-ray beam to pinpoint the origins of resistive switching trends in SNO and disentangle, whether it is a structural or electronic change that drives the modulation in resistance. 
	By studying the x-ray fluorescence spectra near the Ni K-edge supported by \textit{ab initio} simulations, we were able to spatially resolve how proton doping affects the valency of nickel. 
	We also used spatially resolved x-ray nanodiffraction to reveal the subtle changes in the SNO lattice structure and resolve the correlations to the electronic structure. 
	Despite the wide implementation of ionic motion in many materials, including neuromorphic computing hardware \cite{Zhu2020}, the local measurements of light ion concentrations are very challenging \cite{Chen2019}.  
	The experimental results we describe in this work address some of the key questions about the mechanism responsible for resistive changes in nickelate devices doped with light protons.
	
		\begin{figure}
		\includegraphics[width = 0.7\linewidth]{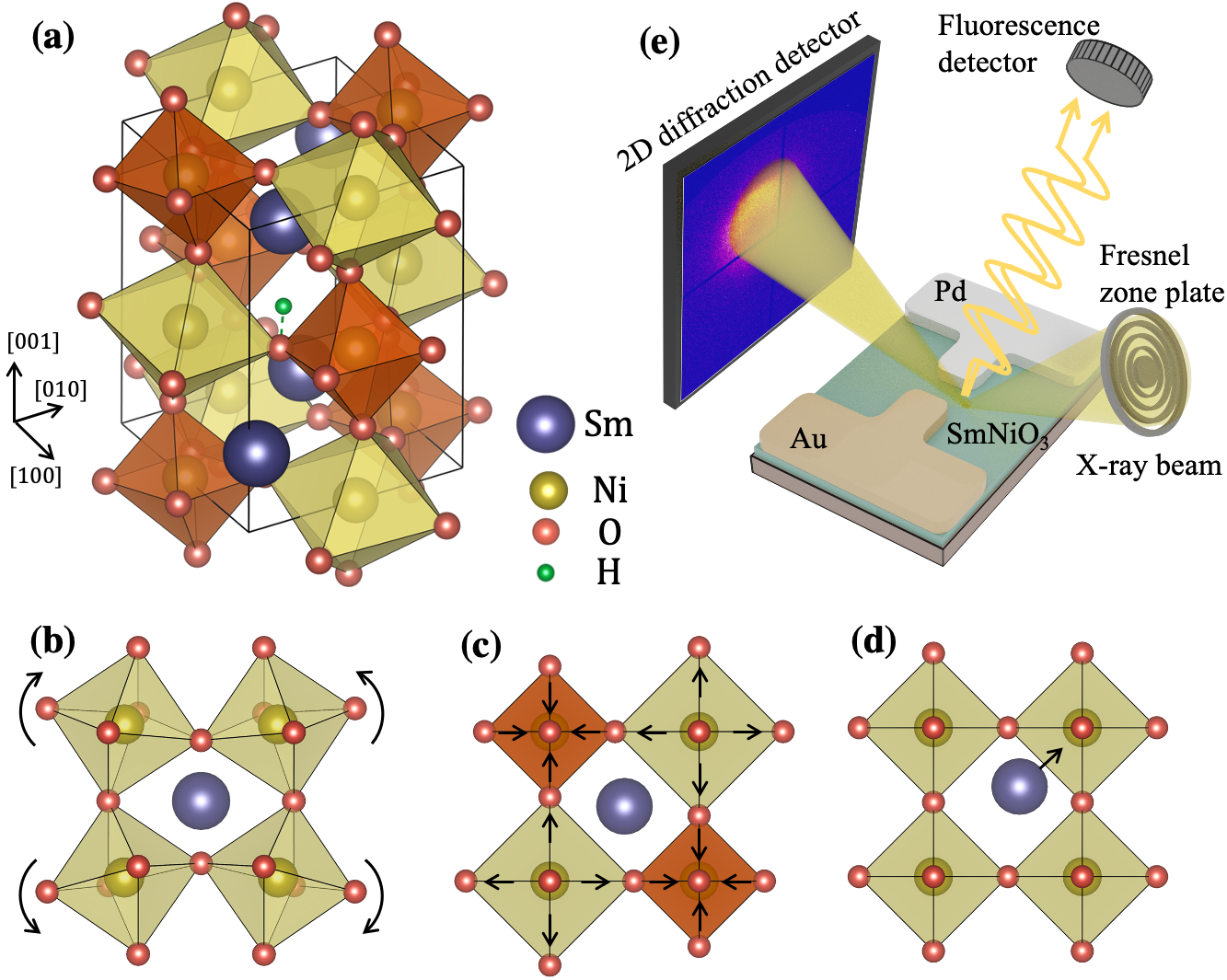}
		\caption{\label{Setup}
			(a) Orthorhombic crystal structure of SNO. The breathing mode is shown by color: expanded $\text{NiO}_6$ octahedra are yellow and contracted are orange.
			(b) Schematic representation of the \ce{NiO6} octahedra rotations (tilt pattern) (c) Breathing mode (d) displacement of the rare-earth cation from the centrosymmetric position inside the cavity between \ce{NiO6} octahedra.
			(e) Scheme of the nanofocused x-ray experiment and the SNO-based device. The focused beam is used for raster scanning of the device where the diffraction signal is recorded by a 2D detector in reflection geometry, and the fluoresce signal is collected by a point energy-resolving detector oriented perpendicular to the sample surface }
	\end{figure}
	
	Pristine SNO has a distorted perovskite structure characterized by corner-connected $\text{NiO}_{6}$ octahedra with $\text{Sm}^{3+}$ ions filling the cavities between the octahedra (Fig.~\ref{Setup}(a)).
	Since the ionic radius of $\text{Sm}^{3+}$ is smaller than the size of the cavity, $\text{NiO}_{6}$ octahedra are tilted \cite{Glazer1972} and distorted and the rare-earth ions are slightly displaced from the central position (Fig.~\ref{Setup}(b)-(d)). 
	The structure can be described by the orthorhombic \textit{Pbnm} symmetry with the unit cell parameters $a_o=5.328$~\AA, $b_o=5.437$~\AA~and $c_o=7.568$~\AA~\cite{Jain2013}. 
	Often the pseudocubic crystal lattice is also used with the unit cell parameters $a_{pc}=b_{pc}=\sqrt{a_o^2+b_o^2}/2\approx3.806$~\AA~and $c_{pc}=c_o/2\approx3.784$~\AA~\cite{Catalano2018,Glazer1972, Glazer1975}.
	The (101) and (202) orthorhombic reflections considered in this work correspond to the $(\tfrac{1}{2}\tfrac{1}{2}\tfrac{1}{2})_{pc}$ and $(111)_{pc}$ reflections in the pseudocubic notation.
	
	Doping of the SNO with hydrogen decreases its electrical conductivity by eight orders of magnitude \cite{Shi2014}.
	Hydrogen doping means that an extra proton H$^+$ is implanted in the crystal lattice together with an extra electron $e^-$ to maintain the electrical neutrality.
	The proposed mechanism has been attributed to the addition of an electron into the system that opens a large gap at the Fermi level \cite{Shi2014, Zhou2016, Ramadoss2016}.
	A similar effect is observed when SNO is doped with other small ions, such as Li$^+$ or Na$^+$ \cite{Shi2014, Sun2018}. 
	However, possible changes in the crystal lattice (e.g., the unit cell parameters \cite{Chen2019, Zhou2016} and the $\text{NiO}_{6}$ tilt pattern \cite{Zhou2016,Liao2018}) have been shown to directly influence the conductivity of the nickelates.
	The goal of this work is to disentangle the effects of electron doping and structural changes with spatial resolution inside a functioning nanodevice.

	\section{Results}
	\subsection{Spectroscopy studies of Ni valence}
	
	\begin{figure*}
		\includegraphics[width = 0.9\linewidth]{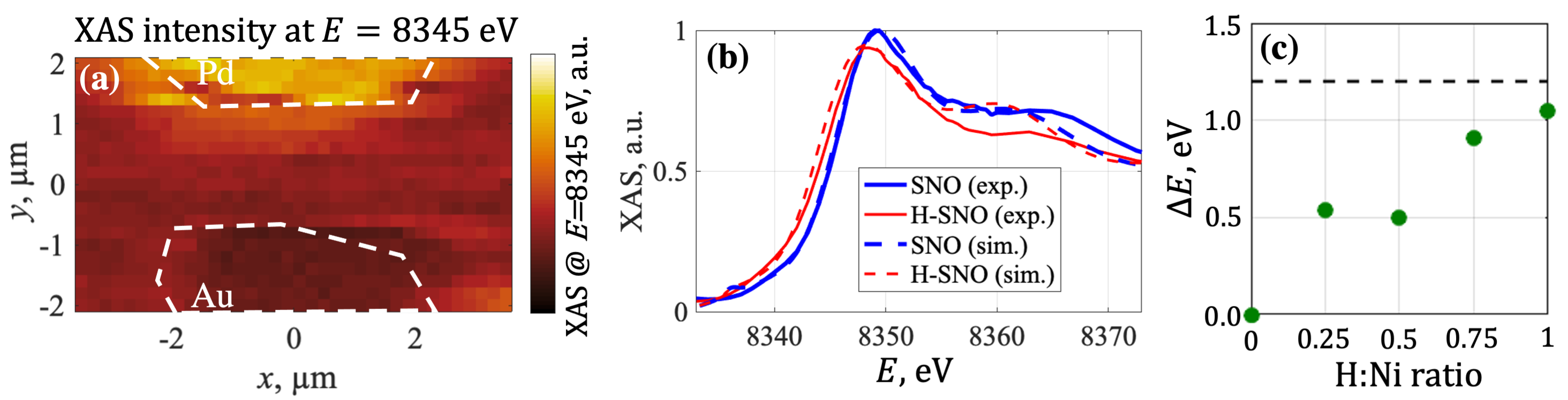}
		\caption{
			(a) Spatially resolved map of the fluorescence signal at $E=8345$~eV. The white dashed lines outline the Pd and Au electrodes. Bright areas next to the Pd electrode correspond to the reduced valence of nickel ions due to the presence of H$^{+}$.
			(b) Experimentally measured normalized XAS spectra at K-edge of Ni in the pristine SNO away from the Pd electrode and the doped H-SNO under the Pd electrode (solid lines) and FEFF-simulated XAS spectra for the doped H-SNO and undoped SNO (dashed lines). The simulated spectra were additionally convoluted with a Gaussian function to match the same energy resolution as in the experiment.
			The experimentally observed shit of the XAS spectrum was reproduced in the DFT-simulations at the 1H:1Ni doping level.
	     	(c) Dependence of the absorption peak shift $\Delta E$ on the concentration of H-dopant for the simulated XAS spectra. Horizontal dashed line marks the experimentally observed value of $\Delta E=1.2$~eV.
		}
		\label{XAS} 
	\end{figure*}
	
	The SNO-based nanodevice \cite{Ramadoss2018,Shi2013,Shi2014, Zhang2020a} and experimental geometry are shown schematically in Fig.~\ref{Setup}(e) \cite{SupplNote}. 
	Simultaneous diffraction and fluorescence measurements were taken while scanning the focused $\sim$30\,nm x-ray beam across the device \cite{SupplNote}. 
	The spatially resolved fluorescence map near the nickel resonance energy $E=8345$~eV across the device is shown in Fig.~\ref{XAS}(a).
	This energy corresponds to the highest slope of the x-ray absorption spectrum (XAS), and thus is most sensitive to shifts of the absorption edge caused by a different electronic valency. 
	While the region of the increased fluorescence signal extends several hundred nanometers outside of the Pd electrode, a clear boundary can be resolved near this electrode.
This indicates diffusion of H-dopants from the area with the highest dopant concentration directly below the Pd electrode to the pristine SNO film away from the electrode.
	A comparison of the fluorescence spectra measured between the two electrodes and the region directly under the Pd electrode is
	shown in Fig.~\ref{XAS}(b). 
	A clear shift in the spectra can be seen between these two regions, suggesting an accumulation of dopant in the vicinity of the Pd electrode.

	The K-edge transition of Ni corresponds to the promotion of the $1s$ core-level electron into the valence $4p$ shells.
	The position of the Ni K-edge peak depends on the number of electrons in the Ni $3d$ shells -- the decrease of the Ni valence caused by H$^{+}$ leads to a shift of the K-edge towards lower energies \cite{Mansour1997,Woolley2011,Gu2014,Zuo2017}. 
	We experimentally resolved the position of the Ni K-edge by identifying the zero point of the second derivative of the two spectra, which gave us an estimate of $1.2\pm0.4$~eV for the energy shift (Fig.~\ref{XAS}(b)).
	
	\subsection{Simulations of SmNiO$_3$ electronic structure}

	To estimate the corresponding change in the SNO electronic structure (i.e. the Ni oxidation state) \cite{Zuo2017, Zhang2018, Kotiuga2019}, we simulated the structure of H-SNO using the Vienna \textit{ab initio} simulation package (VASP) \cite{Kresse1996,Kresse1999} and then calculated the theoretical XAS spectra with the FEFF package \cite{Rehr2010}.
	In our density-functional theory (DFT) simulations, we tested orthorhombic \textit{Pbnm} and monoclinic $P2_1/n$ symmetries of SNO, and performed calculations for pristine SNO and doped H-SNO with different atomic ratios of H:Ni.
	
	Our simulations of the low-temperature monoclinic ($P2_1/n$) phase of insulating SNO with known bond-disproportionation exhibits a small band gap of $E_G\approx0.2$~eV, in agreement with previous publications \cite{Catalano2018}. 
	A gradual increase in H-doping led to the closing of the band-gap at a 1H:2Ni ratio. A further increase in the doping ratio to 1H:1Ni reopens the band gap to 0.97 eV (five times the undoped band gap) as the structural and electronic properties approach the H-SNO orthorhombic phase at 1H:1Ni doping \cite{SupplNote}. 
	This is consistent with the experimental data on H-SNO \cite{Shi2013,Shi2014}. 
	This implies that the monoclinic-orthorhombic structural phase transition may occur upon doping, however 
	no discernible bond disproportionation was observed in our diffraction data. 
	Furthermore, the doping ratio required to open a band gap (1H:1Ni) results in an orthorhombic-like configuration in both phases. 
	Consequently, we primarily focused on the orthorhombic phase for this study.

		To investigate the influence of the dopant concentration on the electronic structure and electrical properties of H-SNO, we performed the simulations for pristine orthorhombic  SNO and four concentration with the 1H:4Ni, 1H:2Ni, 3H:4Ni and 1H:1Ni atomic ratios.
	This was done by placing a single H-dopant in the $2\times2\times2$ supercell, $2\times1\times2$ supercell, conventional unit cell and primitive unit cell, respectively.
	
	The H-dopant inside the SNO unit cell attaches to the oxygen with a bond length of $\approx1$~\text{\AA}, consistent with the expected bond length of OH$^{-}$.
	There are two nonequivalent oxygen ions to which the H-dopant might be chemically bonded: the basal oxygen in the Ni-O (001) plane and the apical oxygen in the Sm-O (001) plane.
	Our simulations show that the bonding with the basal oxygen is energetically more favorable by 0.13~eV.
	This is in agreement with published results \cite{Kotiuga2019, Yoo2018} showing that, in the stable H-SNO configurations, H-dopant is bonded to the oxygen in the Ni-O planes and occupies the void between the rare-earth cations as shown in Fig.~\ref{Setup}(a). 
	
	The integrated spin density analysis from the DFT calculations revealed the gradual reduction of the nickel valence from Ni$^{3+}$ in the pristine SNO to Ni$^{2+}$ in the H-SNO at the highest doping level \cite{SupplNote}. 
	At the doping level of 1H:1Ni, the band gap of $E_G=1.14$~eV opens in H-SNO, indicating the very high resistivity of the hydrogenated H-SNO \cite{Shi2014, Kotiuga2019}, while no band gap is observed at lower doping level (3H:4Ni and below \cite{SupplNote}).
	
	The DFT simulations were performed in two regimes: with the fixed values of the lattice parameters and with volume relaxation.
	In the latter case we observed a $\sim10\%$ increase in the value of $b_o$ lattice parameter under a 1H:2Ni dopant ratio, in line with the previous computational results~\cite{Yoo2020}.
	However, the reduction of nickel oxidation remains consistent with and without volume relaxation. 
	As such, we simulated the XAS spectra of H-SNO with a fixed lattice to maintain consistency with our diffraction results whereby the plane strain in SNO is confined by the interface with the LAO substrate.
	The theoretical XAS spectra for SNO and H-SNO calculated with the FEFF package \cite{Rehr2010} are shown in Fig.~\ref{XAS}(b) and exhibit the same shift $\Delta E_{DFT}\approx$1.05~eV as was observed in the experiment. 

	Using the results of the DFT simulations as an input for FEFF software~\cite{Rehr2010}, we calculated the XAS spectra of the Ni K-edge for various doping level (Fig.~\ref{XAS}(b)) and observed a gradual shift of the absorption edge with the dopant concentration.
	Our simulation confirms that this shift increases with doping concentration as shown in Fig.~\ref{XAS}(c).
		This allowed us to estimate the doping level to be at the atomic ratio of about 1H:1Ni in our experiment.
		Furthermore, the shifting of the absorption peak to a lower energy in H-SNO (Fig.~\ref{XAS}(b)) is qualitatively consistent with the shift in the XAS spectrum for \ce{NiO} (valence state of Ni$^{2+}$) to a lower energy with respect to \ce{Ni2O3} (valence state of Ni$^{3+}$) \cite{SupplNote}.
		Moreover, our DFT simulations indicated that only this high doping level leads to opening of the band gap, while at lower concentrations the nickelate film remains conductive.
		This means that the bright region next to the Pd electrode (Fig.~\ref{XAS}(a)) with the doping level close to 1H:1Ni ratio corresponds to the insulating phase which determine the electrical properties of the nickelate nanodevice \cite{Zhang2020a, Goteti2021}.

	\subsection{Changes in SmNiO$_3$ crystal structure}
	
\begin{figure}
		\includegraphics[width = 0.7\linewidth]{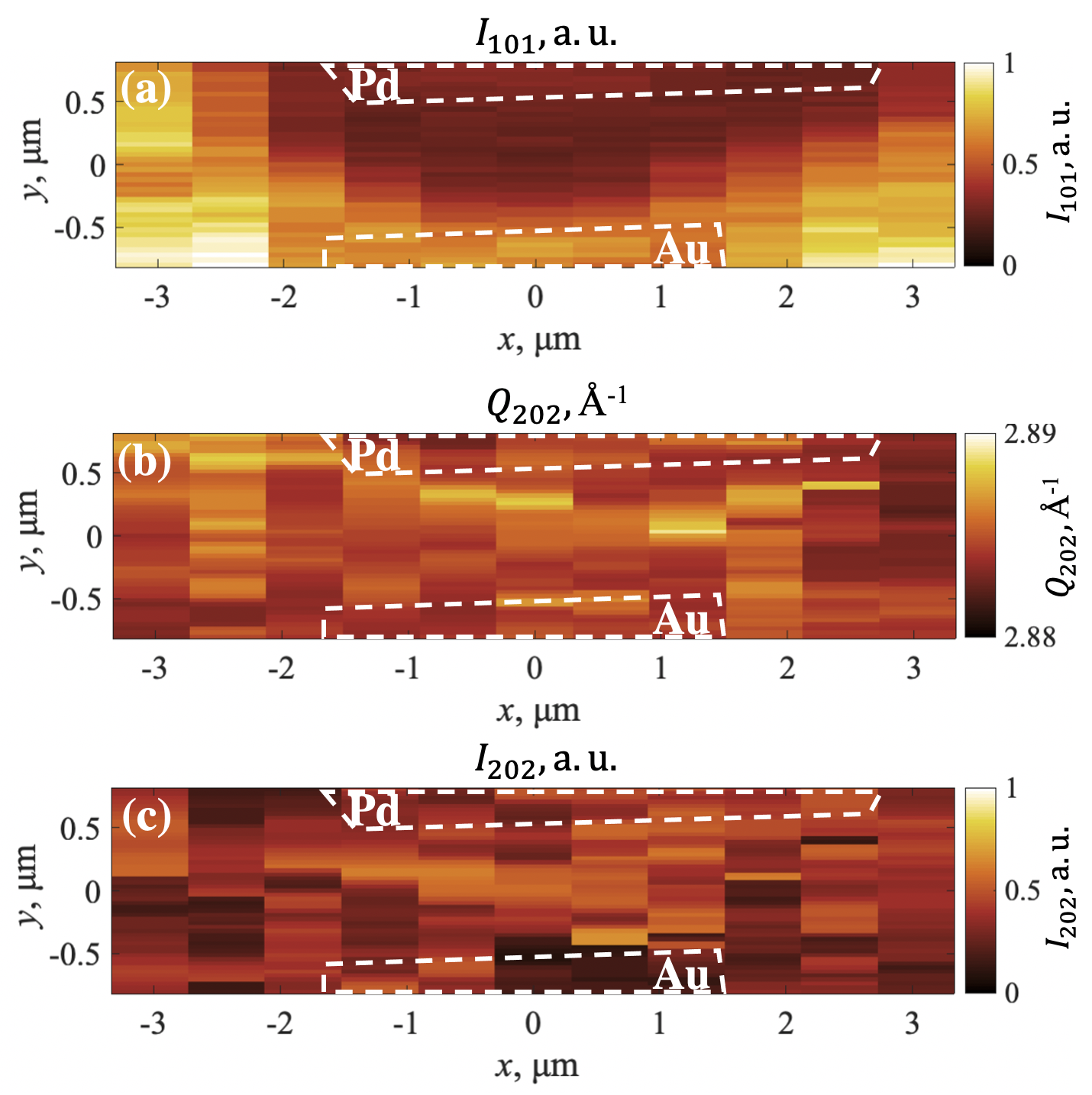}
		\caption{\label{Diffraction}
			(a) Normalized intensity map of the (101) reflection. The white dashed lines outline Pd and Au electrodes. The dark region between the electrodes corresponds to the area where H$^{+}$ doping results in structural changes in the film.
			(b) Spatially-resolved map of the $Q$-position (\AA$^{-1}$) of the (202) reflection.
		    (c) Normalized intensity map of the (202) reflection.
	}
	\end{figure}

	The changes of the crystal structure in the nanodevice upon H$^{+}$ doping can be studied by considering the spatially resolved maps of the diffraction signal.
	In Fig.~\ref{Diffraction}(a) the intensity of the (101) reflection is shown measured on a second device.
	The striking feature of this map is the dark region below the Pd electrode which extends over more than $1~\si{\micro}\text{m}$ towards the Au electrode.
	The intensity in this region is approximately four times smaller than under the Au electrode.
	However, the $Q$-position of the (101)  and (202) reflections changes over less than 0.35~\%  across the device, indicating that the lattice parameters did not change for the doping levels studied here (Fig.~\ref{Diffraction}(b)).
	Also no gradual changes in the doped region can be seen in Fig.~\ref{Diffraction}(c), where the intensity of (202) reflection is shown. 
	Our combined experimental and simulation results strongly suggest a weaker role of the lattice in generating the changes in resistance \cite{Chen2019}.
	
	To demonstrate that a little rearrangement of the atoms within the unit cell can cause a change of the diffraction peak intensity, we calculated the x-ray diffraction from a perovskite structure.
	The intensity of a Bragg reflection is proportional to the squared modulus of the form factor
	\begin{equation}
		\label{DiffractionEq}
		I(\textbf{q})\propto|F(\textbf{q})|^2=\Big\vert \sum_{j} O_j f_j(q)\exp{(i\textbf{q}\textbf{r}_j)}\Big\vert^2 \ ,
	\end{equation}
	where $\textbf{q}$ is a scattering vector, index $j$ numerates atoms in the unit cell (see Fig.~\ref{Setup}(a), $O_j$ is occupancy, $f_j(q)$ is atomic form factor, and $\textbf{r}_j$ is the position of each atom.
	In the ideal perovskite structure (i.e. without tilts and distortions of the NiO$_6$ octahedra as well as displacement of the rare-earth cation), the orthorhombic (101) reflection is forbidden, i.e., the contribution from different atoms in Equation~(\ref{DiffractionEq}) cancel out.
	The deviations from the ideal structure in pristine SNO that result in the non-zero intensity of the (101) reflection are shown in Fig.~\ref{Setup}(b)-(d) and  include $\text{NiO}_{6}$ octahedra rotations \cite{Glazer1972}, $\text{Ni-O}$ bond disproportionation (breathing of the $\text{NiO}_{6}$ octahedra) \cite{Alonso1999,Green2016,Serrano2019}, and displacement of the rare-earth cations \cite{Benedek2013}.
	
	In order to understand the individual contributions to the intensity of the (101) reflection, we considered a model of the SNO structure \cite{SupplNote}.
	The oxygen ions form ideal octahedra around the nickel ions where the length of the $\text{Ni-O}$ bond is $d_{\text{Ni}-\text{0}}=d_{\text{Ni}-\text{Ni}}(1\pm B)/2$; here $d_{\text{Ni}-\text{Ni}}$ is the distance between two $\text{Ni}^{3+}$ ions and $B\approx1.3\%$ is the magnitude of the breathing mode \cite{Varignon2017,Green2016}.
	The alternating expanded and contracted octahedra form a three-dimensional checkerboard pattern \cite{Catalano2018, Green2016}. 
	Furthermore, each $\text{NiO}_{6}$ octahedron is rotated about the [100], [010], and [001] directions in the orthorhombic unit cell \cite{NotationNote} over angles $\alpha$, $\beta$, and $\gamma$ to form the tilt pattern inherent to rare-earth nickelates \cite{Glazer1972}.
	In this orthorhombic notation, the rotations $\alpha= 0\degree$, $\beta=15.26\degree$, and $\gamma=7.9\degree$ correspond to the reported values of the Ni-O-Ni angles $\theta_{ap}=149.5\degree$ and $\theta_{b}=154.2\degree$
	for the apical and basal oxygen atoms, respectively \cite{Jain2013,Fowlie2019}. 
	The displacement of $\text{Sm}^{3+}$ ions from the symmetric position between the $\text{NiO}_{6}$ octahedra was described by three parameters, $d_x$, $d_y$, and $d_z$, corresponding to the shift along the [100], [010], and [001] orthorhombic directions (Fig.~\ref{Setup}(a)). 
	In pristine SNO, the values of $d_x$ and $d_y$ are reported to be 0.06~\AA~and 0.28~\AA, while $d_z$ equals zero \cite{Jain2013,Brahlek2017}.
	
					\begin{figure*}
		\includegraphics[width =1 \linewidth]{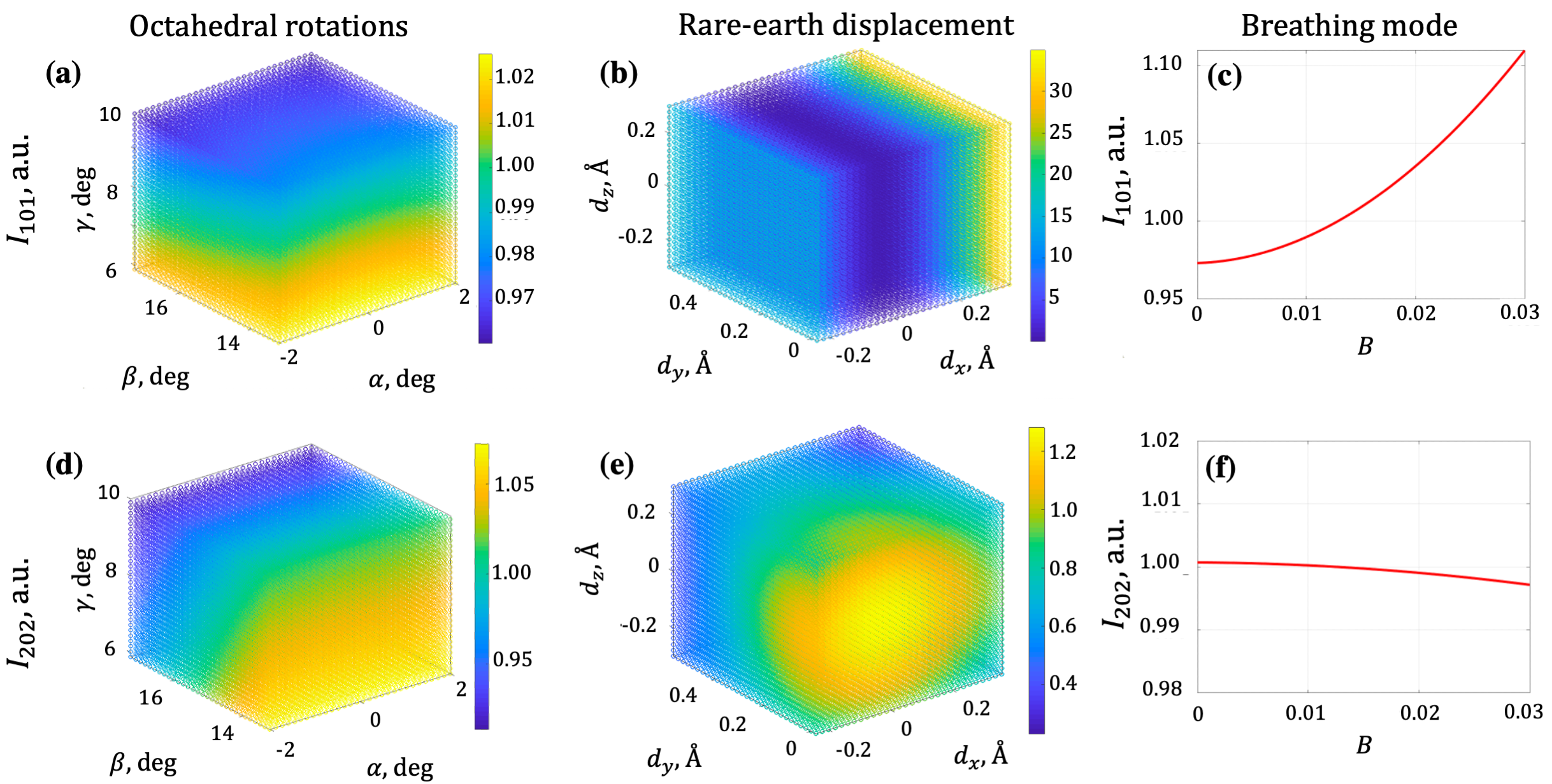}
		\caption{\label{Parameters} 
			(a-c) Simulated intensity the (101) reflection and (d-f )intensity of the (202) reflection as a function of NiO$_{6}$ octahedra tilt angles $\alpha$, $\beta$, $\gamma$  (a,d), the displacement of the Sm$^{3+}$ cation $d_x$,$d_y$,$d_z$ (b,e), and the breathing distortion $B$ (c,f). The intensities $I_{101}$ and $I_{202}$ are normalized to the value in the pristine SNO.	
	}
	\end{figure*}
	
	Because of the twinning ($a_o\approx b_o$), the intensity measurement of the (101) reflection accounts for the contribution from $(101)$, $(10\bar{1})$, $(011)$, and $(01\bar{1})$ \cite{Brahlek2017}.
	Therefore, in our diffraction simulations we calculated the averaged intensity of these four reflections; and the similar procedure was performed for the (202) peak. 
	We varied the values of the above-listed distortions of the orthorhombic crystal lattice around the values reported for pristine SNO by the following amounts: $\Delta\alpha,\Delta\beta,\Delta\gamma=\pm2^\circ$, $B=0-3\%$, and $\Delta d_x,\Delta d_y,\Delta d_z=\pm0.3~$\AA. 
	Our calculations show that the averaged intensity of the (101) reflection is determined mainly by the displacement of the rare-earth cation, while the combined contribution from the breathing mode and tilt pattern constitutes only $~\sim5\%$ of the total intensity  (Fig.~\ref{Parameters}(a)-(c)). 	
	This can be understood since the two latter effects include movement of oxygen atoms that are weak x-ray scatters ($|f_{O^{2-}}|/|f_{Sm^{3+}}|\sim0.1$ at $E=8345$~eV); however, strictly speaking, this argument is valid only for the (101) reflection \cite{Brahlek2017}.
	The intensity of the (202) reflection almost does not change upon considered deviations from the ideal perovskite structure (Fig.~\ref{Parameters}(d)-(f)).

\begin{figure}
		\includegraphics[width = 0.7\linewidth]{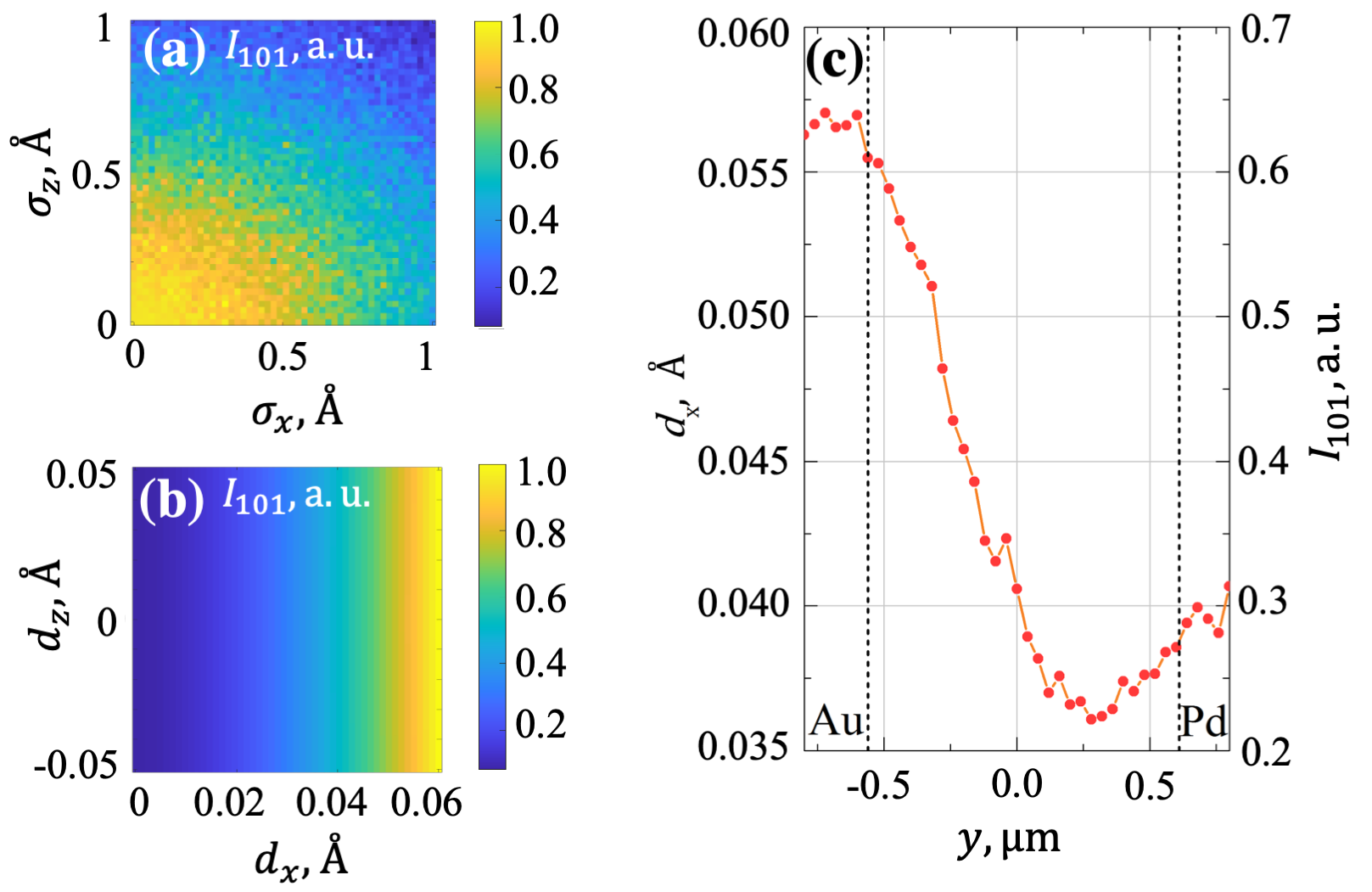}
		\caption{\label{Diffraction_model}
			(a-b) Dependence of the (101) peak intensity on the root-mean-square displacement (a) and mean position of the Sm$^{3+}$ ions (b). The intensity was evaluated by fixing the positions of all ions in the SNO unit cell and allowing only Sm$^{3+}$ to move. For each value of the parameters the result was statistically averaged over $10^4$ realizations. (c) Mean shift of the Sm$^{3+}$ ions and the corresponding change of intensity along the $x=0$ line in the spatially resolved map in Fig.~\ref{Diffraction}(a) using Equation~\ref{DiffractionEq2}. Vertical lines mark Au and Pd electrodes.}
	\end{figure}

	Neglecting the contribution from oxygen ions in Equation~(\ref{DiffractionEq}), the intensity of the averaged (101) reflection depends only on the cation displacements along [100] and [001] directions \cite{SupplNote}
	\begin{equation}
		\label{DiffractionEq2}
		I_{101}\propto\exp\bigg[-\Big(\frac{2\pi\sigma_x}{a_o}\Big)^2-\Big(\frac{2\pi\sigma_z}{c_o}\Big)^2\bigg] \sin^2 \frac{2\pi d_x}{a_o}\cos^2 \frac{2\pi d_z}{c_o}.
	\end{equation}
	Here $\sigma_x$ and $\sigma_z$ are the root-mean-square displacement of the $\text{Sm}^{3+}$ ions along the [100] and [001] directions from the equilibrium positions dictated by $d_x$ and $d_z$.
	In pristine SNO, $\sigma_x\sim\sigma_z\sim0.05$~\AA.
	The dependence of the (101) reflection intensity on $\sigma_x$, $\sigma_z$, $d_x$, and $d_z$ is shown in Fig.~\ref{Diffraction_model}(a)-(b), to illustrate the combined impact of each parameter in Equation~(\ref{DiffractionEq2}). 
	The strongest decrease of the (101) reflection intensity is caused by the  shift of the $\text{Sm}^{3+}$ cations towards the symmetric position along the [100] direction (i.e., $d_x\rightarrow0$). 
	Meanwhile, in our calculations the intensity and the position of the (202) reflection remains practically constant, which coincides with the experimental x-ray data (Fig.~\ref{Diffraction}(b)-(c)).
	
	The one-to-one correspondence also allows one to directly map the intensity of the (101) reflection into the displacement of the rare-earth cation.
	In Fig.~\ref{Diffraction_model}(c), the intensity of the (101) is plotted with the corresponding variation of the displacement parameter $d_x$ shown along the $x=0$ line cut of the intensity map (Fig.~\ref{Diffraction}(a)).
	The proportionality coefficient between intensity and $d_x$ was determined by assigning the maximum intensity of the (101) peak below the Au electrode to the literature values of SNO ($d_x\approx0.06$~\AA). 
	The outstanding feature of this plot is a non-monotonic change of the $d_x$ parameter between the electrodes: the minimum value of $d_x$ is reached approximately 250~nm away from the Pd electrode.  
	This suggests that the diffusion coefficient of H$^{+}$ in SNO may depend on H$^{+}$ concentration, which leads to accumulation and stagnation of dopants next to the catalytic Pd electrode.
	This also shows that H$^{+}$ doping can affect the structure of SNO up to $1~\si{\micro}\text{m}$ away from the Pd electrode, where the protons were initially introduced.
	This information is important for optimization of the electrode shape and scaling of the device and requires a dedicated investigation. 
	Finally, our results are in agreement with own DFT calculations and published works \cite{Kotiuga2019, Yoo2018} which predict that the H-dopant occupies the void next to the rare-earth cation and can cause its displacement 
		sufficient to decrease the intensity of the (101) reflection. 
	
		\section{Discussion}
		
	At first glance, the shift of the rare-earth cation towards the centrosymmetrical position in the unit cell and the decrease of the oxygen octahedra rotations seem to be a counter-intuitive result of the H$^{+}$ implantation, because these structural changes actually make the SNO crystal lattice closer to the ideal perovskite structure.
	On the other hand, the rare-earth nickelates with large rare-earth cations are known to have less distorted crystal structures (i.e., exhibit smaller octahedral tilts), than the nickelates with small cations \cite{Catalano2018}.
	Therefore, implantation of H$^{+}$ next to the rare-earth cation may effectively increase the radius of the latter, which would result in a decrease of the NiO$_6$ tilt angles, causing the system to be more metallic.
	We should also note that, even with our spatial resolution, our probe measures an average structure across the beam footprint. 
	This means that the implanted H$^{+}$ can produce local lattice defects and distortions within a unit cell that cannot be measured by x-ray diffraction (as it is shown in Fig.~\ref{Diffraction_model}(a)).
	A hint that these local distortions are averaged out in diffraction can be seen in the slight broadening of the H-SNO absorption spectra as compared to the pristine SNO (Fig.~\ref{XAS}(b)).
	
	Together, our data suggest that the dramatic changes in resistance upon hydrogenation come from the change in electronic valency, not from structural distortions.
	This is also supported by further DFT simulations, in which we considered separately the structural and electronic effects of H-doping \cite{SupplNote}.
	More specifically, we added an H atom to the SNO and allowed the structure to relax under the following constrains: 
	a) only H doping, no crystal structure relaxation (positions of Sm, Ni and O are fixed);
	b) only structural changes (positions of Sm, Ni and O are as in the H-SNO), but no H doping;
	c) H doping and Sm relaxation (only Sm atoms were allowed to move);
	d) H doping and NiO$_6$ relaxation (only O atoms were allowed to move).
	In result, we found that the pure structural changes do not lead to the band gap opening, while just the H-doping without the structural relaxation lead to the formation of the band gap $E_G\approx$0.5~eV.
	In addition, the relaxation of the  NiO$_6$ octahedra increases the band gap to $E_G\approx$1.0~eV.
	In all cases, the opening of the band gap is accompanied by the reduction of nickel from Ni$^{3+}$ to Ni$^{2+}$ \cite{SupplNote}.
	These simulations confirms that the origin of the insulating properties of H-SNO lies in electron doping, while the structural changes play only the second role. 
	
	These simulations show that mere structural changes do not cause a band gap opening, while only change in nickel valence (without structural relaxation) already causes the opening of a band gap.
	Therefore, the insulating properties of H-SNO arise primary from the change of Ni valence, while subtle changes in the crystal structure (mostly within the \ce{NiO6} octahedra) further increase the band gap to its final value.

	In the diffraction, we are mostly sensitive to the displacement of the Sm$^{3+}$ cations, which we indeed observed in our experiment (Fig.~\ref{Diffraction}(a)), while other structural changes could not be directly detected in our data.
	This gives us only indirect information on the changes in SNO crystal lattice.
	For example, the rare-earth cation displacement is known to be coupled with the tilt of the \ce{NiO6} octahedra \cite{Benedek2013}, so  possibly a change in the \ce{NiO6} rotation also takes place with hydrogenation.
	However, we anticipate that the rotation angles of the NiO$_6$ octahedra are only slightly decreased with hydrogenation, based on the small value of Sm$^{3+}$ displacement that we observe experimentally.
	This further suggests that hydrogenating other transition metal oxides with different crystal structures might also yield changes in resistance.
	
	In summary, we used a nanofocused x-ray beam to perform spatially resolved spectroscopic and structural studies of a H-SNO-based nanodevice.
	As a result of H$^{+}$ doping, we observed the $1.2$~eV shift of the Ni K-edge towards lower energies.
	Our DFT simulations account for this shift by estimating a reduction of the nickel valence from Ni$^{3+}$ to Ni$^{2+}$ at the 1H:1Ni doping level.
	We also observed the decrease of the (101) Bragg peak intensity next to the Pd electrode.
	Structural modeling revealed that this can be caused by a shift of the Sm$^{3+}$ cations over $\sim0.03~\text{\AA}$ towards the centrosymmetrical position inside the voids between the NiO$_6$ octahedra. 
	This displacement of rare earth cations is usually neglected, when the unit cell of the SNO is approximated by pseudocubic structure.
	Combining the x-ray spectroscopy and diffraction data, we elucidate how  H$^{+}$ doping changes the structure and electronic properties of an SNO device. 
	Our methods also pave the way for future x-ray nanoscale studies of devices based on transition metal oxides. 
	
	\section{Methods}
	\subsection{Sample preparation}
	The SNO-based nanodevices studied here consist of a 150~nm thick SNO film epitaxially grown with a high vacuum sputtering system on $(111)_{pc}$-oriented $\text{LaAlO}_{3}$ substrate \cite{Zhang2020a}. 
	This orientation was chosen such as to be able to access Bragg peaks associated with octahedral distortions of the kind (101) \cite{Fowlie2019,Brahlek2017,Hepting2017}.
	Two $5~\si{\micro}\text{m}$-wide electrodes of Pd and Au 
	were fabricated over the film with a lateral gap of $1-2~\si{\micro}\text{m}$ using e-beam lithography.
	The thickness of the electrodes next to the gap was 50~nm. 
	The Pd electrode served as a catalyst to split \ce{H2} molecules and incorporate H$^{+}$ in the SNO film during annealing for 5 minutes at 120 $^{\circ}$C in the H$_2$/N$_2$ mixture (5\%/95\%) \cite{Zhang2020a,Ramadoss2018,Shi2014,Zhou2016}.
	
	\subsection{X-ray nanofocusing experiments}
	The nanofocused x-ray experiments were performed at the 26-ID-C beamline of the Advanced Photon Source (APS) and 3-ID beamline of the National Synchrotron Light Source II (NSLS II) \cite{Yan2018,Nazaretski2017}, using photons with energy of 8315-8385~eV to measure the fluorescence around the Ni K-edge near 8345~eV at room temperature \cite{SupplNote}. 
	At this energy, the resolution was about 0.5~eV achieved by a double-crystal monochromator. 
	The XAS data were taken using a fluorescence detector placed above the device \cite{Pattammattel2020}. 
	The diffraction data were acquired with a two-dimensional photon counting detector with $55~\si{\micro}\text{m}$ pixels oriented perpendicular to the diffracted beam. 
	A Fresnel zone plate was placed upstream from the sample to focus the beam down to $\sim$30~nm at the sample. 
	To collect the spatially resolved diffraction and fluorescence data, the sample was raster-scanned with the nanofocused x-ray beam.
	The Au and Pd contacts served as fiducials to easily locate the device using their fluorescence.
	
	\subsection{DFT simulations}
	
	All DFT calculations \cite{Kresse1996a, Kohn1965} were performed using the Vienna Ab initio Simulation package (VASP) \cite{Kresse1996} within the projector augmented wave (PAW) approach \cite{Blochl1994,SupplNote}. 
	The exchange-correlation effects were modeled using the Perdew-Berke-Ernzerhof (PBE) generalized gradient approximation (GGA) \cite{Perdew1996} functional under the Hubbard correction with $U=2$ eV~\cite{Yoo2018,Anisimov1991}. 
	We used a plane wave cutoff energy of 520~eV with the energies and atomic forces converged to within $10^{-4}$~eV and -0.02 eV/\AA~respectively.
	All analysis and input generation was performed with the aid of the Python Materials Genomics (pymatgen) package \cite{Ong2013}.  
	We performed full relaxation for the pristine and H-doped structures of \ce{SmNiO3} and all calculations were spin-polarized. 
	
	We modeled the pristine and H-doped \ce{SmNiO3} system using the metallic orthorhombic ($Pbnm$) phase. \ce{SmNiO3} is known to be paramagnetic \cite{Catalano2018}, however DFT is only able to simulate magnetically ordered phases (e.g., ferromagnetic (FM) and antiferromagnetic (AFM)) or non-magnetic (NM) phases. As such we investigated both the FM and AFM phases.

	\begin{acknowledgments}

		This work was supported as part of Quantum Materials for Energy Efficient Neuromorphic Computing (Q-MEEN-C), an Energy Frontier Research Center funded by the U.S. Department of Energy (DOE), Office of Science, Basic Energy Sciences (BES), under Award \# DE-SC0019273. X-ray microscopy measurements were supported by the DOE, Office of Science, BES, under Contract \# DE-SC0001805. 
		Research at the Center for Nanoscale Materials and the Advanced Photon Source, both Office of Science user facilities, was supported by the DOE, Office of Science, BES, under contract \# DE-AC02-06CH11357. Research at the Hard X-ray Nanoprobe Beamline of sector 3-ID of the National Synchrotron Light Source II was supported by the DOE Office of Science under contract \# DE-SC0012704 and is operated by Brookhaven National Laboratory.
		
	\end{acknowledgments}

   \newpage
   
   \renewcommand\appendixname{Supplemental Materials}
   \appendix*
   
   	\renewcommand{\thefigure}{S\arabic{figure}}
   \setcounter{figure}{0}
   \renewcommand{\theequation}{S\arabic{equation}}
   \setcounter{equation}{0}
    
	\section{SNO device fabrication and characterization}

\begin{figure}[b]
	\begin{center}
		\includegraphics[width = 0.95\linewidth]{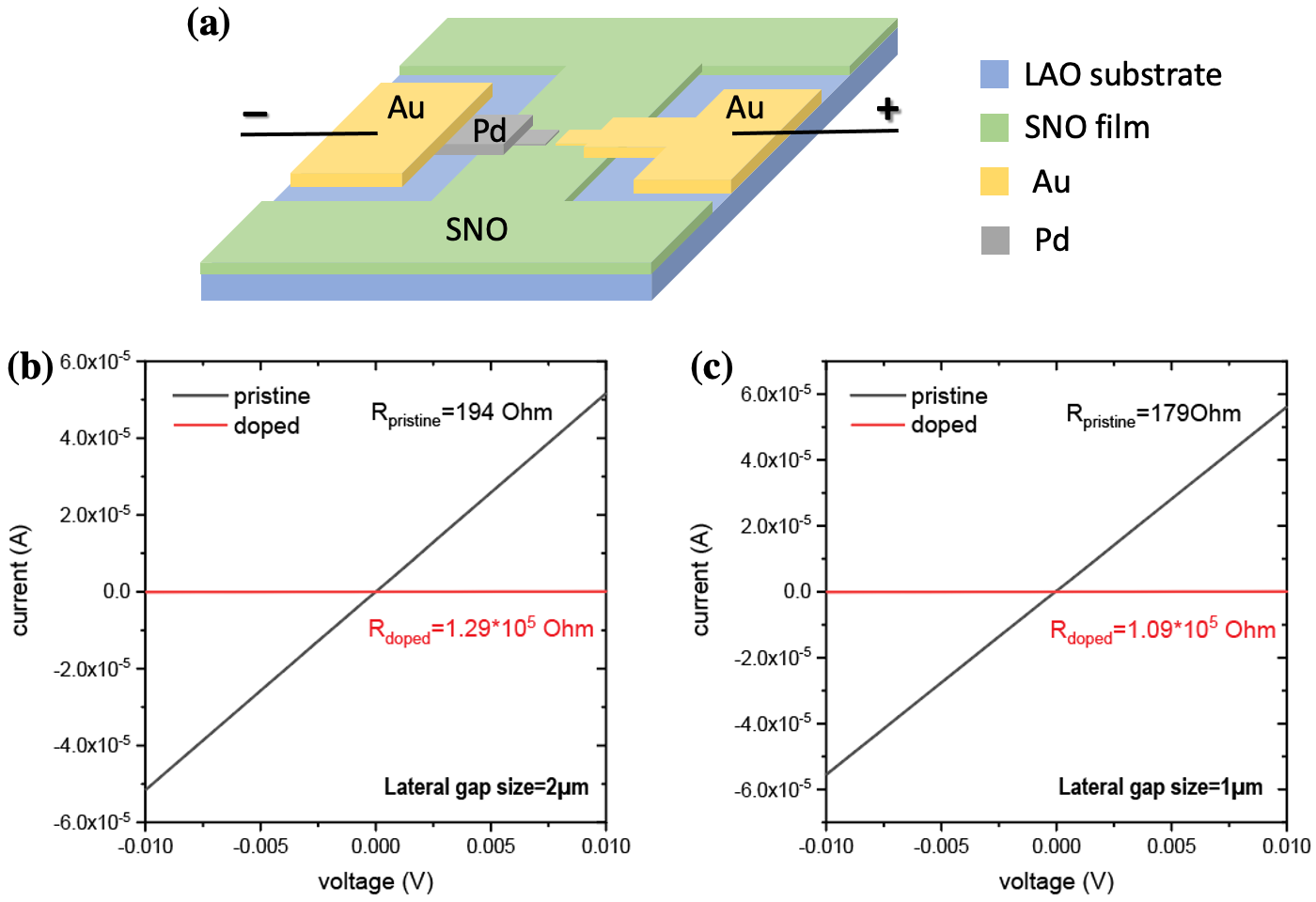}
		\caption{\label{device}
			(a) Scheme of the fabricated SNO device. 
			(b-c) Representative data taken from two different SNO devices showing resistance modulation due to hydrogen doping. (a) Current (I) - voltage (V) characterization between -10mV to +10mV collected from a device with a $2~\si{\micro}\text{m}$ lateral gap size. The fitted slopes indicated the resistance before doping (pristine) was ~194 Ohm and after doping was ~129 kOhm. (b) Current (I) - voltage (V) characterization between -10mV to +10mV collected from a device with a $1~\si{\micro}\text{m}$ lateral gap size. The fitted slopes indicated the resistance before doping (pristine) was ~179 Ohm and after doping was ~109 kOhm. The resistance increases in both cases after hydrogen doping due to the electron transfer from the hydrogen to the nickel orbitals, opening up a transport gap. The device structure and fabrication process are detailed in the main text. The samples were annealed in H$_2$/Ar (5\%/95\%) forming gas at 120 $^\circ$C for 10 minutes for the doping process.
		}
	\end{center}
\end{figure}

\quad The devices studied here consist of a 150~nm SNO film on LAO substrate with the two 50~nm thick contacts of Au and Pd fabricated on top of the SNO film.
The geometry of the device is schematically shown in Figure  \textbf{Figure~\ref{device}a}.
In this work, we used two types of devices with a $1~\si{\micro}\text{m}$ and $2~\si{\micro}\text{m}$ gap between thee Au and Pd contacts.
Both devices have similar electrical properties and exhibit a colossal increase in resistance, as it is shown in \textbf{Figure~\ref{device}b,c}.   

\section{Spatially resolved diffraction studies}

\begin{figure}[b]
	\begin{center}
		\includegraphics[width = 0.75\linewidth]{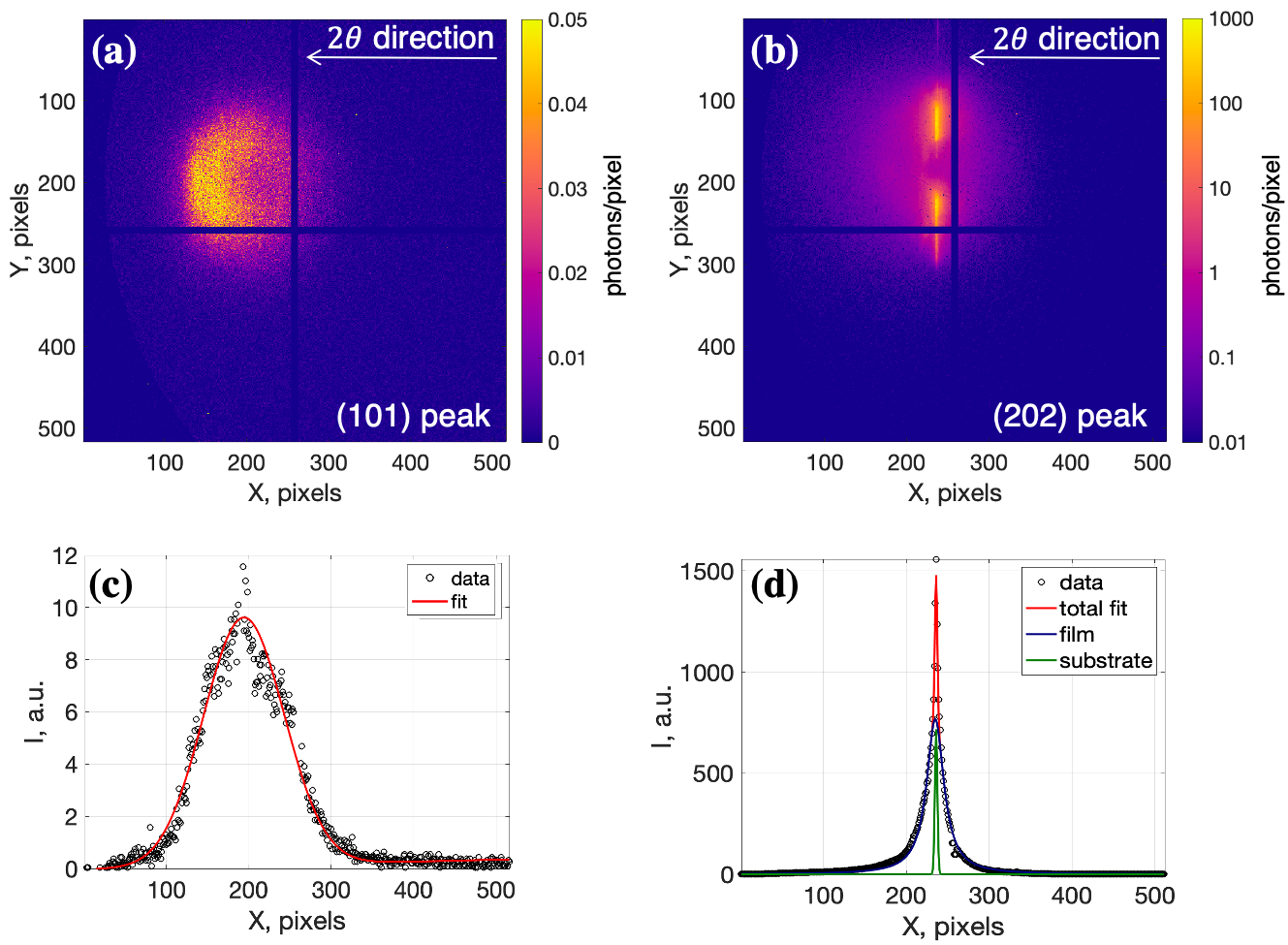}
		\caption{\label{Diffraction_pattern} 
			Averaged diffraction pattern for the (101) reflection (a) and the (202) reflection (b). Projection of the averaged diffraction pattern along the vertical direction is shown in (c,d). The signal from the (101) reflection can be fitted by a single Gaussian function (c), while the signal from the (202) reflection can be approximated by a sum of the Lorentzian and Gaussian functions corresponding to the scattering from the SNO film and the LAO substrate, respectively (d).
		}
	\end{center}
\end{figure}

\quad The SNO-based memory device was scanned with a nanofocused x-ray beam with an exposure time of 10~s for (101) peak and 1~s  for (202) peak.
A Fresnel zone plate was placed upstream from the sample to focus the beam down to $\sim$30~nm at the sample. 
The footprint of the x-ray beam for the (101) peak was approximately $170\times30$~nm$^2$ (hor. $\times$ vert.), and for the (202) peak -- $90\times30$~nm$^2$.
At the 26-ID-C beamline (APS) the zone plate optics was moved in order to ensure stability of the sample during the scan.
At the 3-ID beamline (NSLS-II) the sample was moved with respect to the x-ray beam, while the stability was ensured through the high stiffness design of the microscope sample stage.
The XAS data were taken using a fluorescence detector placed above the device \cite{Pattammattel2020}. 
The diffraction data were acquired with a two-dimensional photon counting detector with $55~\si{\micro}\text{m}$ pixels oriented perpendicular to the diffracted beam. 
The averaged diffraction patterns are shown in \textbf{Figure~\ref{Diffraction_pattern}a,b}.
The scattering angle $2\theta$ increases from right to left and the dispersion along the vertical direction is due to the Fresnel zone plate used for focusing.
The projected signal along the $2\theta$-direction is shown in \textbf{Figure~\ref{Diffraction_pattern}c,d.}

\begin{figure}
	\begin{center}
		\includegraphics[width = 0.9\linewidth]{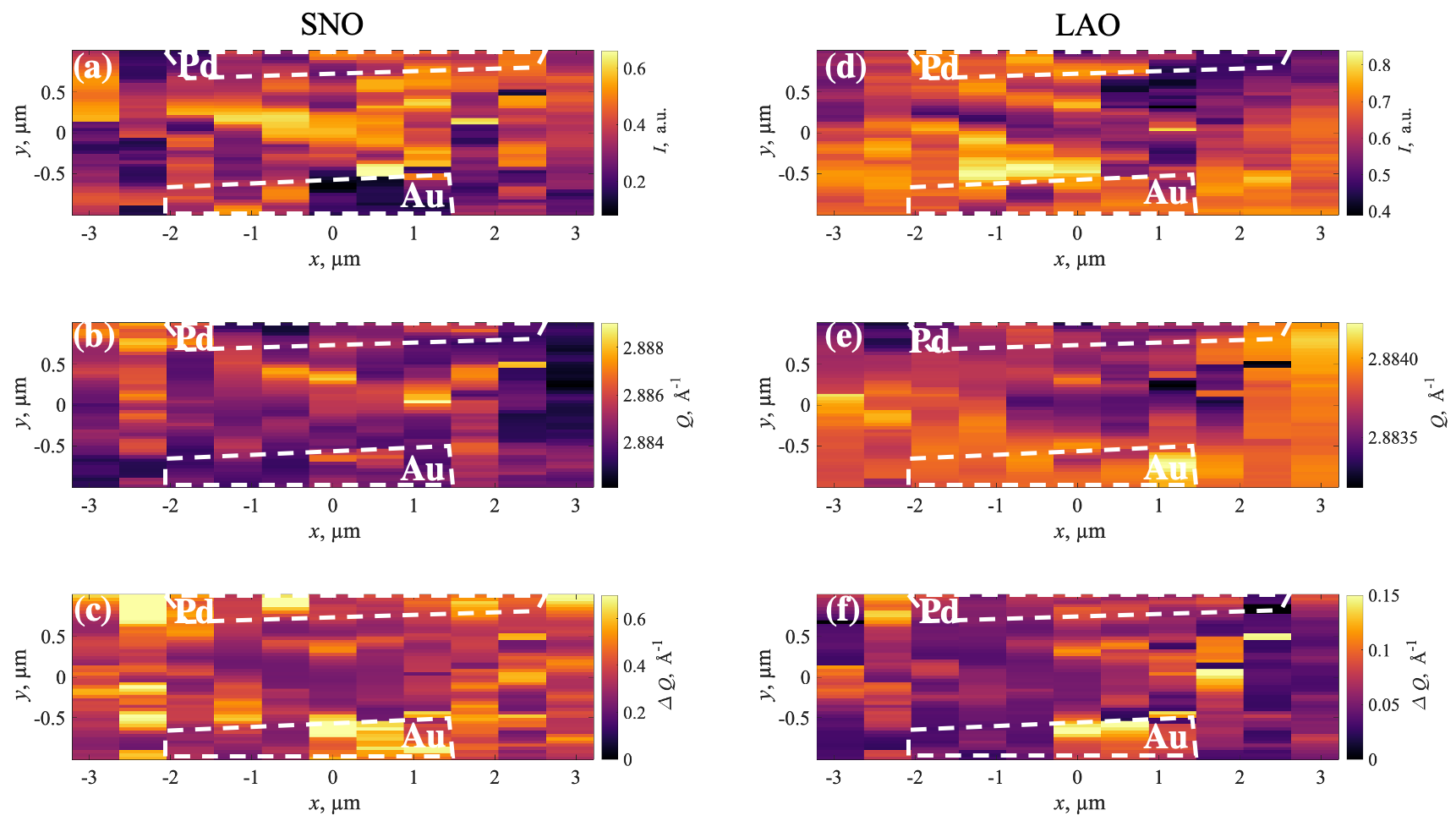}
		\caption{\label{Spatial_maps} 
			(a-c) Intensity (a), $Q$-position (b), and the width (c) of the (202) reflection from the SNO film.
			(d-f) Intensity (d), $Q$-position (e), and the width (f) of the LAO (111) substrate reflection. The electrodes are outlined with dashed lines.
		}
	\end{center}
\end{figure}

In \textbf{Figure~\ref{Spatial_maps}} spatially resolved maps of the intensity, position and width of the (202) reflection from the  \ce{SmNiO3} (SNO) film and the $(111)_{pc}$ reflection from the \ce{LaAlO3} (LAO) substrate are shown.
These maps show no changes of the unit cell parameters induced by H$^{+}$ doping. In contrast to that, the intensity of the (101) reflection is about four times dimmer in the proton-doped region next to the Pd electrode (see \textbf{Figure 3a} in the main text).
The variations of intensity in \textbf{Figure~\ref{Spatial_maps}a,d} are attributed to the heterogeneity of the substrate and almost complete overlap between the film and substrate peaks, which did not allow to reliably separate two signals (see \textbf{Figure~\ref{Diffraction_pattern}c,d}).

\begin{figure*}
	\begin{center}
		\includegraphics[width = \linewidth]{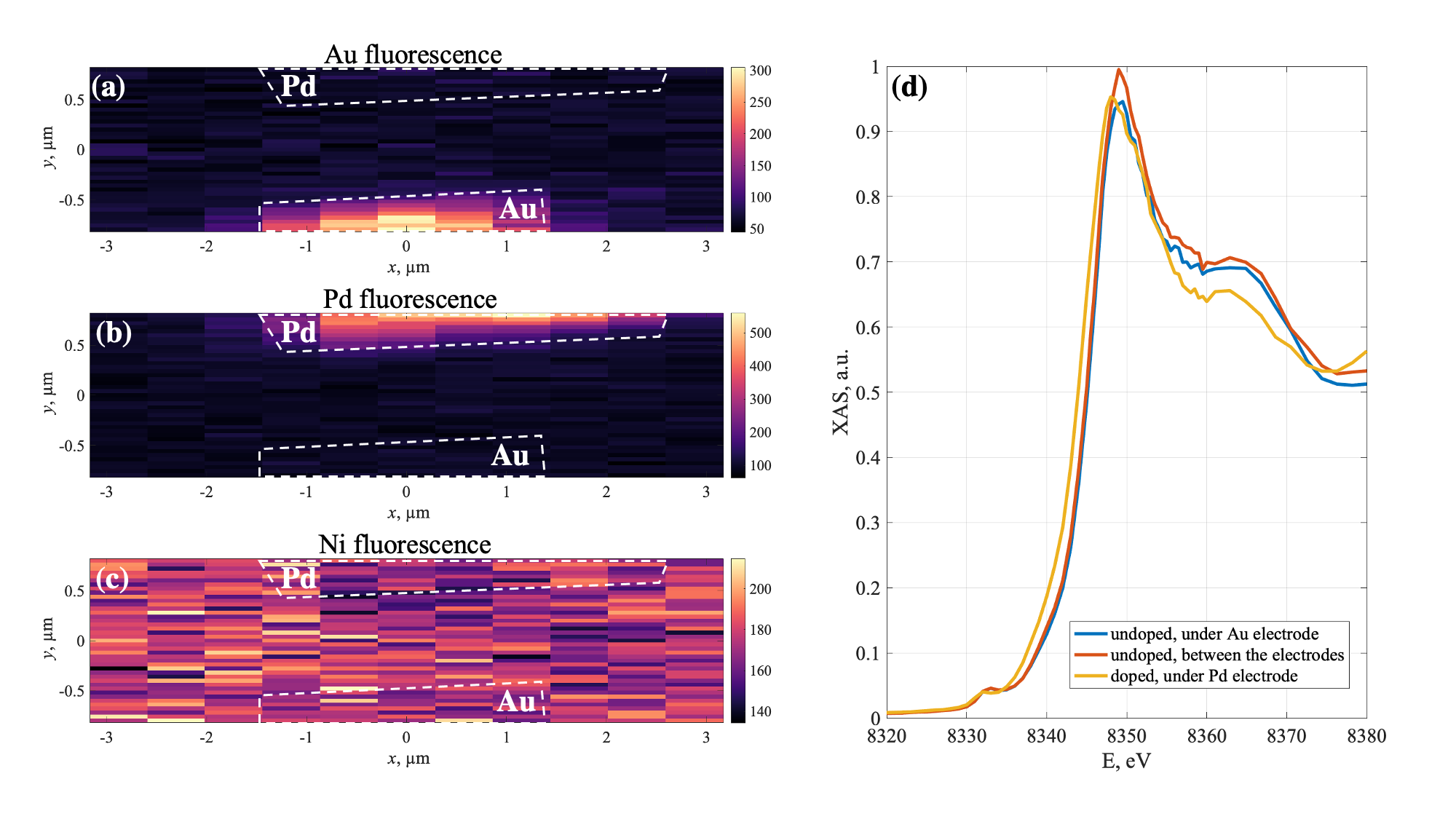}
		\caption{\label{Electrodes} 
			(a-b) Fluorescence signal from Au (a) and Pd (b) indicates position of electrodes. (c) Non-resonant fluorescent signal from Ni shows the uniform SNO film across the device.
			(d) XAS spectra at Ni K-edge taken under the Au and Pd electrodes and in the middle of the gap between the electrodes. 
		}
	\end{center}
\end{figure*}

The position of electrodes was determined using the Au and Pd fluorescence signal, as shown in \textbf{Figure~\ref{Electrodes}a,b}.
The thickness of the electrodes next to the gap was 50~nm. 
The Ni fluorescence map at $E=8395$~eV (away from the K-edge resonance, hence not sensitive to the valence state of nickel) shows that the SNO film is uniform.
A small absorption of the Ni fluorescence signal under the electrodes can be noticeable in \textbf{Figure~\ref{Electrodes}d}, but it has a neglectable effect on the position of the Ni K-edge.

\newpage
\section{Simulation of x-ray diffraction}
\subsection{\ce{SmNiO3} perovskite crystal lattice}

\quad An orthorhombic unit cell of a perovskite SNO lattice was constructed in several steps.
In our analysis, we assumed that Ni atoms do not move with hydrogen doping, so the positions of the twelve Ni atoms was fixed within a unit cell.
Then four Sm atoms were placed inside a unit cell and allowed to be shifted from the centrosymmetric positions over $d_x$, $d_y$, and $d_z$ in such a way, that the orthorhombic \textit{Pbnm} symmetry was preserved.
Finally, we placed six O atoms in the corners of octahedra centered at every Ni atom (twenty oxygen atoms in total, if counting only atoms within a unit cell) and took the tilt and breathing distortion of the \ce{NiO6} octahedra into account.
To achieve that, the position of each oxygen atom was calculated according to the following procedure: 

\begin{enumerate}
	\item For each oxygen atom, a nesting nickel atom was selected, around which the rotations will be performed.
	\item The position of the oxygen atom was calculated as
	\begin{equation}
		\label{Coordinates1}
		\textbf{r}_\text{O}=\textbf{r}_{\text{Ni}}+\tilde{\textbf{u}}
		\ ,
	\end{equation}
	where $\textbf{r}_\text{O}=(x_\text{O},y_\text{O},z_\text{O})$ are coordinates of the oxygen atom and $\textbf{r}_\text{Ni}=(x_\text{Ni},y_\text{Ni},z_\text{Ni})$ are coordinates of the corresponding nickel atom.
	The relative coordinates $\tilde{\textbf{u}}=(\tilde{x},\tilde{y},\tilde{z})$ were calculated as
	\begin{equation}
		\label{Coordinates2}
		\tilde{\textbf{u}}=(1\pm B)\cdot \hat{R}(\pm\alpha,\pm\beta,\pm\gamma)\textbf{u}_0
		\ .
	\end{equation}
	Here $\textbf{u}_0=(x_0, y_0, z_0)$ are the relative coordinates of the oxygen atom in an ideal perovskite structure (in which each oxygen atom is located exactly in the middle between two nickel atoms), $\hat{R}(\alpha,\beta,\gamma)$ is an operator of rotation over the angles $\alpha$, $\beta$ and $\gamma$ about $x$-, $y$- and $z$-axis, correspondingly, and the pre-factor $(1 \pm B)$ takes the breathing distortion with a magnitude $B$ into account (plus sign corresponds to the expanded NiO$_6$ octahedra, and minus sign -- to the contracted).
	\item The signs of rotation angles $\alpha$, $\beta$ and $\gamma$ for each octahedra are selected in such a way that the final structure corresponds to the $a^{-}a^{-}c^{+}$ tilt pattern in Glazer notation \cite{Glazer1972}.
\end{enumerate}

We assigned an occupation number $O_j$ to each atom, which takes into account that the atom might be shared between several neighboring unit cells.
For example, $O_j=1$ for an atom that is completely inside a unit cell, while $O_j=1/4$ for an atom placed on the edge of a unit cell , since this atom is shared between four adjacent unit cells.

\begin{figure}
	\includegraphics[width = \linewidth]{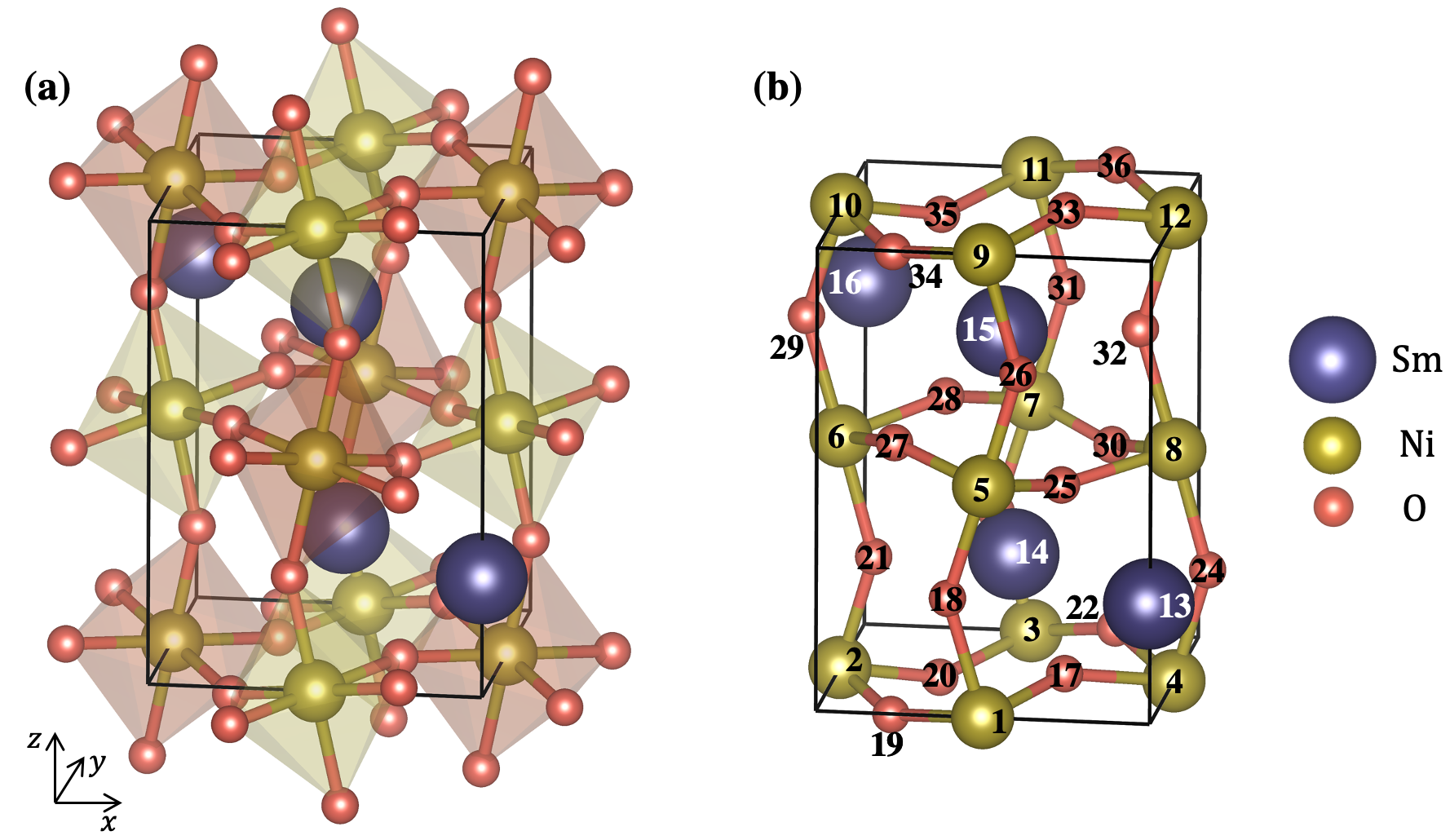}
	\caption{\label{Structure_Compare} 
		(a) Orthorhombic unit cell of \ce{SmNiO3} taken from the literature \cite{Jain2013}. For better visualization, \ce{NiO6} octahedra are shown around each Ni atom. The contracted octahedra are orange, and expanded are yellow (the breathing mode). 
		(b) An identical unit cell of \ce{SmNiO3} consisting of 36 atoms used for simulations of x-ray diffraction (see atomic coordinates in \textbf{Table~\ref{Atomic_coordinates_tab}}). Only the atoms within a unit cell are shown. The numbers corresponds to the index $j$.
	}
\end{figure}

The parameters of the structural model are summarized in \textbf{Table~\ref{Atomic_coordinates_tab}}.
The first column contains an index of an atom $j$, the second column -- type of an atom. Columns 3--5 contains coordinates $x$, $y$ and $z$ of an atom, and column 6 -- occupation number $O_j$. For each Ni atom the signs of rotation angles $\alpha$, $\beta$ and $\gamma$ and the breathing distortion are specified in columns 7 and 8, respectively (this information is duplicated for each oxygen atom).
The values of $\tilde{\textbf{u}}=(\tilde{x},\tilde{y},\tilde{z})$ for each oxygen atom are calculated with \textbf{Equation~\ref{Coordinates2}}, using the undistorted coordinates  $\textbf{u}_0=(x_0, y_0, z_0)$ specified in columns 9--11.
The crystal structure of pristine SNO constructed with our model using the parameter values $\alpha= 0\degree$, $\beta=15.26\degree$, and $\gamma=7.9\degree$, $d_x=0.06$~\AA, $d_y=0.28$~\AA, $d_z=0$~\AA, and $B=0.013$ coincides with the literature data \cite{Jain2013,Alonso1999} (see \textbf{Figure~\ref{Structure_Compare}}).

\begin{table}[]
	\begin{center}
		\begin{tabular}{|l|l|l|l|l|l|c|c|l|l|l|}
			\hline
			\#$j$  & Atom & $x_j$       & $y_j$       & $z_j$       & $O_j$ & Rotation & Breathing & $x_0$& $y_0$& $z_0$ \\ \hline
			1  & Ni   & $a_o/2$ & 0       & 0       & 1/4 & -- -- + & + &  \multicolumn{3}{l|}{}  \\ \cline{1-8}
			2  & Ni   & 0       & $b_o/2$ & 0       & 1/4  & + + -- & -- &  \multicolumn{3}{l|}{}  \\ \cline{1-8}
			3  & Ni   & $a_o/2$ & $b_o$   & 0       & 1/4  & -- -- + & +  &  \multicolumn{3}{l|}{}  \\ \cline{1-8}
			4  & Ni   & $a_o$   & $b_o/2$ & 0       & 1/4  & + + --  & --  &  \multicolumn{3}{l|}{}  \\ \cline{1-8}
			5  & Ni   & $a_o/2$ & 0       & $c_o/2$ & 1/2 & + + + & -- &  \multicolumn{3}{l|}{}  \\ \cline{1-8}
			6  & Ni   & 0       & $b_o/2$ & $c_o/2$ & 1/2 & -- -- -- & +  &  \multicolumn{3}{l|}{}  \\ \cline{1-8}
			7  & Ni   & $a_o/2$ & $b_o$   & $c_o/2$ & 1/2 & + + + & --   &  \multicolumn{3}{l|}{}  \\ \cline{1-8}
			8  & Ni   & $a_o$   & $b_o/2$ & $c_o/2$ & 1/2 & -- -- -- & +  &  \multicolumn{3}{l|}{}  \\ \cline{1-8}
			9  & Ni   & $a_o/2$ & 0       & $c_o$   & 1/4 & -- -- +& + &  \multicolumn{3}{l|}{}  \\ \cline{1-8}
			10 & Ni   & 0       & $b_o/2$ & $c_o$   & 1/4 & + + -- & --   &  \multicolumn{3}{l|}{}  \\ \cline{1-8}
			11 & Ni   & $a_o/2$ & $b_o$   & $c_o$   & 1/4  & -- -- + & +  &  \multicolumn{3}{l|}{}  \\ \cline{1-8}
			12 & Ni   & $a_o$   & $b_o/2$ & $c_o$   & 1/4  & + + -- & --  &  \multicolumn{3}{l|}{}  \\ \cline{1-11}
			13 & Sm   & $a_o-d_x$        &   $d_y$      &   $c_0/4+d_z$      &    1 & \multicolumn{5}{l|}{}  \\ \cline{1-6}
			14 & Sm   &  $a_o/2+d_x$       &   $b_o/2+d_y$      &    $c_0/4+d_z$     &    1 & \multicolumn{5}{l|}{}  \\ \cline{1-6}
			15 & Sm   &    $a_o/2-d_x$     &    $b_o/2-d_y$     &    $3c_0/4-d_z$      &    1  & \multicolumn{5}{l|}{} \\ \cline{1-6}
			16 & Sm   &    $d_x$     &    $b_o-d_y$     &    $3c_0/4-d_z$     &   1  & \multicolumn{5}{l|}{} \\ \hline
			17 & O (1)   &    $a_o/2+\tilde{x}$     &    $\tilde{y}$     &    $\tilde{z}$    &   1/2  & -- -- + & + & $a_o/4$ & $b_o/4$& 0 \\ \hline
			18 & O (1)  &    $a_o/2+\tilde{x}$     &    $\tilde{y}$     &    $\tilde{z}$    &   1/2 & -- -- +& +  & 0 & 0  & $c_o/4$ \\ \hline
			19 & O (2)  &    $\tilde{x}$    &    $b_o/2+\tilde{y}$     &   $\tilde{z}$     &  1/2  & + + -- & --  & $a_o/4$& $-b_o/4$& 0  \\ \hline
			20 & O (2)  &    $\tilde{x}$    &    $b_o/2+\tilde{y}$    &   $\tilde{z}$      &  1/2  & + + -- & --   & $a_o/4$& $b_o/4$& 0 \\ \hline
			21 & O (2) &    $\tilde{x}$   &   $b_o/2+\tilde{y}$      &   $\tilde{z}$      &   1/2 & + + -- & --   & 0 & 0  & $c_o/4$ \\ \hline
			22 & O (3)   &    $a_o/2+\tilde{x}$     &    $b_o+\tilde{y}$     &    $\tilde{z}$    &   1/2 & -- -- + & +  & $a_o/4$& $-b_o/4$& 0  \\ \hline
			23 & O (3)    &   $a_o/2+\tilde{x}$      &    $b_o+\tilde{y}$     &   $\tilde{z}$     &   1/2  & -- -- +&+  & 0 & 0  & $c_o/4$ \\ \hline
			24 & O (4)   &    $a_o+\tilde{x}$     &   $b_o/2+\tilde{y}$      &   $\tilde{z}$      &    1/2 & + + --& -- & 0 & 0  & $c_o/4$ \\ \hline
			25 & O (5)   &    $a_o/2+\tilde{x}$     &  $\tilde{y}$       &   $c_o/2+\tilde{z}$     &   1  & + + + &-- & $a_o/4$& $b_o/4$& 0  \\ \hline
			26 & O (5)   &    $a_o/2+\tilde{x}$     &  $\tilde{y}$       &  $c_o/2+\tilde{z}$      & 1/2 & + + + & --    & 0 & 0 & $c_o/4$ \\ \hline
			27 & O  (6)  &    $\tilde{x}$    &  $b_o/2+\tilde{y}$    &   $c_o/2+\tilde{z}$    &  1  &  -- -- -- & +    & $a_o/4$& $-b_o/4$& 0 \\ \hline
			28 & O (6)   &    $\tilde{x}$    &  $b_o/2+\tilde{y}$   &   $c_o/2+\tilde{z}$     &  1  &  -- -- -- & +  & $a_o/4$& $b_o/4$& 0 \\ \hline
			29 & O  (6)  &    $\tilde{x}$     &  $b_o/2+\tilde{y}$    &   $c_o/2+\tilde{z}$     &  1/2  &  -- -- -- & +  & 0 & 0 & $c_o/4$  \\ \hline
			30 & O (7)   &    $a_o/2+\tilde{x}$     &   $b_o+\tilde{y}$      &    $c_o/2+\tilde{z}$     & 1  &  + + + & --    & $a_o/4$& $b_o/4$& 0 \\ \hline
			31 & O (7)   &  $a_o/2+\tilde{x}$       & $b_o+\tilde{y}$         &   $c_o/2+\tilde{z}$ & 1/2   & + + +  & --  & 0 &0 &$c_o/4$ \\ \hline
			32 & O (8)   &    $a_o+\tilde{x}$    &    $b_o/2+\tilde{y}$     &   $c_o/2+\tilde{z}$  & 1/2  &-- -- -- & +   & 0 & 0 & $c_o/4$\\ \hline
			33 & O  (9)  &     $a_o/2+\tilde{x}$    &   $\tilde{y}$      &   $c_o+\tilde{z}$     & 1/2  & -- -- + & +   & $a_o/4$& $b_o/4$& 0\\ \hline
			34 & O (10)   &    $\tilde{x}$     &   $b_o/2+\tilde{y}$      &  $c_o+\tilde{z}$      & 1/2  & + + -- & -- & $a_o/4$& $-b_o/4$& 0  \\ \hline
			35 & O (10)   &    $\tilde{x}$     &   $b_o/2+\tilde{y}$     &  $c_o+\tilde{z}$      & 1/2  & + + -- & --  & $a_o/4$& $b_o/4$& 0 \\ \hline
			36 & O (11)   &     $a_o/2+\tilde{x}$    & $b_o+\tilde{y}$  &  $c_o+\tilde{z}$   &  1/2 & -- -- +& +  & $a_o/4$& $-b_o/4$& 0 \\ \hline
		\end{tabular}
	\end{center}
	\caption{\label{Atomic_coordinates_tab} Coordinates of the atoms in \ce{SmNiO3} orthorhombic unit cell with lattice parameters $a_o=5.328$~\AA, $b_o=5.437$~\AA~and $c_o=7.568$~\AA~\cite{Jain2013}.
		The numbers in brackets next to oxygen atoms indicate an index of Ni atom, relative to which the coordinates of the oxygen atoms are calculated (see \textbf{Equation~\ref{Coordinates1}} and \textbf{Figure~\ref{Structure_Compare}b}).
		The coordinates $\tilde{\textbf{u}}=(\tilde{x},\tilde{y},\tilde{z})$ for each oxygen atom are calculated according to \textbf{Equation~\ref{Coordinates2}}, in which the signs of rotation angles $\alpha$, $\beta$ and $\gamma$ and the breathing distortion are specified in columns 7 and 8, and the values $(x_0, y_0, z_0)$ for the undistorted lattice are specified in columns 9--11.}
\end{table}

\subsection{Rotation operator $\hat{R}(\alpha,\beta,\gamma)$}

\quad In this work, we assumed that the tilt of the NiO$_6$ octahedron can be described as a rotation of a rigid body.
This allowed us to calculate the coordinates of each O atom by stretching the Ni--O bond (breathing mode) and rotating it relative to its position in the pristine SNO structure.  
It is known that rotations are not commutative, therefore the rotation operator $\hat{R}(\alpha,\beta,\gamma)$ can not be represented as three consecutive rotations around $x$-, $y$- and $z$-axis, because the final result will depend on the order of rotations.
However for small angles, the rotation operators can be interchanged with the final result remaining constant up to the second-order terms \cite{Fowlie2019}.
Therefore, the rotation operator $\hat{R}(\alpha,\beta,\gamma)$ can be represented as a product of small commutative (Abelian) rotations around three orthogonal axes
\begin{equation}
	\label{Rotations1}
	\hat{R}(\alpha,\beta,\gamma)=\prod_{i=1}^N \hat{R}_x\Big(\frac{\alpha}{N}\Big) \hat{R}_y\Big(\frac{\beta}{N}\Big)
	\hat{R}_z\Big(\frac{\gamma}{N}\Big)
	\ .
\end{equation}
Here $N$ is a large number (in our simulations we used $N=100$), and  $\hat{R}_x(\theta)$, $\hat{R}_y(\theta)$, and $\hat{R}_z(\theta)$ are operators of rotation about $x$-, $y$- and $z$-axis, respectively, with matrices
\begin{align}
	\label{Rotations2}
	\hat{R}_x(\theta)=
	\begin{bmatrix}
		1 & 0 & 0\\
		0 & \cos\theta & -\sin\theta \\
		0 & \sin\theta & \cos\theta
	\end{bmatrix}
	\ , \\
	\hat{R}_y(\theta)=
	\begin{bmatrix}
		\cos\theta & 0 & \sin\theta\\
		0 & 1 & 0 \\
		-\sin\theta & 0 & \cos\theta
	\end{bmatrix}
	\ , \\
	\hat{R}_z(\theta)=
	\begin{bmatrix}
		\cos\theta & -\sin\theta & 0\\
		\sin\theta & \cos\theta & 0 \\
		0 & 0 & 1
	\end{bmatrix}
	\ .
\end{align}

We want to emphasize here that since we use the orthorhombic symmetry, all rotations in this work are performed about orthorhombic axes, and not about traditionally used pseudocubic axes \cite{Catalano2018}.

\subsection{Simulation of x-ray diffraction}

The intensity of the diffracted beam $I_{hkl}$ was calculated as the squared modulus of the form factor $F(\textbf{q})$
\begin{equation}
	\label{Diffraction1}
	I_{hkl}\propto|F(\textbf{q})|^2 \ ,
\end{equation}
where
\begin{equation}
	\label{Diffraction2}
	F(\textbf{q})=\sum_{j=1}^{36} O_j f_j(q,E)\exp{(i\textbf{q}\textbf{r}_j)} \ ,
\end{equation}
and index $j$ runs over all atoms in the unit cell (see \textbf{Table~\ref{Atomic_coordinates_tab}}).
The atomic form factors $f_j(q,E)$ at x-ray energy $E=8345$~eV were calculated for Sm$^{3+}$, Ni$^{3+}$ and O$^{2-}$ ions using Cromer-Mann parameterization \cite{Brown2006} and tabulated energy-dependent corrections \cite{Chantler1995}.
The components of the scattering vector $\textbf{q}=(q_x,q_y,q_z)$ for the reflection $(hkl)$ equal to $(2\pi h/a_o, 2\pi k/b_o, 2\pi l/c_o)$.
In this work, we considered two groups of symmetry equivalent reflections: (101) and (202) -- that corresponds to the $(\tfrac{1}{2} \tfrac{1}{2} \tfrac{1}{2})_{pc}$ and (111)$_{pc}$ reflections in pseudocubic notation, correspondingly.
Due to orthorhombic twinning (the values of the lattice parameters are close, but not exactly equal $a_o\approx b_o\approx c_o/\sqrt{2}$), it is impossible to distinguish between the individual reflections within each class, so the intensity of the (101) reflection was evaluated as the averaged value of $(101)$, $(10\bar{1})$, $(011)$, and $(01\bar{1})$ intensities \cite{Brahlek2017}.
In a similar way, the intensity of the (202) is actually averaged over $(202)$, $(20\bar{2})$, $(022)$, and $(02\bar{2})$ reflections.

Direct evaluation of the (101) and (202) peaks intensity showed that the first peak is approximately 250 times weaker and its intensity strongly depends on $d_x$ component of the rare-earth cation displacement (see \textbf{Figure 4} in the main text).
At the same time the intensity of the (202) reflection is not strongly affected by these distortions (i.e., the octahedral rotations, breathing mode and the displacement of rare-earth cations).
Neglecting the weak contribution from Ni and O atoms in \textbf{Equation~\ref{Diffraction2}}, the form factors can be simplified to
\begin{align}
	\label{Diffraction3}
	&F_{101,10\bar{1}}\propto\sin\frac{2\pi d_x}{a_o}\cos\frac{2\pi d_z}{c_o} \ , \\
	&F_{011,01\bar{1}}=0 \ , \\
	&F_{202,20\bar{2}}\propto\cos\frac{4\pi d_x}{a_o} \cos\frac{4\pi d_z}{c_o} \ , \\ 
	&F_{022,02\bar{2}}\propto\cos\Big(\frac{4\pi d_y}{b_o}\pm\frac{4\pi d_z}{c_o}\Big) \ . 
\end{align}

One can also study how the local defects influence the intensity of the x-ray reflection, i.e., what happens if the values of Sm$^{3+}$ displacements $d_x$, $d_y$ and $d_z$ are not the same for all unit cells.
The direct evaluation shows that if the displacements $d_x$, $d_y$ and $d_z$ follow a normal distribution with standard deviation $\sigma_x$, $\sigma_y$ and $\sigma_z$, respectively, the intensity of the averaged (due to twinning) reflections can be estimated as
\begin{align}
	\label{Diffraction4}
	I_{101}&\propto\exp\bigg[-\Big(\frac{2\pi\sigma_x}{a_o}\Big)^2-\Big(\frac{2\pi\sigma_z}{c_o}\Big)^2\bigg] \sin^2 \frac{2\pi d_x}{a_o}\cos^2 \frac{2\pi d_z}{c_o} \ , \\
	I_{202}&\propto\exp\bigg[-\Big(\frac{4\pi\sigma_x}{a_o}\Big)^2-\Big(\frac{4\pi\sigma_z}{c_o}\Big)^2\bigg] \cos^2 \frac{4\pi d_x}{a_o} \cos^2 \frac{4\pi d_z}{c_o}+ \nonumber  \\
	&+ \exp\bigg[-\Big(\frac{4\pi\sigma_y}{b_o}\Big)^2-\Big(\frac{4\pi\sigma_z}{c_o}\Big)^2\bigg] \cos^2 \Big(\frac{4\pi d_y}{b_o}+\frac{4\pi d_z}{c_o}\Big)+ \label{Diffraction5}\\
	&+ \exp\bigg[-\Big(\frac{4\pi\sigma_y}{b_o}\Big)^2-\Big(\frac{4\pi\sigma_z}{c_o}\Big)^2\bigg] \cos^2 \Big(\frac{4\pi d_y}{b_o}-\frac{4\pi d_z}{c_o}\Big) 
	\ . \nonumber
\end{align}
These results agree with the direct calculations with \textbf{Equations~\ref{Diffraction1}} and \textbf{\ref{Diffraction2}}, shown in \textbf{Figure 4b,e} in the main text:
the intensity of the (101) reflection can be estimated as $I_{101}\propto d_x^2$ for a small cation displacement.
\textbf{Equations~\ref{Diffraction4}} and \textbf{\ref{Diffraction5}} show how the experimentally measured intensities of Bragg reflections are related to the parameters of our structural model.


\section{DFT-simulations of SNO electronic structure}

\subsection{Magnetic considerations}

SNO is known to be paramagnetic \cite{Catalano2018}, however, the DFT calculations are limited to magnetically ordered (e.g., ferromagnetic or antiferromagnetic) or nonmagnetic systems. 
As such, one needs to select a proxy magnetic configuration based on structural consistency with experimental observations and stability of the doped H-SNO structure. 
We started by relaxing the orthorhombic phase of SNO shown in \textbf{Figure~\ref{Structure_Compare}} under a nonmagnetic (NM), ferromagnetic (FM) and antiferromagnetic (AFM) configuration \cite{Yan2018}. 

\begin{figure*}
	\includegraphics[width = 0.45\linewidth]{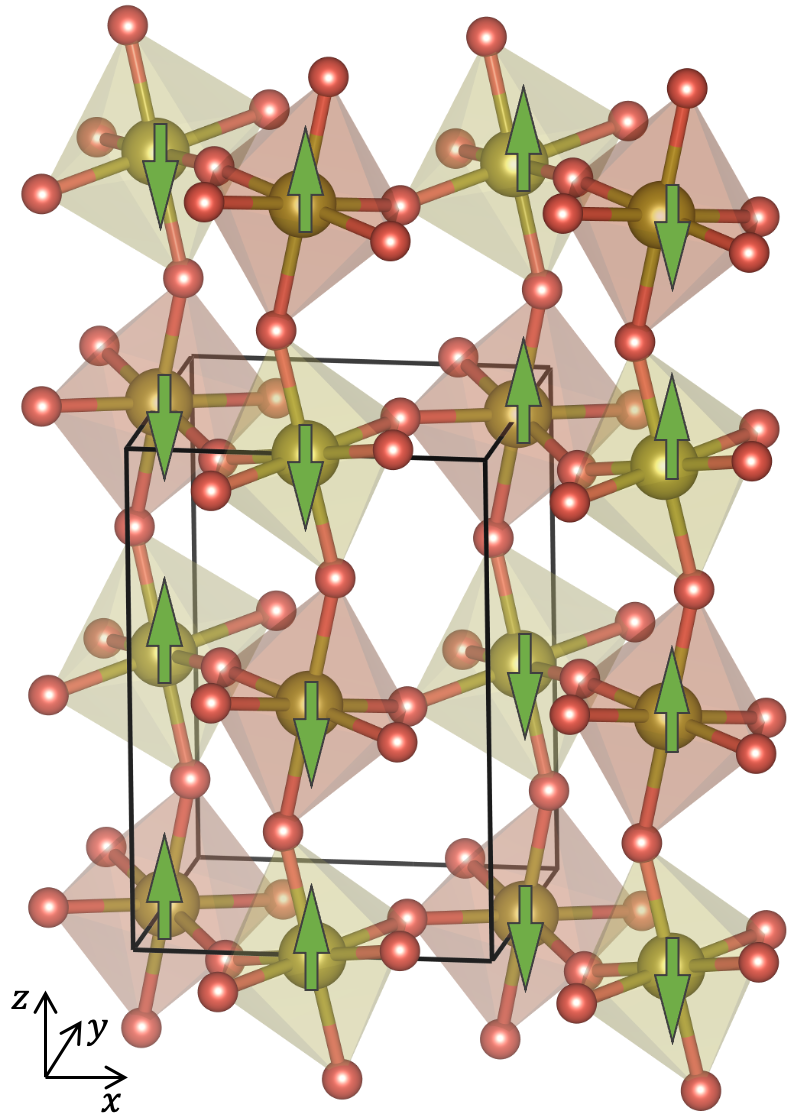}
	\caption{\label{AFM} 
		Sketch of the spin configuration of the T-AFM phase characterized by the particular $\uparrow\uparrow\downarrow\downarrow$ stacking along all thee crystallographic directions \cite{Varignon2017}. The rare-earth cations are not shown for clarity, black lines represent the orthorhombic unit cell.} 
\end{figure*}

We used the Interstitialcy Finding Tool (InFiT) \cite{Zimmermann2017} to locate all possible interstitial sites for H under a 1H:32Ni ratio in SNO.
We found that the formation energy of H-SNO in the FM phase was 70 meV more stable than in the T-AFM phase (\textbf{Figure~\ref{AFM}}).
In NM phase (zero-spin configuration), no band gap opening was observed with H-doping (see the next section), so this phase can not be used to study the doping effects.  
Hence all investigations of the effect of H-doping concentration on the oxidation state and x-ray absorption spectrum (XAS) were done in the FM phase.

\subsection{Opening of the band gap with hydrogen doping}

\begin{figure*}
	\includegraphics[width = 0.75\linewidth]{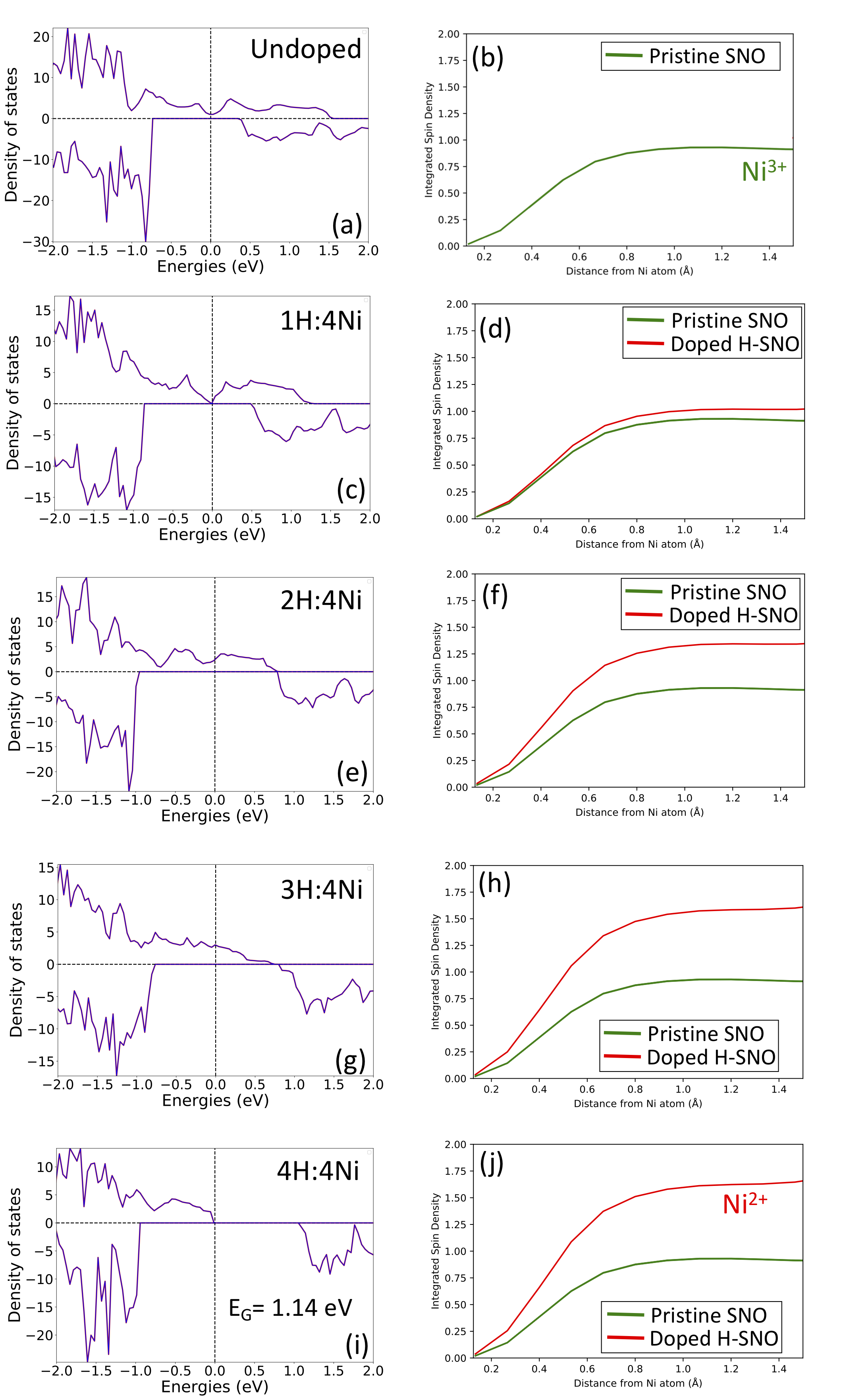}
	\caption{\label{DoS2} 
		The DoS (left column) and integrated spin density (right column) of H-doped orthorhombic SNO at various doping concentrations, namely no doping (a-b), 1H:4Ni (c-d), 1H:2Ni (e-f), 3H:4Ni (g-h), and 1H:1Ni (i-j). The integrated spin density of undoped orthorhombic SNO is show in green as a reference for all doping levels.}
\end{figure*}

We tested the effect of H concentration on th density of states (DoS) in orthorhombic SNO (\textbf{Figure~\ref{DoS2}}). 
We observed that at concentrations below 1H:1Ni, orthorhombic SNO remains metallic, however at a 1H:1Ni ratio, a large band gap of 1.14 eV appears which is consistent with previous studies \cite{Yoo2018}. 
We observed that the integrated spin density of Ni in H-SNO gradually increases relative to the undoped system, which indicates that the Ni oxidation state increases with H concentration. 
The converged spin density of undoped SNO at 0.92 is associated with the Ni$^{3+}$ cation. 
The integrated spin density associated with the 1H:1Ni concentration has a converged integrated spin density nearly doubled (1.7) that of the undoped SNO which is associated with the Ni$^{2+}$ cation (\textbf{Figure~\ref{DoS2}b}). 
This supports the notion that a change in transition states is required for the opening of a band gap and thus formation of an insulating phase. 
In \textbf{Figure~\ref{DoS2}}, we considered several possible interstitial sites for each concentration and only showed the results for the most energetically stable configuration.

\begin{figure*}[b]
	\includegraphics[width = \linewidth]{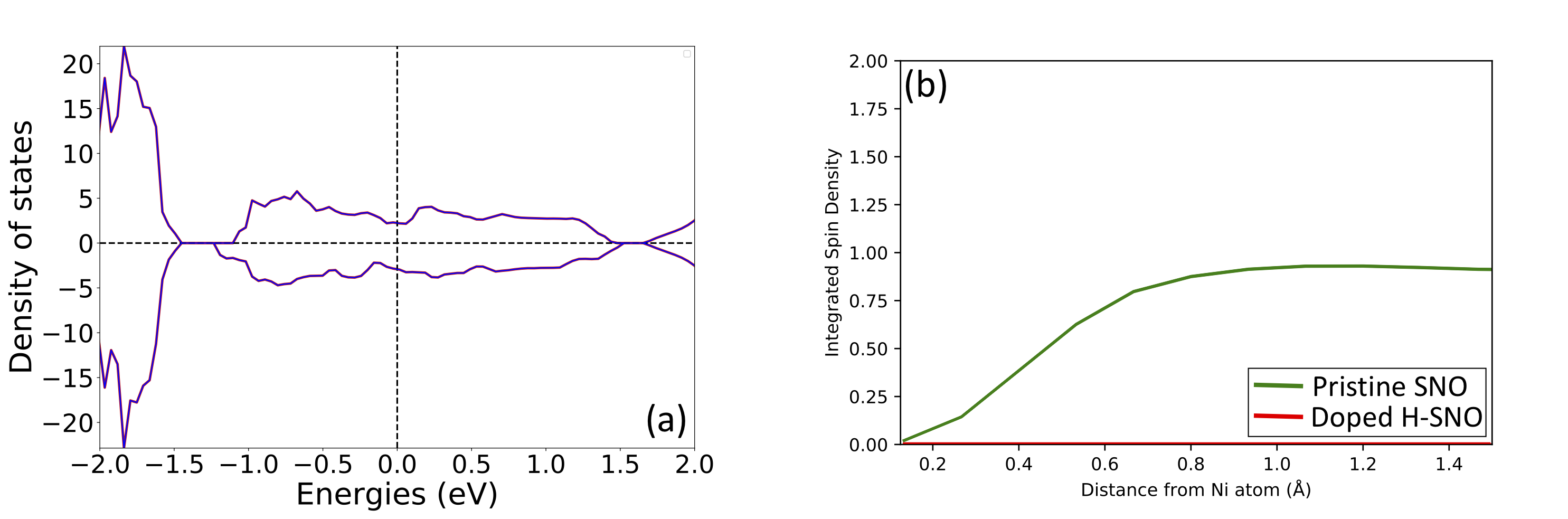}
	\caption{\label{DoS4} 
		The DoS (a) and integrated spin density (b) of H-doped orthorhombic SNO at zero-spin configuration (NM phase).}
\end{figure*} 

Next we evaluated the effect of valence change. We already calculated the DoS for a 1H:1Ni doped SNO structure under a non-zero spin configuration (\textbf{Figure~\ref{DoS2}i,j}) which results in a change of Ni valence from Ni$^{3+}$ to Ni$^{2+}$. We also calculated the DoS of the same structure at a zero spin configuration (which corresponds to NM phase) which results in a change of Ni valence from Ni$^{3+}$ to Ni$^{4+}$ in \textbf{Figure~\ref{DoS4}a-b}. These calculations were done with the lattice and sites fixed at their relaxed undoped positions to isolate the valence effect from any structural distortion. Since we observed no band gap opening in NM phase (\textbf{Figure~\ref{DoS4}a}), this phase was excluded from further analysis.
From \textbf{Figure~\ref{DoS2}}  and \textbf{Figure~\ref{DoS4}} we conclude, that the band gap opens only if the Ni valence decreases from 3+ (pristine SNO) to almost 2+ (1H:1Ni doping). 

\begin{figure*}
	\includegraphics[width = \linewidth]{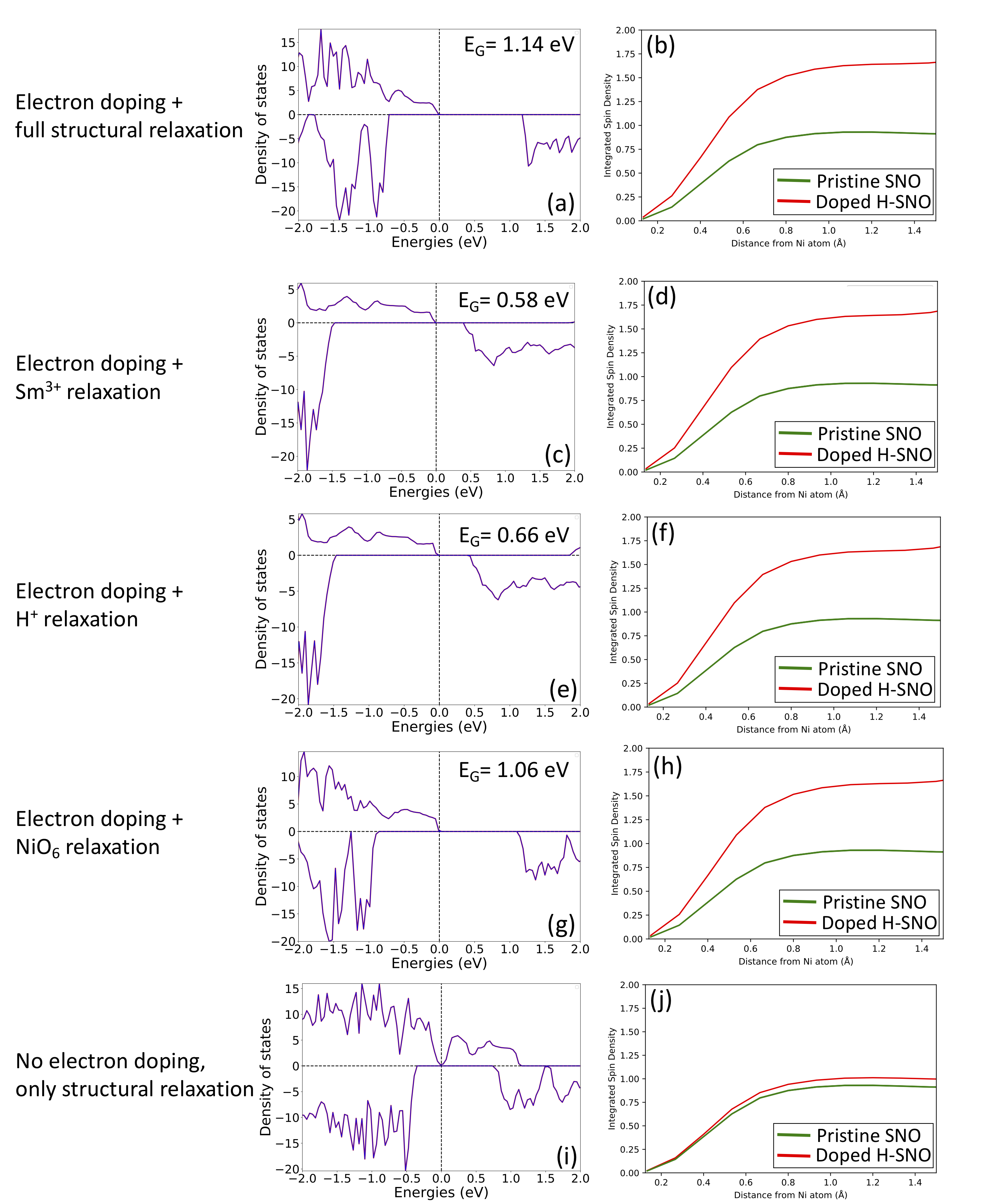}
	\caption{\label{DoS3} 
		The DoS (left column) and integrated spin density (right column) of H-doped orthorhombic SNO at 1H:1Ni concentration with full volume and H-site relaxation (a-b), with only Sm site relaxation (c-d), only H site relaxation (e-f), and only \ce{NiO6} octahedra relaxation. Panels (i-j) show the results of the structural relaxation of H-SNO only, without electron doping from H-dopant.}
\end{figure*}

To decouple the structural and electronic effects of the H-doping on the band gap opening, we performed several calculations with some artificially introduced constrains.
In \textbf{Figure~\ref{DoS3}a-b} the band gap opening is shown for 1H:1Ni doping when all atoms in the H-SNO structure are allowed to move in order to reach the most energetically stable configurations.
This leads to the opening of the large band gap $E_G=1.14$~eV, as we described earlier.
However, if one only allows the Sm$^{3+}$ ions (\textbf{Figure~\ref{DoS3}c,d}) or H$^+$ (\textbf{Figure~\ref{DoS3}e,f}) to relax, the band gap appears to be almost twice smaller.
Contrary to that, if one only allows \ce{NiO6} octahedra to relax (\textbf{Figure~\ref{DoS3}g,h}), the resulting band gap again becomes almost fully open.
Specifically, the polyhedron volumes of \ce{NiO6} increase from their undoped volume of 10.2~\AA$^3$ to 10.8~\AA$^3$ as a result of the decrease in the nickel oxidation state. 
This leads us to conclusion: the band gap opens in any case, when the level of H-doping is high enough, however, only combined effect of H-doping and structural changes in the \ce{NiO6} octahedra can explain the final large value of $E_G\approx1$~eV in the H-SNO.
To assess the effect of pure structural distortion on the band gap opening,  we performed the similar computations for the fully relaxed H-SNO structure without adding any H-dopants (\textbf{Figure~\ref{DoS3}i,j}).
This structure shows no band gap, indicating that the prime contribution to the insulating properties of H-SNO comes from the electron doping and not from the structural changes. 

\subsection{Comparison of XAS spectrum to known reference compounds}

\begin{figure*}
	\includegraphics[width = 0.5\linewidth]{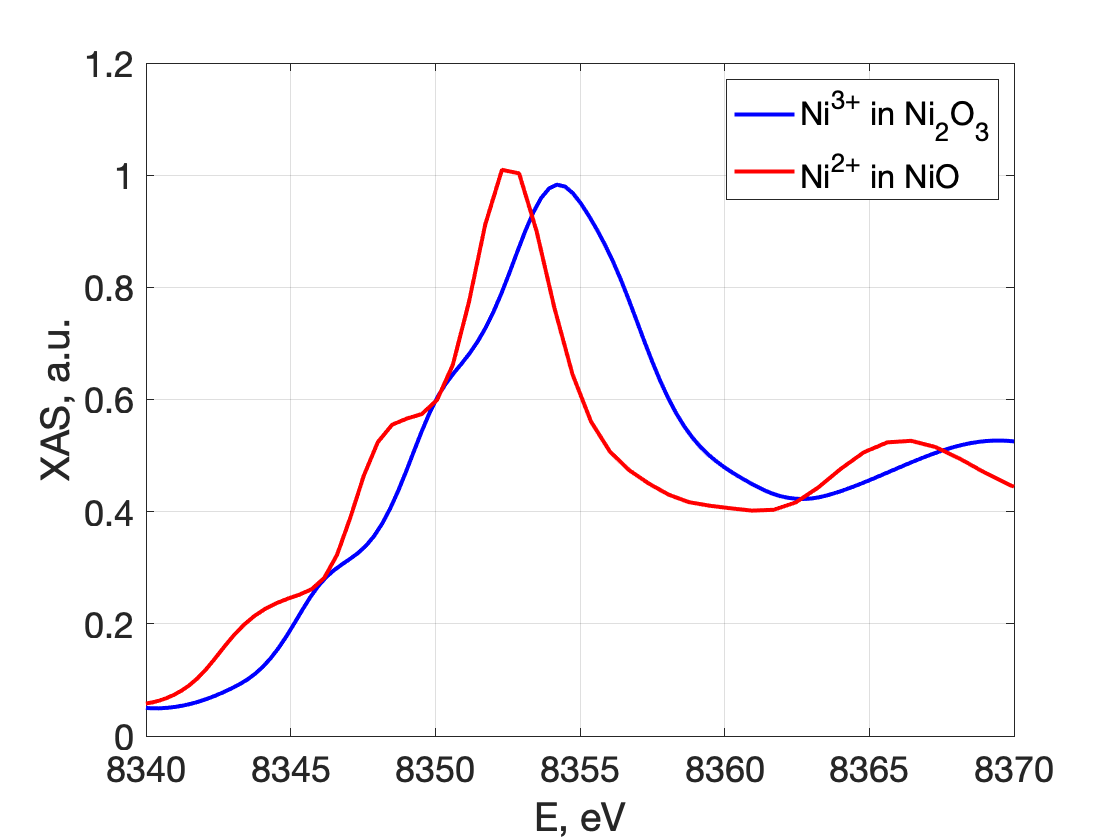}
	\caption{\label{SNO_vs_Ni_O_XAS} 
		FEFF-simulated normalized XAS spectra at the K-edge of Ni$^{3+}$ in \ce{Ni2O3} (blue) and Ni$^{2+}$ in \ce{NiO} (red). The spectra were obtained from the Materials Project database \cite{Ong2015, Zheng2018}.}
\end{figure*}

To directly explore the effect of Ni reduction in the XAS spectra, we presented the FEFF-simulated spectra for two reference compounds, \ce{NiO} (Ni$^{2+}$) and \ce{Ni2O3} (Ni$^{3+}$). 
As shown in Figure~\ref{SNO_vs_Ni_O_XAS}, 
the K-edge of Ni in NiO oxide is approximately 1.75 eV lower than in Ni$_{2}$O$_{3}$ oxide.
The same shift of approximately 1.2 eV was observed between the SNO a H-SNO in our experimental data, which suggests the reduction of nickel from Ni$^{3+}$ in the pristine SNO to Ni$^{2+}$ in the fully doped H-SNO.
This shift can be interpreted as a cation losing its ability to attract electrons as the valence state decreases, thereby requiring less energy to excite electrons to emit X-rays.

\begin{figure*}[h!]
	\includegraphics[width = 1\linewidth]{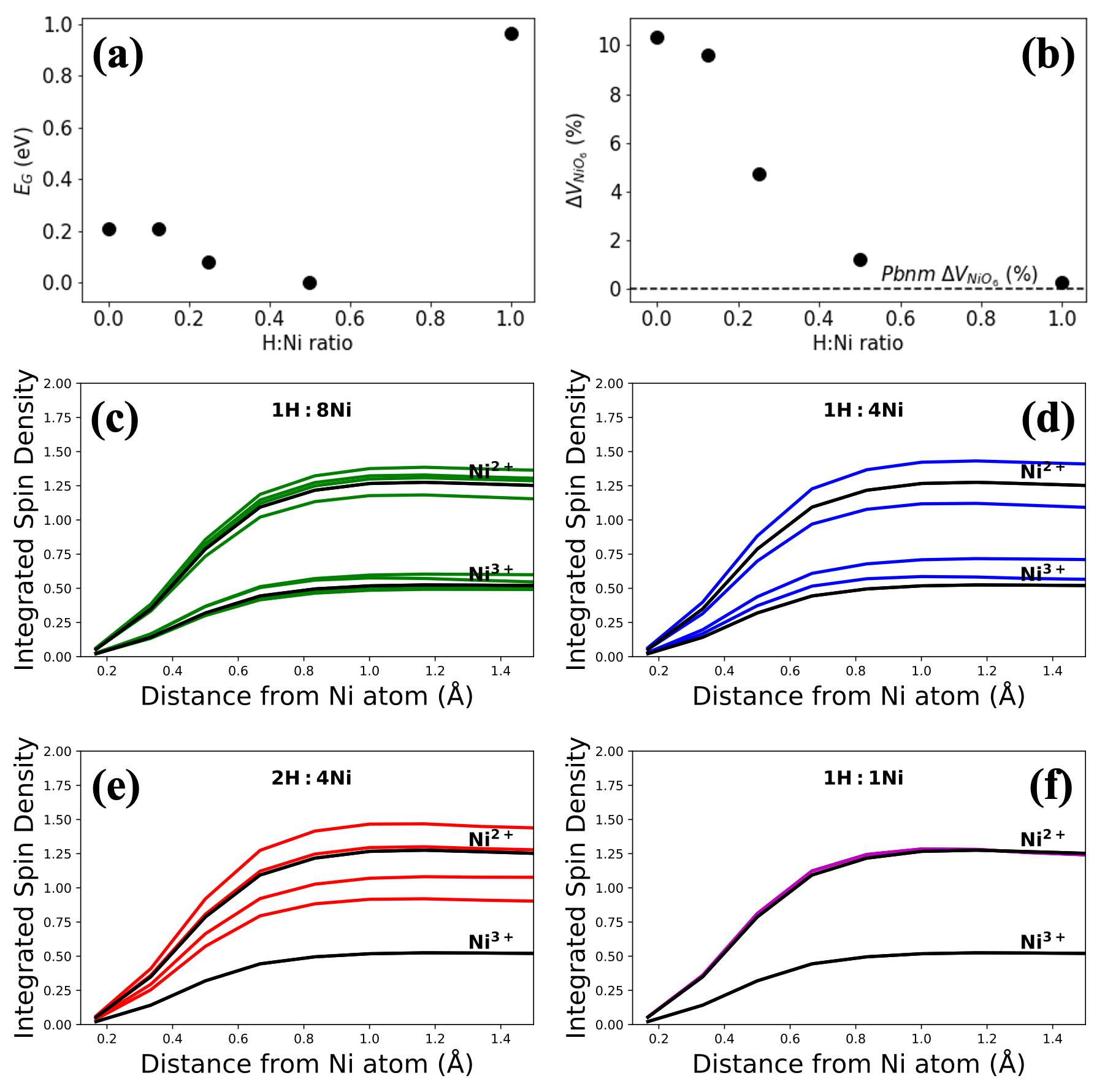}
	\caption{\label{LT_analysis} 
		The band gap $E_G$ (a) and bond-disproportionation (b) of monoclinic insulating \ce{SmNiO3} in the T-AFM phase is plotted against the H-doping level. Here, bond-disproportionation is defined as the percentage difference in volume of \ce{NiO6} polyhedrons with opposite spins (see \textbf{Figure~\ref{AFM}}). (c-f) The integrated spin density of H-doped monoclinic SNO at a ratio of 1H:8Ni (c), 1H:4Ni (d), 2H:4Ni (e) and 1H:1Ni (f). Color lines represent different nickel atoms in the doped H-SNO and black lines show the integrated spin density of the undoped SNO (all calculations are performed for the monoclinic phase).} 
\end{figure*}

\subsection{Hydrogen doping in the monoclinic phase}


\ce{SmNiO3} can also exist in an insulating monoclinic phase ($P2_1/n$), which differs from the  metallic orthorhombic phase (\textit{Pbnm}) by a small Ni-O bond disproportionation in the form of a breathing-mode configuration (with a magnitude of $\sim1-2$\% in Ni-O bond length, which leads to the \ce{NiO6} octahedra volume expansion/contraction of $\sim10$\%).
We performed DFT simulations to assess the effect of H-doping on the electronic properties of the monoclinic phase under a T-AFM spin configuration (\textbf{Figure~\ref{AFM}}).
The results are shown in \textbf{Figure~\ref{LT_analysis}a} where the band gap $E_G$ is plotted as a function of the H-dopant concentration. 
The obtained value of the band gap $E_G\approx0.2$~eV for the pristine monoclinic SNO agrees with the published data \cite{Catalano2018}, however as the H:Ni ratio increases, we observe a decrease in $E_G$ until the gap disappears completely at a ratio of 1H:2Ni. 
This coincides with the gradual dissipation of breathing-mode (\textbf{Figure~\ref{LT_analysis}b}), as the difference in \ce{NiO6} volumes disappears, and the mixed valence states of Ni$^{2/3+}$ (show in the black lines) converges to Ni$^{2+}$ (\textbf{Figure~\ref{LT_analysis}c-e}). Increasing the ratio to 1H:1Ni in the monoclinic phase yields similar results to 1H:1Ni doping in the orthorhombic phase with all Ni-cations converging to a valence state of Ni$^{2+}$ (\textbf{Figure~\ref{LT_analysis}f}) and the volume of all \ce{NiO6} polyhedrons becoming uniform. 
Consequently, the band gap $E_G$ in the monoclinic phase increases to the value of 0.97 eV, approximately five times the undoped band gap of the monoclinic phase. 
The absence of any noticeable band gap prior to the reduction of all nickel ions to the 2+ valence state is a further evidence that the decrease in electrical conductivity is connected to the valence state.
This and the fact that our experimental data did not show any evidences of the  significant bond disproportionation in both SNO and H-SNO justifies the focus on the orthorhombic phase in our DFT simulations.

	\bibliography{SNO_References}

\begin{thebibliography}{56}%
\makeatletter
\providecommand \@ifxundefined [1]{%
 \@ifx{#1\undefined}
}%
\providecommand \@ifnum [1]{%
 \ifnum #1\expandafter \@firstoftwo
 \else \expandafter \@secondoftwo
 \fi
}%
\providecommand \@ifx [1]{%
 \ifx #1\expandafter \@firstoftwo
 \else \expandafter \@secondoftwo
 \fi
}%
\providecommand \natexlab [1]{#1}%
\providecommand \enquote  [1]{``#1''}%
\providecommand \bibnamefont  [1]{#1}%
\providecommand \bibfnamefont [1]{#1}%
\providecommand \citenamefont [1]{#1}%
\providecommand \href@noop [0]{\@secondoftwo}%
\providecommand \href [0]{\begingroup \@sanitize@url \@href}%
\providecommand \@href[1]{\@@startlink{#1}\@@href}%
\providecommand \@@href[1]{\endgroup#1\@@endlink}%
\providecommand \@sanitize@url [0]{\catcode `\\12\catcode `\$12\catcode
  `\&12\catcode `\#12\catcode `\^12\catcode `\_12\catcode `\%12\relax}%
\providecommand \@@startlink[1]{}%
\providecommand \@@endlink[0]{}%
\providecommand \url  [0]{\begingroup\@sanitize@url \@url }%
\providecommand \@url [1]{\endgroup\@href {#1}{\urlprefix }}%
\providecommand \urlprefix  [0]{URL }%
\providecommand \Eprint [0]{\href }%
\providecommand \doibase [0]{http://dx.doi.org/}%
\providecommand \selectlanguage [0]{\@gobble}%
\providecommand \bibinfo  [0]{\@secondoftwo}%
\providecommand \bibfield  [0]{\@secondoftwo}%
\providecommand \translation [1]{[#1]}%
\providecommand \BibitemOpen [0]{}%
\providecommand \bibitemStop [0]{}%
\providecommand \bibitemNoStop [0]{.\EOS\space}%
\providecommand \EOS [0]{\spacefactor3000\relax}%
\providecommand \BibitemShut  [1]{\csname bibitem#1\endcsname}%
\let\auto@bib@innerbib\@empty
\bibitem [{\citenamefont {Ramanathan}(2018)}]{Ramanathan2018}%
  \BibitemOpen
  \bibfield  {author} {\bibinfo {author} {\bibfnamefont {S.}~\bibnamefont
  {Ramanathan}},\ }\href {\doibase DOI: 10.1557/mrs.2018.147} {\bibfield
  {journal} {\bibinfo  {journal} {MRS Bull.}\ }\textbf {\bibinfo {volume}
  {43}},\ \bibinfo {pages} {534} (\bibinfo {year} {2018})}\BibitemShut
  {NoStop}%
\bibitem [{\citenamefont {Roy}\ \emph {et~al.}(2019)\citenamefont {Roy},
  \citenamefont {Jaiswal},\ and\ \citenamefont {Panda}}]{Roy2019}%
  \BibitemOpen
  \bibfield  {author} {\bibinfo {author} {\bibfnamefont {K.}~\bibnamefont
  {Roy}}, \bibinfo {author} {\bibfnamefont {A.}~\bibnamefont {Jaiswal}}, \ and\
  \bibinfo {author} {\bibfnamefont {P.}~\bibnamefont {Panda}},\ }\href
  {\doibase 10.1038/s41586-019-1677-2} {\bibfield  {journal} {\bibinfo
  {journal} {Nature}\ }\textbf {\bibinfo {volume} {575}},\ \bibinfo {pages}
  {607} (\bibinfo {year} {2019})}\BibitemShut {NoStop}%
\bibitem [{\citenamefont {Zhang}\ \emph
  {et~al.}(2020{\natexlab{a}})\citenamefont {Zhang}, \citenamefont {Panda},
  \citenamefont {Lin}, \citenamefont {Kalcheim}, \citenamefont {Wang},
  \citenamefont {Freeland}, \citenamefont {Fong}, \citenamefont {Priya},
  \citenamefont {Schuller}, \citenamefont {Sankaranarayanan}, \citenamefont
  {Roy},\ and\ \citenamefont {Ramanathan}}]{Zhang2020}%
  \BibitemOpen
  \bibfield  {author} {\bibinfo {author} {\bibfnamefont {H.-T.}\ \bibnamefont
  {Zhang}}, \bibinfo {author} {\bibfnamefont {P.}~\bibnamefont {Panda}},
  \bibinfo {author} {\bibfnamefont {J.}~\bibnamefont {Lin}}, \bibinfo {author}
  {\bibfnamefont {Y.}~\bibnamefont {Kalcheim}}, \bibinfo {author}
  {\bibfnamefont {K.}~\bibnamefont {Wang}}, \bibinfo {author} {\bibfnamefont
  {J.~W.}\ \bibnamefont {Freeland}}, \bibinfo {author} {\bibfnamefont {D.~D.}\
  \bibnamefont {Fong}}, \bibinfo {author} {\bibfnamefont {S.}~\bibnamefont
  {Priya}}, \bibinfo {author} {\bibfnamefont {I.~K.}\ \bibnamefont {Schuller}},
  \bibinfo {author} {\bibfnamefont {S.~K. R.~S.}\ \bibnamefont
  {Sankaranarayanan}}, \bibinfo {author} {\bibfnamefont {K.}~\bibnamefont
  {Roy}}, \ and\ \bibinfo {author} {\bibfnamefont {S.}~\bibnamefont
  {Ramanathan}},\ }\href {\doibase 10.1063/1.5113574} {\bibfield  {journal}
  {\bibinfo  {journal} {Appl. Phys. Rev.}\ }\textbf {\bibinfo {volume} {7}},\
  \bibinfo {pages} {11309} (\bibinfo {year} {2020}{\natexlab{a}})}\BibitemShut
  {NoStop}%
\bibitem [{\citenamefont {del Valle}\ \emph {et~al.}(2018)\citenamefont {del
  Valle}, \citenamefont {Ram{\'{i}}rez}, \citenamefont {Rozenberg},\ and\
  \citenamefont {Schuller}}]{DelValle2018}%
  \BibitemOpen
  \bibfield  {author} {\bibinfo {author} {\bibfnamefont {J.}~\bibnamefont {del
  Valle}}, \bibinfo {author} {\bibfnamefont {J.~G.}\ \bibnamefont
  {Ram{\'{i}}rez}}, \bibinfo {author} {\bibfnamefont {M.~J.}\ \bibnamefont
  {Rozenberg}}, \ and\ \bibinfo {author} {\bibfnamefont {I.~K.}\ \bibnamefont
  {Schuller}},\ }\href {\doibase 10.1063/1.5047800} {\bibfield  {journal}
  {\bibinfo  {journal} {J. Appl. Phys.}\ }\textbf {\bibinfo {volume} {124}},\
  \bibinfo {pages} {211101} (\bibinfo {year} {2018})}\BibitemShut {NoStop}%
\bibitem [{\citenamefont {Zhang}\ \emph
  {et~al.}(2020{\natexlab{b}})\citenamefont {Zhang}, \citenamefont {Park},
  \citenamefont {Zaluzhnyy}, \citenamefont {Wang}, \citenamefont {Wadekar},
  \citenamefont {Manna}, \citenamefont {Andrawis}, \citenamefont {Sprau},
  \citenamefont {Sun}, \citenamefont {Zhang}, \citenamefont {Huang},
  \citenamefont {Zhou}, \citenamefont {Zhang}, \citenamefont {Narayanan},
  \citenamefont {Srinivasan}, \citenamefont {Hua}, \citenamefont {Nazaretski},
  \citenamefont {Huang}, \citenamefont {Yan}, \citenamefont {Ge}, \citenamefont
  {Chu}, \citenamefont {Cherukara}, \citenamefont {Holt}, \citenamefont
  {Krishnamurthy}, \citenamefont {Shpyrko}, \citenamefont {Sankaranarayanan},
  \citenamefont {Frano}, \citenamefont {Roy},\ and\ \citenamefont
  {Ramanathan}}]{Zhang2020a}%
  \BibitemOpen
  \bibfield  {author} {\bibinfo {author} {\bibfnamefont {H.-T.}\ \bibnamefont
  {Zhang}}, \bibinfo {author} {\bibfnamefont {T.~J.}\ \bibnamefont {Park}},
  \bibinfo {author} {\bibfnamefont {I.~A.}\ \bibnamefont {Zaluzhnyy}}, \bibinfo
  {author} {\bibfnamefont {Q.}~\bibnamefont {Wang}}, \bibinfo {author}
  {\bibfnamefont {S.~N.}\ \bibnamefont {Wadekar}}, \bibinfo {author}
  {\bibfnamefont {S.}~\bibnamefont {Manna}}, \bibinfo {author} {\bibfnamefont
  {R.}~\bibnamefont {Andrawis}}, \bibinfo {author} {\bibfnamefont {P.~O.}\
  \bibnamefont {Sprau}}, \bibinfo {author} {\bibfnamefont {Y.}~\bibnamefont
  {Sun}}, \bibinfo {author} {\bibfnamefont {Z.}~\bibnamefont {Zhang}}, \bibinfo
  {author} {\bibfnamefont {C.}~\bibnamefont {Huang}}, \bibinfo {author}
  {\bibfnamefont {H.}~\bibnamefont {Zhou}}, \bibinfo {author} {\bibfnamefont
  {Z.}~\bibnamefont {Zhang}}, \bibinfo {author} {\bibfnamefont
  {B.}~\bibnamefont {Narayanan}}, \bibinfo {author} {\bibfnamefont
  {G.}~\bibnamefont {Srinivasan}}, \bibinfo {author} {\bibfnamefont
  {N.}~\bibnamefont {Hua}}, \bibinfo {author} {\bibfnamefont {E.}~\bibnamefont
  {Nazaretski}}, \bibinfo {author} {\bibfnamefont {X.}~\bibnamefont {Huang}},
  \bibinfo {author} {\bibfnamefont {H.}~\bibnamefont {Yan}}, \bibinfo {author}
  {\bibfnamefont {M.}~\bibnamefont {Ge}}, \bibinfo {author} {\bibfnamefont
  {Y.~S.}\ \bibnamefont {Chu}}, \bibinfo {author} {\bibfnamefont {M.~J.}\
  \bibnamefont {Cherukara}}, \bibinfo {author} {\bibfnamefont {M.~V.}\
  \bibnamefont {Holt}}, \bibinfo {author} {\bibfnamefont {M.}~\bibnamefont
  {Krishnamurthy}}, \bibinfo {author} {\bibfnamefont {O.~G.}\ \bibnamefont
  {Shpyrko}}, \bibinfo {author} {\bibfnamefont {S.~K.}\ \bibnamefont
  {Sankaranarayanan}}, \bibinfo {author} {\bibfnamefont {A.}~\bibnamefont
  {Frano}}, \bibinfo {author} {\bibfnamefont {K.}~\bibnamefont {Roy}}, \ and\
  \bibinfo {author} {\bibfnamefont {S.}~\bibnamefont {Ramanathan}},\ }\href
  {\doibase 10.1038/s41467-020-16105-y} {\bibfield  {journal} {\bibinfo
  {journal} {Nat. Commun.}\ }\textbf {\bibinfo {volume} {11}},\ \bibinfo
  {pages} {2245} (\bibinfo {year} {2020}{\natexlab{b}})}\BibitemShut {NoStop}%
\bibitem [{\citenamefont {Shi}\ \emph {et~al.}(2013)\citenamefont {Shi},
  \citenamefont {Ha}, \citenamefont {Zhou}, \citenamefont {Schoofs},\ and\
  \citenamefont {Ramanathan}}]{Shi2013}%
  \BibitemOpen
  \bibfield  {author} {\bibinfo {author} {\bibfnamefont {J.}~\bibnamefont
  {Shi}}, \bibinfo {author} {\bibfnamefont {S.~D.}\ \bibnamefont {Ha}},
  \bibinfo {author} {\bibfnamefont {Y.}~\bibnamefont {Zhou}}, \bibinfo {author}
  {\bibfnamefont {F.}~\bibnamefont {Schoofs}}, \ and\ \bibinfo {author}
  {\bibfnamefont {S.}~\bibnamefont {Ramanathan}},\ }\href {\doibase
  10.1038/ncomms3676} {\bibfield  {journal} {\bibinfo  {journal} {Nat.
  Commun.}\ }\textbf {\bibinfo {volume} {4}},\ \bibinfo {pages} {2676}
  (\bibinfo {year} {2013})}\BibitemShut {NoStop}%
\bibitem [{\citenamefont {Zhu}\ \emph {et~al.}(2020)\citenamefont {Zhu},
  \citenamefont {Zhang}, \citenamefont {Yang},\ and\ \citenamefont
  {Huang}}]{Zhu2020}%
  \BibitemOpen
  \bibfield  {author} {\bibinfo {author} {\bibfnamefont {J.}~\bibnamefont
  {Zhu}}, \bibinfo {author} {\bibfnamefont {T.}~\bibnamefont {Zhang}}, \bibinfo
  {author} {\bibfnamefont {Y.}~\bibnamefont {Yang}}, \ and\ \bibinfo {author}
  {\bibfnamefont {R.}~\bibnamefont {Huang}},\ }\href {\doibase
  10.1063/1.5118217} {\bibfield  {journal} {\bibinfo  {journal} {Appl. Phys.
  Rev.}\ }\textbf {\bibinfo {volume} {7}},\ \bibinfo {pages} {011312} (\bibinfo
  {year} {2020})}\BibitemShut {NoStop}%
\bibitem [{\citenamefont {Ramadoss}\ \emph {et~al.}(2018)\citenamefont
  {Ramadoss}, \citenamefont {Zuo}, \citenamefont {Sun}, \citenamefont {Zhang},
  \citenamefont {Lin}, \citenamefont {Bhaskar}, \citenamefont {Shin},
  \citenamefont {Alam}, \citenamefont {Guha}, \citenamefont {Weinstein},\ and\
  \citenamefont {Ramanathan}}]{Ramadoss2018}%
  \BibitemOpen
  \bibfield  {author} {\bibinfo {author} {\bibfnamefont {K.}~\bibnamefont
  {Ramadoss}}, \bibinfo {author} {\bibfnamefont {F.}~\bibnamefont {Zuo}},
  \bibinfo {author} {\bibfnamefont {Y.}~\bibnamefont {Sun}}, \bibinfo {author}
  {\bibfnamefont {Z.}~\bibnamefont {Zhang}}, \bibinfo {author} {\bibfnamefont
  {J.}~\bibnamefont {Lin}}, \bibinfo {author} {\bibfnamefont {U.}~\bibnamefont
  {Bhaskar}}, \bibinfo {author} {\bibfnamefont {S.}~\bibnamefont {Shin}},
  \bibinfo {author} {\bibfnamefont {M.~A.}\ \bibnamefont {Alam}}, \bibinfo
  {author} {\bibfnamefont {S.}~\bibnamefont {Guha}}, \bibinfo {author}
  {\bibfnamefont {D.}~\bibnamefont {Weinstein}}, \ and\ \bibinfo {author}
  {\bibfnamefont {S.}~\bibnamefont {Ramanathan}},\ }\href {\doibase
  10.1109/LED.2018.2865776} {\bibfield  {journal} {\bibinfo  {journal} {IEEE
  Electron Device Lett.}\ }\textbf {\bibinfo {volume} {39}},\ \bibinfo {pages}
  {1500} (\bibinfo {year} {2018})}\BibitemShut {NoStop}%
\bibitem [{\citenamefont {Staub}\ \emph {et~al.}(2002)\citenamefont {Staub},
  \citenamefont {Meijer}, \citenamefont {Fauth}, \citenamefont {Allenspach},
  \citenamefont {Bednorz}, \citenamefont {Karpinski}, \citenamefont {Kazakov},
  \citenamefont {Paolasini},\ and\ \citenamefont {d'Acapito}}]{Staub2002}%
  \BibitemOpen
  \bibfield  {author} {\bibinfo {author} {\bibfnamefont {U.}~\bibnamefont
  {Staub}}, \bibinfo {author} {\bibfnamefont {G.~I.}\ \bibnamefont {Meijer}},
  \bibinfo {author} {\bibfnamefont {F.}~\bibnamefont {Fauth}}, \bibinfo
  {author} {\bibfnamefont {R.}~\bibnamefont {Allenspach}}, \bibinfo {author}
  {\bibfnamefont {J.~G.}\ \bibnamefont {Bednorz}}, \bibinfo {author}
  {\bibfnamefont {J.}~\bibnamefont {Karpinski}}, \bibinfo {author}
  {\bibfnamefont {S.~M.}\ \bibnamefont {Kazakov}}, \bibinfo {author}
  {\bibfnamefont {L.}~\bibnamefont {Paolasini}}, \ and\ \bibinfo {author}
  {\bibfnamefont {F.}~\bibnamefont {d'Acapito}},\ }\href {\doibase
  10.1103/PhysRevLett.88.126402} {\bibfield  {journal} {\bibinfo  {journal}
  {Phys. Rev. Lett.}\ }\textbf {\bibinfo {volume} {88}},\ \bibinfo {pages}
  {126402} (\bibinfo {year} {2002})}\BibitemShut {NoStop}%
\bibitem [{\citenamefont {Lu}\ \emph {et~al.}(2016)\citenamefont {Lu},
  \citenamefont {Frano}, \citenamefont {Bluschke}, \citenamefont {Hepting},
  \citenamefont {Macke}, \citenamefont {Strempfer}, \citenamefont {Wochner},
  \citenamefont {Cristiani}, \citenamefont {Logvenov}, \citenamefont
  {Habermeier}, \citenamefont {Haverkort}, \citenamefont {Keimer},\ and\
  \citenamefont {Benckiser}}]{Lu2016}%
  \BibitemOpen
  \bibfield  {author} {\bibinfo {author} {\bibfnamefont {Y.}~\bibnamefont
  {Lu}}, \bibinfo {author} {\bibfnamefont {A.}~\bibnamefont {Frano}}, \bibinfo
  {author} {\bibfnamefont {M.}~\bibnamefont {Bluschke}}, \bibinfo {author}
  {\bibfnamefont {M.}~\bibnamefont {Hepting}}, \bibinfo {author} {\bibfnamefont
  {S.}~\bibnamefont {Macke}}, \bibinfo {author} {\bibfnamefont
  {J.}~\bibnamefont {Strempfer}}, \bibinfo {author} {\bibfnamefont
  {P.}~\bibnamefont {Wochner}}, \bibinfo {author} {\bibfnamefont
  {G.}~\bibnamefont {Cristiani}}, \bibinfo {author} {\bibfnamefont
  {G.}~\bibnamefont {Logvenov}}, \bibinfo {author} {\bibfnamefont {H.-U.}\
  \bibnamefont {Habermeier}}, \bibinfo {author} {\bibfnamefont {M.~W.}\
  \bibnamefont {Haverkort}}, \bibinfo {author} {\bibfnamefont {B.}~\bibnamefont
  {Keimer}}, \ and\ \bibinfo {author} {\bibfnamefont {E.}~\bibnamefont
  {Benckiser}},\ }\href {\doibase 10.1103/PhysRevB.93.165121} {\bibfield
  {journal} {\bibinfo  {journal} {Phys. Rev. B}\ }\textbf {\bibinfo {volume}
  {93}},\ \bibinfo {pages} {165121} (\bibinfo {year} {2016})}\BibitemShut
  {NoStop}%
\bibitem [{\citenamefont {Chen}\ \emph {et~al.}(2019)\citenamefont {Chen},
  \citenamefont {Mao}, \citenamefont {Ge}, \citenamefont {Wang}, \citenamefont
  {Ke}, \citenamefont {Wang}, \citenamefont {Wang}, \citenamefont
  {D{\"{o}}beli}, \citenamefont {Geng}, \citenamefont {Matsuzaki},
  \citenamefont {Shi},\ and\ \citenamefont {Jiang}}]{Chen2019}%
  \BibitemOpen
  \bibfield  {author} {\bibinfo {author} {\bibfnamefont {J.}~\bibnamefont
  {Chen}}, \bibinfo {author} {\bibfnamefont {W.}~\bibnamefont {Mao}}, \bibinfo
  {author} {\bibfnamefont {B.}~\bibnamefont {Ge}}, \bibinfo {author}
  {\bibfnamefont {J.}~\bibnamefont {Wang}}, \bibinfo {author} {\bibfnamefont
  {X.}~\bibnamefont {Ke}}, \bibinfo {author} {\bibfnamefont {V.}~\bibnamefont
  {Wang}}, \bibinfo {author} {\bibfnamefont {Y.}~\bibnamefont {Wang}}, \bibinfo
  {author} {\bibfnamefont {M.}~\bibnamefont {D{\"{o}}beli}}, \bibinfo {author}
  {\bibfnamefont {W.}~\bibnamefont {Geng}}, \bibinfo {author} {\bibfnamefont
  {H.}~\bibnamefont {Matsuzaki}}, \bibinfo {author} {\bibfnamefont
  {J.}~\bibnamefont {Shi}}, \ and\ \bibinfo {author} {\bibfnamefont
  {Y.}~\bibnamefont {Jiang}},\ }\href {\doibase 10.1038/s41467-019-08613-3}
  {\bibfield  {journal} {\bibinfo  {journal} {Nat. Commun.}\ }\textbf {\bibinfo
  {volume} {10}},\ \bibinfo {pages} {694} (\bibinfo {year} {2019})}\BibitemShut
  {NoStop}%
\bibitem [{\citenamefont {Glazer}(1972)}]{Glazer1972}%
  \BibitemOpen
  \bibfield  {author} {\bibinfo {author} {\bibfnamefont {A.~M.}\ \bibnamefont
  {Glazer}},\ }\href {\doibase 10.1107/S0567740872007976} {\bibfield  {journal}
  {\bibinfo  {journal} {Acta Cryst. B}\ }\textbf {\bibinfo {volume} {28}},\
  \bibinfo {pages} {3384} (\bibinfo {year} {1972})}\BibitemShut {NoStop}%
\bibitem [{\citenamefont {Jain}\ \emph {et~al.}(2013)\citenamefont {Jain},
  \citenamefont {Ong}, \citenamefont {Hautier}, \citenamefont {Chen},
  \citenamefont {Richards}, \citenamefont {Dacek}, \citenamefont {Cholia},
  \citenamefont {Gunter}, \citenamefont {Skinner}, \citenamefont {Ceder},\ and\
  \citenamefont {Persson}}]{Jain2013}%
  \BibitemOpen
  \bibfield  {author} {\bibinfo {author} {\bibfnamefont {A.}~\bibnamefont
  {Jain}}, \bibinfo {author} {\bibfnamefont {S.~P.}\ \bibnamefont {Ong}},
  \bibinfo {author} {\bibfnamefont {G.}~\bibnamefont {Hautier}}, \bibinfo
  {author} {\bibfnamefont {W.}~\bibnamefont {Chen}}, \bibinfo {author}
  {\bibfnamefont {W.~D.}\ \bibnamefont {Richards}}, \bibinfo {author}
  {\bibfnamefont {S.}~\bibnamefont {Dacek}}, \bibinfo {author} {\bibfnamefont
  {S.}~\bibnamefont {Cholia}}, \bibinfo {author} {\bibfnamefont
  {D.}~\bibnamefont {Gunter}}, \bibinfo {author} {\bibfnamefont
  {D.}~\bibnamefont {Skinner}}, \bibinfo {author} {\bibfnamefont
  {G.}~\bibnamefont {Ceder}}, \ and\ \bibinfo {author} {\bibfnamefont {K.~A.}\
  \bibnamefont {Persson}},\ }\href {\doibase 10.1063/1.4812323} {\bibfield
  {journal} {\bibinfo  {journal} {APL Mater.}\ }\textbf {\bibinfo {volume}
  {1}},\ \bibinfo {pages} {11002} (\bibinfo {year} {2013})}\BibitemShut
  {NoStop}%
\bibitem [{\citenamefont {Catalano}\ \emph {et~al.}(2018)\citenamefont
  {Catalano}, \citenamefont {Gibert}, \citenamefont {Fowlie}, \citenamefont
  {{\'{I}}{\~{n}}iguez}, \citenamefont {Triscone},\ and\ \citenamefont
  {Kreisel}}]{Catalano2018}%
  \BibitemOpen
  \bibfield  {author} {\bibinfo {author} {\bibfnamefont {S.}~\bibnamefont
  {Catalano}}, \bibinfo {author} {\bibfnamefont {M.}~\bibnamefont {Gibert}},
  \bibinfo {author} {\bibfnamefont {J.}~\bibnamefont {Fowlie}}, \bibinfo
  {author} {\bibfnamefont {J.}~\bibnamefont {{\'{I}}{\~{n}}iguez}}, \bibinfo
  {author} {\bibfnamefont {J.-M.}\ \bibnamefont {Triscone}}, \ and\ \bibinfo
  {author} {\bibfnamefont {J.}~\bibnamefont {Kreisel}},\ }\href {\doibase
  10.1088/1361-6633/aaa37a} {\bibfield  {journal} {\bibinfo  {journal} {Rep.
  Prog. Phys.}\ }\textbf {\bibinfo {volume} {81}},\ \bibinfo {pages} {46501}
  (\bibinfo {year} {2018})}\BibitemShut {NoStop}%
\bibitem [{\citenamefont {Glazer}(1975)}]{Glazer1975}%
  \BibitemOpen
  \bibfield  {author} {\bibinfo {author} {\bibfnamefont {A.~M.}\ \bibnamefont
  {Glazer}},\ }\href {\doibase 10.1107/S0567739475001635} {\bibfield  {journal}
  {\bibinfo  {journal} {Acta Cryst. A}\ }\textbf {\bibinfo {volume} {31}},\
  \bibinfo {pages} {756} (\bibinfo {year} {1975})}\BibitemShut {NoStop}%
\bibitem [{\citenamefont {Shi}\ \emph {et~al.}(2014)\citenamefont {Shi},
  \citenamefont {Zhou},\ and\ \citenamefont {Ramanathan}}]{Shi2014}%
  \BibitemOpen
  \bibfield  {author} {\bibinfo {author} {\bibfnamefont {J.}~\bibnamefont
  {Shi}}, \bibinfo {author} {\bibfnamefont {Y.}~\bibnamefont {Zhou}}, \ and\
  \bibinfo {author} {\bibfnamefont {S.}~\bibnamefont {Ramanathan}},\ }\href
  {\doibase 10.1038/ncomms5860} {\bibfield  {journal} {\bibinfo  {journal}
  {Nat. Commun.}\ }\textbf {\bibinfo {volume} {5}},\ \bibinfo {pages} {4860}
  (\bibinfo {year} {2014})}\BibitemShut {NoStop}%
\bibitem [{\citenamefont {Zhou}\ \emph {et~al.}(2016)\citenamefont {Zhou},
  \citenamefont {Guan}, \citenamefont {Zhou}, \citenamefont {Ramadoss},
  \citenamefont {Adam}, \citenamefont {Liu}, \citenamefont {Lee}, \citenamefont
  {Shi}, \citenamefont {Tsuchiya}, \citenamefont {Fong},\ and\ \citenamefont
  {Ramanathan}}]{Zhou2016}%
  \BibitemOpen
  \bibfield  {author} {\bibinfo {author} {\bibfnamefont {Y.}~\bibnamefont
  {Zhou}}, \bibinfo {author} {\bibfnamefont {X.}~\bibnamefont {Guan}}, \bibinfo
  {author} {\bibfnamefont {H.}~\bibnamefont {Zhou}}, \bibinfo {author}
  {\bibfnamefont {K.}~\bibnamefont {Ramadoss}}, \bibinfo {author}
  {\bibfnamefont {S.}~\bibnamefont {Adam}}, \bibinfo {author} {\bibfnamefont
  {H.}~\bibnamefont {Liu}}, \bibinfo {author} {\bibfnamefont {S.}~\bibnamefont
  {Lee}}, \bibinfo {author} {\bibfnamefont {J.}~\bibnamefont {Shi}}, \bibinfo
  {author} {\bibfnamefont {M.}~\bibnamefont {Tsuchiya}}, \bibinfo {author}
  {\bibfnamefont {D.~D.}\ \bibnamefont {Fong}}, \ and\ \bibinfo {author}
  {\bibfnamefont {S.}~\bibnamefont {Ramanathan}},\ }\href {\doibase
  10.1038/nature17653} {\bibfield  {journal} {\bibinfo  {journal} {Nature}\
  }\textbf {\bibinfo {volume} {534}},\ \bibinfo {pages} {231} (\bibinfo {year}
  {2016})}\BibitemShut {NoStop}%
\bibitem [{\citenamefont {Ramadoss}\ \emph {et~al.}(2016)\citenamefont
  {Ramadoss}, \citenamefont {Mandal}, \citenamefont {Dai}, \citenamefont {Wan},
  \citenamefont {Zhou}, \citenamefont {Rokhinson}, \citenamefont {Chen},
  \citenamefont {Hu},\ and\ \citenamefont {Ramanathan}}]{Ramadoss2016}%
  \BibitemOpen
  \bibfield  {author} {\bibinfo {author} {\bibfnamefont {K.}~\bibnamefont
  {Ramadoss}}, \bibinfo {author} {\bibfnamefont {N.}~\bibnamefont {Mandal}},
  \bibinfo {author} {\bibfnamefont {X.}~\bibnamefont {Dai}}, \bibinfo {author}
  {\bibfnamefont {Z.}~\bibnamefont {Wan}}, \bibinfo {author} {\bibfnamefont
  {Y.}~\bibnamefont {Zhou}}, \bibinfo {author} {\bibfnamefont {L.}~\bibnamefont
  {Rokhinson}}, \bibinfo {author} {\bibfnamefont {Y.~P.}\ \bibnamefont {Chen}},
  \bibinfo {author} {\bibfnamefont {J.}~\bibnamefont {Hu}}, \ and\ \bibinfo
  {author} {\bibfnamefont {S.}~\bibnamefont {Ramanathan}},\ }\href {\doibase
  10.1103/PhysRevB.94.235124} {\bibfield  {journal} {\bibinfo  {journal} {Phys.
  Rev. B}\ }\textbf {\bibinfo {volume} {94}},\ \bibinfo {pages} {235124}
  (\bibinfo {year} {2016})}\BibitemShut {NoStop}%
\bibitem [{\citenamefont {Sun}\ \emph {et~al.}(2018)\citenamefont {Sun},
  \citenamefont {Kotiuga}, \citenamefont {Lim}, \citenamefont {Narayanan},
  \citenamefont {Cherukara}, \citenamefont {Zhang}, \citenamefont {Dong},
  \citenamefont {Kou}, \citenamefont {Sun}, \citenamefont {Lu}, \citenamefont
  {Waluyo}, \citenamefont {Hunt}, \citenamefont {Tanaka}, \citenamefont
  {Hattori}, \citenamefont {Gamage}, \citenamefont {Abate}, \citenamefont
  {Pol}, \citenamefont {Zhou}, \citenamefont {Sankaranarayanan}, \citenamefont
  {Yildiz}, \citenamefont {Rabe},\ and\ \citenamefont {Ramanathan}}]{Sun2018}%
  \BibitemOpen
  \bibfield  {author} {\bibinfo {author} {\bibfnamefont {Y.}~\bibnamefont
  {Sun}}, \bibinfo {author} {\bibfnamefont {M.}~\bibnamefont {Kotiuga}},
  \bibinfo {author} {\bibfnamefont {D.}~\bibnamefont {Lim}}, \bibinfo {author}
  {\bibfnamefont {B.}~\bibnamefont {Narayanan}}, \bibinfo {author}
  {\bibfnamefont {M.}~\bibnamefont {Cherukara}}, \bibinfo {author}
  {\bibfnamefont {Z.}~\bibnamefont {Zhang}}, \bibinfo {author} {\bibfnamefont
  {Y.}~\bibnamefont {Dong}}, \bibinfo {author} {\bibfnamefont {R.}~\bibnamefont
  {Kou}}, \bibinfo {author} {\bibfnamefont {C.-J.}\ \bibnamefont {Sun}},
  \bibinfo {author} {\bibfnamefont {Q.}~\bibnamefont {Lu}}, \bibinfo {author}
  {\bibfnamefont {I.}~\bibnamefont {Waluyo}}, \bibinfo {author} {\bibfnamefont
  {A.}~\bibnamefont {Hunt}}, \bibinfo {author} {\bibfnamefont {H.}~\bibnamefont
  {Tanaka}}, \bibinfo {author} {\bibfnamefont {A.~N.}\ \bibnamefont {Hattori}},
  \bibinfo {author} {\bibfnamefont {S.}~\bibnamefont {Gamage}}, \bibinfo
  {author} {\bibfnamefont {Y.}~\bibnamefont {Abate}}, \bibinfo {author}
  {\bibfnamefont {V.~G.}\ \bibnamefont {Pol}}, \bibinfo {author} {\bibfnamefont
  {H.}~\bibnamefont {Zhou}}, \bibinfo {author} {\bibfnamefont {S.~K. R.~S.}\
  \bibnamefont {Sankaranarayanan}}, \bibinfo {author} {\bibfnamefont
  {B.}~\bibnamefont {Yildiz}}, \bibinfo {author} {\bibfnamefont {K.~M.}\
  \bibnamefont {Rabe}}, \ and\ \bibinfo {author} {\bibfnamefont
  {S.}~\bibnamefont {Ramanathan}},\ }\href {\doibase 10.1073/pnas.1805029115}
  {\bibfield  {journal} {\bibinfo  {journal} {Proc. Natl. Acad. Sci. U.S.A.}\
  }\textbf {\bibinfo {volume} {115}},\ \bibinfo {pages} {9672} (\bibinfo {year}
  {2018})}\BibitemShut {NoStop}%
\bibitem [{\citenamefont {Liao}\ \emph {et~al.}(2018)\citenamefont {Liao},
  \citenamefont {Gauquelin}, \citenamefont {Green}, \citenamefont
  {M{\"{u}}ller-Caspary}, \citenamefont {Lobato}, \citenamefont {Li},
  \citenamefont {{Van Aert}}, \citenamefont {Verbeeck}, \citenamefont
  {Huijben}, \citenamefont {Grisolia}, \citenamefont {Rouco}, \citenamefont
  {{El Hage}}, \citenamefont {Villegas}, \citenamefont {Mercy}, \citenamefont
  {Bibes}, \citenamefont {Ghosez}, \citenamefont {Sawatzky}, \citenamefont
  {Rijnders},\ and\ \citenamefont {Koster}}]{Liao2018}%
  \BibitemOpen
  \bibfield  {author} {\bibinfo {author} {\bibfnamefont {Z.}~\bibnamefont
  {Liao}}, \bibinfo {author} {\bibfnamefont {N.}~\bibnamefont {Gauquelin}},
  \bibinfo {author} {\bibfnamefont {R.~J.}\ \bibnamefont {Green}}, \bibinfo
  {author} {\bibfnamefont {K.}~\bibnamefont {M{\"{u}}ller-Caspary}}, \bibinfo
  {author} {\bibfnamefont {I.}~\bibnamefont {Lobato}}, \bibinfo {author}
  {\bibfnamefont {L.}~\bibnamefont {Li}}, \bibinfo {author} {\bibfnamefont
  {S.}~\bibnamefont {{Van Aert}}}, \bibinfo {author} {\bibfnamefont
  {J.}~\bibnamefont {Verbeeck}}, \bibinfo {author} {\bibfnamefont
  {M.}~\bibnamefont {Huijben}}, \bibinfo {author} {\bibfnamefont {M.~N.}\
  \bibnamefont {Grisolia}}, \bibinfo {author} {\bibfnamefont {V.}~\bibnamefont
  {Rouco}}, \bibinfo {author} {\bibfnamefont {R.}~\bibnamefont {{El Hage}}},
  \bibinfo {author} {\bibfnamefont {J.~E.}\ \bibnamefont {Villegas}}, \bibinfo
  {author} {\bibfnamefont {A.}~\bibnamefont {Mercy}}, \bibinfo {author}
  {\bibfnamefont {M.}~\bibnamefont {Bibes}}, \bibinfo {author} {\bibfnamefont
  {P.}~\bibnamefont {Ghosez}}, \bibinfo {author} {\bibfnamefont {G.~A.}\
  \bibnamefont {Sawatzky}}, \bibinfo {author} {\bibfnamefont {G.}~\bibnamefont
  {Rijnders}}, \ and\ \bibinfo {author} {\bibfnamefont {G.}~\bibnamefont
  {Koster}},\ }\href {\doibase 10.1073/pnas.1807457115} {\bibfield  {journal}
  {\bibinfo  {journal} {Proc. Natl. Acad. Sci. U.S.A.}\ }\textbf {\bibinfo
  {volume} {115}},\ \bibinfo {pages} {9515} (\bibinfo {year}
  {2018})}\BibitemShut {NoStop}%
\bibitem [{Sup()}]{SupplNote}%
  \BibitemOpen
  \href@noop {} {}\bibinfo {note} {See Supplemental Material for supporting
  details on experiment, simulations of x-ray diffraction, and DFT
  calculations.}\BibitemShut {Stop}%
\bibitem [{\citenamefont {{Mansour N.}}\ and\ \citenamefont {{Melendres
  A.}}(1997)}]{Mansour1997}%
  \BibitemOpen
  \bibfield  {author} {\bibinfo {author} {\bibfnamefont {A.}~\bibnamefont
  {{Mansour N.}}}\ and\ \bibinfo {author} {\bibfnamefont {C.}~\bibnamefont
  {{Melendres A.}}},\ }\href {https://doi.org/10.1051/jp4:19972178} {\bibfield
  {journal} {\bibinfo  {journal} {J. Phys. IV France}\ }\textbf {\bibinfo
  {volume} {7}},\ \bibinfo {pages} {1171} (\bibinfo {year} {1997})}\BibitemShut
  {NoStop}%
\bibitem [{\citenamefont {Woolley}\ \emph {et~al.}(2011)\citenamefont
  {Woolley}, \citenamefont {Illy}, \citenamefont {Ryan},\ and\ \citenamefont
  {Skinner}}]{Woolley2011}%
  \BibitemOpen
  \bibfield  {author} {\bibinfo {author} {\bibfnamefont {R.~J.}\ \bibnamefont
  {Woolley}}, \bibinfo {author} {\bibfnamefont {B.~N.}\ \bibnamefont {Illy}},
  \bibinfo {author} {\bibfnamefont {M.~P.}\ \bibnamefont {Ryan}}, \ and\
  \bibinfo {author} {\bibfnamefont {S.~J.}\ \bibnamefont {Skinner}},\ }\href
  {\doibase 10.1039/C1JM14320D} {\bibfield  {journal} {\bibinfo  {journal} {J.
  Mater. Chem.}\ }\textbf {\bibinfo {volume} {21}},\ \bibinfo {pages} {18592}
  (\bibinfo {year} {2011})}\BibitemShut {NoStop}%
\bibitem [{\citenamefont {Gu}\ \emph {et~al.}(2014)\citenamefont {Gu},
  \citenamefont {Wang},\ and\ \citenamefont {Wang}}]{Gu2014}%
  \BibitemOpen
  \bibfield  {author} {\bibinfo {author} {\bibfnamefont {W.}~\bibnamefont
  {Gu}}, \bibinfo {author} {\bibfnamefont {H.}~\bibnamefont {Wang}}, \ and\
  \bibinfo {author} {\bibfnamefont {K.}~\bibnamefont {Wang}},\ }\href {\doibase
  10.1039/C4DT00308J} {\bibfield  {journal} {\bibinfo  {journal} {Dalton
  Trans.}\ }\textbf {\bibinfo {volume} {43}},\ \bibinfo {pages} {6406}
  (\bibinfo {year} {2014})}\BibitemShut {NoStop}%
\bibitem [{\citenamefont {Zuo}\ \emph {et~al.}(2017)\citenamefont {Zuo},
  \citenamefont {Panda}, \citenamefont {Kotiuga}, \citenamefont {Li},
  \citenamefont {Kang}, \citenamefont {Mazzoli}, \citenamefont {Zhou},
  \citenamefont {Barbour}, \citenamefont {Wilkins}, \citenamefont {Narayanan},
  \citenamefont {Cherukara}, \citenamefont {Zhang}, \citenamefont
  {Sankaranarayanan}, \citenamefont {Comin}, \citenamefont {Rabe},
  \citenamefont {Roy},\ and\ \citenamefont {Ramanathan}}]{Zuo2017}%
  \BibitemOpen
  \bibfield  {author} {\bibinfo {author} {\bibfnamefont {F.}~\bibnamefont
  {Zuo}}, \bibinfo {author} {\bibfnamefont {P.}~\bibnamefont {Panda}}, \bibinfo
  {author} {\bibfnamefont {M.}~\bibnamefont {Kotiuga}}, \bibinfo {author}
  {\bibfnamefont {J.}~\bibnamefont {Li}}, \bibinfo {author} {\bibfnamefont
  {M.}~\bibnamefont {Kang}}, \bibinfo {author} {\bibfnamefont {C.}~\bibnamefont
  {Mazzoli}}, \bibinfo {author} {\bibfnamefont {H.}~\bibnamefont {Zhou}},
  \bibinfo {author} {\bibfnamefont {A.}~\bibnamefont {Barbour}}, \bibinfo
  {author} {\bibfnamefont {S.}~\bibnamefont {Wilkins}}, \bibinfo {author}
  {\bibfnamefont {B.}~\bibnamefont {Narayanan}}, \bibinfo {author}
  {\bibfnamefont {M.}~\bibnamefont {Cherukara}}, \bibinfo {author}
  {\bibfnamefont {Z.}~\bibnamefont {Zhang}}, \bibinfo {author} {\bibfnamefont
  {S.~K. R.~S.}\ \bibnamefont {Sankaranarayanan}}, \bibinfo {author}
  {\bibfnamefont {R.}~\bibnamefont {Comin}}, \bibinfo {author} {\bibfnamefont
  {K.~M.}\ \bibnamefont {Rabe}}, \bibinfo {author} {\bibfnamefont
  {K.}~\bibnamefont {Roy}}, \ and\ \bibinfo {author} {\bibfnamefont
  {S.}~\bibnamefont {Ramanathan}},\ }\href {\doibase
  10.1038/s41467-017-00248-6} {\bibfield  {journal} {\bibinfo  {journal} {Nat.
  Commun.}\ }\textbf {\bibinfo {volume} {8}},\ \bibinfo {pages} {240} (\bibinfo
  {year} {2017})}\BibitemShut {NoStop}%
\bibitem [{\citenamefont {Zhang}\ \emph {et~al.}(2018)\citenamefont {Zhang},
  \citenamefont {Schwanz}, \citenamefont {Narayanan}, \citenamefont {Kotiuga},
  \citenamefont {Dura}, \citenamefont {Cherukara}, \citenamefont {Zhou},
  \citenamefont {Freeland}, \citenamefont {Li}, \citenamefont {Sutarto},
  \citenamefont {He}, \citenamefont {Wu}, \citenamefont {Zhu}, \citenamefont
  {Sun}, \citenamefont {Ramadoss}, \citenamefont {Nonnenmann}, \citenamefont
  {Yu}, \citenamefont {Comin}, \citenamefont {Rabe}, \citenamefont
  {Sankaranarayanan},\ and\ \citenamefont {Ramanathan}}]{Zhang2018}%
  \BibitemOpen
  \bibfield  {author} {\bibinfo {author} {\bibfnamefont {Z.}~\bibnamefont
  {Zhang}}, \bibinfo {author} {\bibfnamefont {D.}~\bibnamefont {Schwanz}},
  \bibinfo {author} {\bibfnamefont {B.}~\bibnamefont {Narayanan}}, \bibinfo
  {author} {\bibfnamefont {M.}~\bibnamefont {Kotiuga}}, \bibinfo {author}
  {\bibfnamefont {J.~A.}\ \bibnamefont {Dura}}, \bibinfo {author}
  {\bibfnamefont {M.}~\bibnamefont {Cherukara}}, \bibinfo {author}
  {\bibfnamefont {H.}~\bibnamefont {Zhou}}, \bibinfo {author} {\bibfnamefont
  {J.~W.}\ \bibnamefont {Freeland}}, \bibinfo {author} {\bibfnamefont
  {J.}~\bibnamefont {Li}}, \bibinfo {author} {\bibfnamefont {R.}~\bibnamefont
  {Sutarto}}, \bibinfo {author} {\bibfnamefont {F.}~\bibnamefont {He}},
  \bibinfo {author} {\bibfnamefont {C.}~\bibnamefont {Wu}}, \bibinfo {author}
  {\bibfnamefont {J.}~\bibnamefont {Zhu}}, \bibinfo {author} {\bibfnamefont
  {Y.}~\bibnamefont {Sun}}, \bibinfo {author} {\bibfnamefont {K.}~\bibnamefont
  {Ramadoss}}, \bibinfo {author} {\bibfnamefont {S.~S.}\ \bibnamefont
  {Nonnenmann}}, \bibinfo {author} {\bibfnamefont {N.}~\bibnamefont {Yu}},
  \bibinfo {author} {\bibfnamefont {R.}~\bibnamefont {Comin}}, \bibinfo
  {author} {\bibfnamefont {K.~M.}\ \bibnamefont {Rabe}}, \bibinfo {author}
  {\bibfnamefont {S.~K. R.~S.}\ \bibnamefont {Sankaranarayanan}}, \ and\
  \bibinfo {author} {\bibfnamefont {S.}~\bibnamefont {Ramanathan}},\ }\href
  {\doibase 10.1038/nature25008} {\bibfield  {journal} {\bibinfo  {journal}
  {Nature}\ }\textbf {\bibinfo {volume} {553}},\ \bibinfo {pages} {68}
  (\bibinfo {year} {2018})}\BibitemShut {NoStop}%
\bibitem [{\citenamefont {Kotiuga}\ and\ \citenamefont
  {Rabe}(2019)}]{Kotiuga2019}%
  \BibitemOpen
  \bibfield  {author} {\bibinfo {author} {\bibfnamefont {M.}~\bibnamefont
  {Kotiuga}}\ and\ \bibinfo {author} {\bibfnamefont {K.~M.}\ \bibnamefont
  {Rabe}},\ }\href {\doibase 10.1103/PhysRevMaterials.3.115002} {\bibfield
  {journal} {\bibinfo  {journal} {Phys. Rev. Mater.}\ }\textbf {\bibinfo
  {volume} {3}},\ \bibinfo {pages} {115002} (\bibinfo {year} {2019})},\ \Eprint
  {http://arxiv.org/abs/1909.03425} {1909.03425} \BibitemShut {NoStop}%
\bibitem [{\citenamefont {Kresse}\ and\ \citenamefont
  {Furthm{\"{u}}ller}(1996{\natexlab{a}})}]{Kresse1996}%
  \BibitemOpen
  \bibfield  {author} {\bibinfo {author} {\bibfnamefont {G.}~\bibnamefont
  {Kresse}}\ and\ \bibinfo {author} {\bibfnamefont {J.}~\bibnamefont
  {Furthm{\"{u}}ller}},\ }\href {\doibase 10.1103/PhysRevB.54.11169} {\bibfield
   {journal} {\bibinfo  {journal} {Phys. Rev. B}\ }\textbf {\bibinfo {volume}
  {54}},\ \bibinfo {pages} {11169} (\bibinfo {year}
  {1996}{\natexlab{a}})}\BibitemShut {NoStop}%
\bibitem [{\citenamefont {Kresse}\ and\ \citenamefont
  {Joubert}(1999)}]{Kresse1999}%
  \BibitemOpen
  \bibfield  {author} {\bibinfo {author} {\bibfnamefont {G.}~\bibnamefont
  {Kresse}}\ and\ \bibinfo {author} {\bibfnamefont {D.}~\bibnamefont
  {Joubert}},\ }\href {\doibase 10.1103/PhysRevB.59.1758} {\bibfield  {journal}
  {\bibinfo  {journal} {Phys. Rev. B}\ }\textbf {\bibinfo {volume} {59}},\
  \bibinfo {pages} {1758} (\bibinfo {year} {1999})}\BibitemShut {NoStop}%
\bibitem [{\citenamefont {Rehr}\ \emph {et~al.}(2010)\citenamefont {Rehr},
  \citenamefont {Kas}, \citenamefont {Vila}, \citenamefont {Prange},\ and\
  \citenamefont {Jorissen}}]{Rehr2010}%
  \BibitemOpen
  \bibfield  {author} {\bibinfo {author} {\bibfnamefont {J.~J.}\ \bibnamefont
  {Rehr}}, \bibinfo {author} {\bibfnamefont {J.~J.}\ \bibnamefont {Kas}},
  \bibinfo {author} {\bibfnamefont {F.~D.}\ \bibnamefont {Vila}}, \bibinfo
  {author} {\bibfnamefont {M.~P.}\ \bibnamefont {Prange}}, \ and\ \bibinfo
  {author} {\bibfnamefont {K.}~\bibnamefont {Jorissen}},\ }\href {\doibase
  10.1039/B926434E} {\bibfield  {journal} {\bibinfo  {journal} {Phys. Chem.
  Chem. Phys.}\ }\textbf {\bibinfo {volume} {12}},\ \bibinfo {pages} {5503}
  (\bibinfo {year} {2010})}\BibitemShut {NoStop}%
\bibitem [{\citenamefont {Yoo}\ and\ \citenamefont {Liao}(2018)}]{Yoo2018}%
  \BibitemOpen
  \bibfield  {author} {\bibinfo {author} {\bibfnamefont {P.}~\bibnamefont
  {Yoo}}\ and\ \bibinfo {author} {\bibfnamefont {P.}~\bibnamefont {Liao}},\
  }\href {\doibase 10.1039/C8ME00002F} {\bibfield  {journal} {\bibinfo
  {journal} {Mol. Syst. Des. Eng.}\ }\textbf {\bibinfo {volume} {3}},\ \bibinfo
  {pages} {264} (\bibinfo {year} {2018})}\BibitemShut {NoStop}%
\bibitem [{\citenamefont {Yoo}\ and\ \citenamefont {Liao}(2020)}]{Yoo2020}%
  \BibitemOpen
  \bibfield  {author} {\bibinfo {author} {\bibfnamefont {P.}~\bibnamefont
  {Yoo}}\ and\ \bibinfo {author} {\bibfnamefont {P.}~\bibnamefont {Liao}},\
  }\href {\doibase 10.1039/C9CP06522A} {\bibfield  {journal} {\bibinfo
  {journal} {Phys. Chem. Chem. Phys.}\ }\textbf {\bibinfo {volume} {22}},\
  \bibinfo {pages} {6888} (\bibinfo {year} {2020})}\BibitemShut {NoStop}%
\bibitem [{\citenamefont {Goteti}\ \emph {et~al.}(2021)\citenamefont {Goteti},
  \citenamefont {Zaluzhnyy}, \citenamefont {Ramanathan}, \citenamefont
  {Dynes},\ and\ \citenamefont {Frano}}]{Goteti2021}%
  \BibitemOpen
  \bibfield  {author} {\bibinfo {author} {\bibfnamefont {U.}~\bibnamefont
  {Goteti}}, \bibinfo {author} {\bibfnamefont {I.~A.}\ \bibnamefont
  {Zaluzhnyy}}, \bibinfo {author} {\bibfnamefont {S.}~\bibnamefont
  {Ramanathan}}, \bibinfo {author} {\bibfnamefont {R.~C.}\ \bibnamefont
  {Dynes}}, \ and\ \bibinfo {author} {\bibfnamefont {A.}~\bibnamefont
  {Frano}},\ }\href {\doibase 10.1073/pnas.2103934118} {\bibfield  {journal}
  {\bibinfo  {journal} {Proc. Natl. Acad. Sci. U.S.A.}\ } (\bibinfo {year}
  {2021}),\ 10.1073/pnas.2103934118}\BibitemShut {NoStop}%
\bibitem [{\citenamefont {Alonso}\ \emph {et~al.}(1999)\citenamefont {Alonso},
  \citenamefont {Garc\'{\i}a-Mu\~noz}, \citenamefont {Fern\'andez-D\'{\i}az},
  \citenamefont {Aranda}, \citenamefont {Mart\'{\i}nez-Lope},\ and\
  \citenamefont {Casais}}]{Alonso1999}%
  \BibitemOpen
  \bibfield  {author} {\bibinfo {author} {\bibfnamefont {J.~A.}\ \bibnamefont
  {Alonso}}, \bibinfo {author} {\bibfnamefont {J.~L.}\ \bibnamefont
  {Garc\'{\i}a-Mu\~noz}}, \bibinfo {author} {\bibfnamefont {M.~T.}\
  \bibnamefont {Fern\'andez-D\'{\i}az}}, \bibinfo {author} {\bibfnamefont
  {M.~A.~G.}\ \bibnamefont {Aranda}}, \bibinfo {author} {\bibfnamefont {M.~J.}\
  \bibnamefont {Mart\'{\i}nez-Lope}}, \ and\ \bibinfo {author} {\bibfnamefont
  {M.~T.}\ \bibnamefont {Casais}},\ }\href {\doibase
  10.1103/PhysRevLett.82.3871} {\bibfield  {journal} {\bibinfo  {journal}
  {Phys. Rev. Lett.}\ }\textbf {\bibinfo {volume} {82}},\ \bibinfo {pages}
  {3871} (\bibinfo {year} {1999})}\BibitemShut {NoStop}%
\bibitem [{\citenamefont {Green}\ \emph {et~al.}(2016)\citenamefont {Green},
  \citenamefont {Haverkort},\ and\ \citenamefont {Sawatzky}}]{Green2016}%
  \BibitemOpen
  \bibfield  {author} {\bibinfo {author} {\bibfnamefont {R.~J.}\ \bibnamefont
  {Green}}, \bibinfo {author} {\bibfnamefont {M.~W.}\ \bibnamefont
  {Haverkort}}, \ and\ \bibinfo {author} {\bibfnamefont {G.~A.}\ \bibnamefont
  {Sawatzky}},\ }\href {\doibase 10.1103/PhysRevB.94.195127} {\bibfield
  {journal} {\bibinfo  {journal} {Phys. Rev. B}\ }\textbf {\bibinfo {volume}
  {94}},\ \bibinfo {pages} {195127} (\bibinfo {year} {2016})}\BibitemShut
  {NoStop}%
\bibitem [{\citenamefont {Serrano-S\'{a}nchez}\ \emph
  {et~al.}(2019)\citenamefont {Serrano-S\'{a}nchez}, \citenamefont {Fauth},
  \citenamefont {Martínez},\ and\ \citenamefont {Alonso}}]{Serrano2019}%
  \BibitemOpen
  \bibfield  {author} {\bibinfo {author} {\bibfnamefont {F.}~\bibnamefont
  {Serrano-S\'{a}nchez}}, \bibinfo {author} {\bibfnamefont {F.}~\bibnamefont
  {Fauth}}, \bibinfo {author} {\bibfnamefont {J.~L.}\ \bibnamefont
  {Martínez}}, \ and\ \bibinfo {author} {\bibfnamefont {J.~A.}\ \bibnamefont
  {Alonso}},\ }\href {\doibase 10.1021/acs.inorgchem.9b02013} {\bibfield
  {journal} {\bibinfo  {journal} {Inorg. Chem.}\ }\textbf {\bibinfo {volume}
  {58}},\ \bibinfo {pages} {11828} (\bibinfo {year} {2019})}\BibitemShut
  {NoStop}%
\bibitem [{\citenamefont {Benedek}\ and\ \citenamefont
  {Fennie}(2013)}]{Benedek2013}%
  \BibitemOpen
  \bibfield  {author} {\bibinfo {author} {\bibfnamefont {N.~A.}\ \bibnamefont
  {Benedek}}\ and\ \bibinfo {author} {\bibfnamefont {C.~J.}\ \bibnamefont
  {Fennie}},\ }\href {\doibase 10.1021/jp402046t} {\bibfield  {journal}
  {\bibinfo  {journal} {J. Phys. Chem. C}\ }\textbf {\bibinfo {volume} {117}},\
  \bibinfo {pages} {13339} (\bibinfo {year} {2013})}\BibitemShut {NoStop}%
\bibitem [{\citenamefont {Varignon}\ \emph {et~al.}(2017)\citenamefont
  {Varignon}, \citenamefont {Grisolia}, \citenamefont {{\'{I}}{\~{n}}iguez},
  \citenamefont {Barth{\'{e}}l{\'{e}}my},\ and\ \citenamefont
  {Bibes}}]{Varignon2017}%
  \BibitemOpen
  \bibfield  {author} {\bibinfo {author} {\bibfnamefont {J.}~\bibnamefont
  {Varignon}}, \bibinfo {author} {\bibfnamefont {M.~N.}\ \bibnamefont
  {Grisolia}}, \bibinfo {author} {\bibfnamefont {J.}~\bibnamefont
  {{\'{I}}{\~{n}}iguez}}, \bibinfo {author} {\bibfnamefont {A.}~\bibnamefont
  {Barth{\'{e}}l{\'{e}}my}}, \ and\ \bibinfo {author} {\bibfnamefont
  {M.}~\bibnamefont {Bibes}},\ }\href {\doibase 10.1038/s41535-017-0024-9}
  {\bibfield  {journal} {\bibinfo  {journal} {npj Quantum Mater.}\ }\textbf
  {\bibinfo {volume} {2}},\ \bibinfo {pages} {21} (\bibinfo {year}
  {2017})}\BibitemShut {NoStop}%
\bibitem [{Not()}]{NotationNote}%
  \BibitemOpen
  \href@noop {} {}\bibinfo {note} {Historically these rotations were introduces
  for the pseudocubic symmetry. The same tillt of the NiO$_6$ octahedra can be
  obtained by rotation over 10.8$^\circ$, 10.8$^\circ$, and 7.0$^\circ$ around
  the [100]$_{pc}$, [010]$_{pc}$, and [001]$_{pc}$ pseudocubic directions,
  respectively.}\BibitemShut {Stop}%
\bibitem [{\citenamefont {Fowlie}\ \emph {et~al.}(2019)\citenamefont {Fowlie},
  \citenamefont {Lichtensteiger}, \citenamefont {Gibert}, \citenamefont
  {Meley}, \citenamefont {Willmott},\ and\ \citenamefont
  {Triscone}}]{Fowlie2019}%
  \BibitemOpen
  \bibfield  {author} {\bibinfo {author} {\bibfnamefont {J.}~\bibnamefont
  {Fowlie}}, \bibinfo {author} {\bibfnamefont {C.}~\bibnamefont
  {Lichtensteiger}}, \bibinfo {author} {\bibfnamefont {M.}~\bibnamefont
  {Gibert}}, \bibinfo {author} {\bibfnamefont {H.}~\bibnamefont {Meley}},
  \bibinfo {author} {\bibfnamefont {P.}~\bibnamefont {Willmott}}, \ and\
  \bibinfo {author} {\bibfnamefont {J.-M.}\ \bibnamefont {Triscone}},\ }\href
  {\doibase 10.1021/acs.nanolett.9b01772} {\bibfield  {journal} {\bibinfo
  {journal} {Nano Lett.}\ }\textbf {\bibinfo {volume} {19}},\ \bibinfo {pages}
  {4188} (\bibinfo {year} {2019})}\BibitemShut {NoStop}%
\bibitem [{\citenamefont {Brahlek}\ \emph {et~al.}(2017)\citenamefont
  {Brahlek}, \citenamefont {Choquette}, \citenamefont {Smith}, \citenamefont
  {Engel-Herbert},\ and\ \citenamefont {May}}]{Brahlek2017}%
  \BibitemOpen
  \bibfield  {author} {\bibinfo {author} {\bibfnamefont {M.}~\bibnamefont
  {Brahlek}}, \bibinfo {author} {\bibfnamefont {A.~K.}\ \bibnamefont
  {Choquette}}, \bibinfo {author} {\bibfnamefont {C.~R.}\ \bibnamefont
  {Smith}}, \bibinfo {author} {\bibfnamefont {R.}~\bibnamefont
  {Engel-Herbert}}, \ and\ \bibinfo {author} {\bibfnamefont {S.~J.}\
  \bibnamefont {May}},\ }\href {\doibase 10.1063/1.4974362} {\bibfield
  {journal} {\bibinfo  {journal} {J. Appl. Phys.}\ }\textbf {\bibinfo {volume}
  {121}},\ \bibinfo {pages} {45303} (\bibinfo {year} {2017})}\BibitemShut
  {NoStop}%
\bibitem [{\citenamefont {Hepting}(2017)}]{Hepting2017}%
  \BibitemOpen
  \bibfield  {author} {\bibinfo {author} {\bibfnamefont {M.}~\bibnamefont
  {Hepting}},\ }\href {\doibase 10.1007/978-3-319-60531-9} {\emph {\bibinfo
  {title} {{Ordering Phenomena in Rare-Earth Nickelate Heterostructures}}}},\
  Springer Theses\ (\bibinfo  {publisher} {Springer International Publishing},\
  \bibinfo {address} {Cham},\ \bibinfo {year} {2017})\BibitemShut {NoStop}%
\bibitem [{\citenamefont {Yan}\ \emph {et~al.}(2018)\citenamefont {Yan},
  \citenamefont {Bouet}, \citenamefont {Zhou}, \citenamefont {Huang},
  \citenamefont {Nazaretski}, \citenamefont {Xu}, \citenamefont {Cocco},
  \citenamefont {Chiu}, \citenamefont {Brinkman},\ and\ \citenamefont
  {Chu}}]{Yan2018}%
  \BibitemOpen
  \bibfield  {author} {\bibinfo {author} {\bibfnamefont {H.}~\bibnamefont
  {Yan}}, \bibinfo {author} {\bibfnamefont {N.}~\bibnamefont {Bouet}}, \bibinfo
  {author} {\bibfnamefont {J.}~\bibnamefont {Zhou}}, \bibinfo {author}
  {\bibfnamefont {X.}~\bibnamefont {Huang}}, \bibinfo {author} {\bibfnamefont
  {E.}~\bibnamefont {Nazaretski}}, \bibinfo {author} {\bibfnamefont
  {W.}~\bibnamefont {Xu}}, \bibinfo {author} {\bibfnamefont {A.~P.}\
  \bibnamefont {Cocco}}, \bibinfo {author} {\bibfnamefont {W.~K.~S.}\
  \bibnamefont {Chiu}}, \bibinfo {author} {\bibfnamefont {K.~S.}\ \bibnamefont
  {Brinkman}}, \ and\ \bibinfo {author} {\bibfnamefont {Y.~S.}\ \bibnamefont
  {Chu}},\ }\href {\doibase 10.1088/2399-1984/aab25d} {\bibfield  {journal}
  {\bibinfo  {journal} {Nano Futures}\ }\textbf {\bibinfo {volume} {2}},\
  \bibinfo {pages} {011001} (\bibinfo {year} {2018})}\BibitemShut {NoStop}%
\bibitem [{\citenamefont {Nazaretski}\ \emph {et~al.}(2017)\citenamefont
  {Nazaretski}, \citenamefont {Yan}, \citenamefont {Lauer}, \citenamefont
  {Bouet}, \citenamefont {Huang}, \citenamefont {Xu}, \citenamefont {Zhou},
  \citenamefont {Shu}, \citenamefont {Hwu},\ and\ \citenamefont
  {Chu}}]{Nazaretski2017}%
  \BibitemOpen
  \bibfield  {author} {\bibinfo {author} {\bibfnamefont {E.}~\bibnamefont
  {Nazaretski}}, \bibinfo {author} {\bibfnamefont {H.}~\bibnamefont {Yan}},
  \bibinfo {author} {\bibfnamefont {K.}~\bibnamefont {Lauer}}, \bibinfo
  {author} {\bibfnamefont {N.}~\bibnamefont {Bouet}}, \bibinfo {author}
  {\bibfnamefont {X.}~\bibnamefont {Huang}}, \bibinfo {author} {\bibfnamefont
  {W.}~\bibnamefont {Xu}}, \bibinfo {author} {\bibfnamefont {J.}~\bibnamefont
  {Zhou}}, \bibinfo {author} {\bibfnamefont {D.}~\bibnamefont {Shu}}, \bibinfo
  {author} {\bibfnamefont {Y.}~\bibnamefont {Hwu}}, \ and\ \bibinfo {author}
  {\bibfnamefont {Y.~S.}\ \bibnamefont {Chu}},\ }\href {\doibase
  10.1107/S1600577517011183} {\bibfield  {journal} {\bibinfo  {journal} {J.
  Synchrotron Radiat.}\ }\textbf {\bibinfo {volume} {24}},\ \bibinfo {pages}
  {1113} (\bibinfo {year} {2017})}\BibitemShut {NoStop}%
\bibitem [{\citenamefont {Pattammattel}\ \emph {et~al.}(2020)\citenamefont
  {Pattammattel}, \citenamefont {Tappero}, \citenamefont {Ge}, \citenamefont
  {Chu}, \citenamefont {Huang}, \citenamefont {Gao},\ and\ \citenamefont
  {Yan}}]{Pattammattel2020}%
  \BibitemOpen
  \bibfield  {author} {\bibinfo {author} {\bibfnamefont {A.}~\bibnamefont
  {Pattammattel}}, \bibinfo {author} {\bibfnamefont {R.}~\bibnamefont
  {Tappero}}, \bibinfo {author} {\bibfnamefont {M.}~\bibnamefont {Ge}},
  \bibinfo {author} {\bibfnamefont {Y.~S.}\ \bibnamefont {Chu}}, \bibinfo
  {author} {\bibfnamefont {X.}~\bibnamefont {Huang}}, \bibinfo {author}
  {\bibfnamefont {Y.}~\bibnamefont {Gao}}, \ and\ \bibinfo {author}
  {\bibfnamefont {H.}~\bibnamefont {Yan}},\ }\href {\doibase
  10.1126/sciadv.abb3615} {\bibfield  {journal} {\bibinfo  {journal} {Sci.
  Adv.}\ }\textbf {\bibinfo {volume} {6}},\ \bibinfo {pages} {eabb3615}
  (\bibinfo {year} {2020})}\BibitemShut {NoStop}%
\bibitem [{\citenamefont {Kresse}\ and\ \citenamefont
  {Furthm{\"{u}}ller}(1996{\natexlab{b}})}]{Kresse1996a}%
  \BibitemOpen
  \bibfield  {author} {\bibinfo {author} {\bibfnamefont {G.}~\bibnamefont
  {Kresse}}\ and\ \bibinfo {author} {\bibfnamefont {J.}~\bibnamefont
  {Furthm{\"{u}}ller}},\ }\href {\doibase
  https://doi.org/10.1103/PhysRevB.54.11169} {\bibfield  {journal} {\bibinfo
  {journal} {Phys. Rev. B}\ }\textbf {\bibinfo {volume} {54}},\ \bibinfo
  {pages} {11169} (\bibinfo {year} {1996}{\natexlab{b}})}\BibitemShut {NoStop}%
\bibitem [{\citenamefont {Kohn}\ and\ \citenamefont {Sham}(1965)}]{Kohn1965}%
  \BibitemOpen
  \bibfield  {author} {\bibinfo {author} {\bibfnamefont {W.}~\bibnamefont
  {Kohn}}\ and\ \bibinfo {author} {\bibfnamefont {L.~J.}\ \bibnamefont
  {Sham}},\ }\href {\doibase 10.1103/PhysRev.140.A1133} {\bibfield  {journal}
  {\bibinfo  {journal} {Phys. Rev.}\ }\textbf {\bibinfo {volume} {140}},\
  \bibinfo {pages} {1133} (\bibinfo {year} {1965})}\BibitemShut {NoStop}%
\bibitem [{\citenamefont {Bl{\"{o}}chl}(1994)}]{Blochl1994}%
  \BibitemOpen
  \bibfield  {author} {\bibinfo {author} {\bibfnamefont {P.~E.}\ \bibnamefont
  {Bl{\"{o}}chl}},\ }\href {\doibase 10.1103/PhysRevB.50.17953} {\bibfield
  {journal} {\bibinfo  {journal} {Phys. Rev. B}\ }\textbf {\bibinfo {volume}
  {50}},\ \bibinfo {pages} {17953} (\bibinfo {year} {1994})}\BibitemShut
  {NoStop}%
\bibitem [{\citenamefont {Perdew}\ \emph {et~al.}(1996)\citenamefont {Perdew},
  \citenamefont {Burke},\ and\ \citenamefont {Ernzerhof}}]{Perdew1996}%
  \BibitemOpen
  \bibfield  {author} {\bibinfo {author} {\bibfnamefont {J.~P.}\ \bibnamefont
  {Perdew}}, \bibinfo {author} {\bibfnamefont {K.}~\bibnamefont {Burke}}, \
  and\ \bibinfo {author} {\bibfnamefont {M.}~\bibnamefont {Ernzerhof}},\ }\href
  {\doibase 10.1103/PhysRevLett.77.3865} {\bibfield  {journal} {\bibinfo
  {journal} {Phys. Rev. Lett.}\ }\textbf {\bibinfo {volume} {77}},\ \bibinfo
  {pages} {3865} (\bibinfo {year} {1996})}\BibitemShut {NoStop}%
\bibitem [{\citenamefont {Anisimov}\ \emph {et~al.}(1991)\citenamefont
  {Anisimov}, \citenamefont {Zaanen},\ and\ \citenamefont
  {Andersen}}]{Anisimov1991}%
  \BibitemOpen
  \bibfield  {author} {\bibinfo {author} {\bibfnamefont {V.~I.}\ \bibnamefont
  {Anisimov}}, \bibinfo {author} {\bibfnamefont {J.}~\bibnamefont {Zaanen}}, \
  and\ \bibinfo {author} {\bibfnamefont {O.~K.}\ \bibnamefont {Andersen}},\
  }\href {\doibase 10.1103/PhysRevB.44.943} {\bibfield  {journal} {\bibinfo
  {journal} {Phys. Rev. B}\ }\textbf {\bibinfo {volume} {44}},\ \bibinfo
  {pages} {943} (\bibinfo {year} {1991})}\BibitemShut {NoStop}%
\bibitem [{\citenamefont {Ong}\ \emph {et~al.}(2013)\citenamefont {Ong},
  \citenamefont {Richards}, \citenamefont {Jain}, \citenamefont {Hautier},
  \citenamefont {Kocher}, \citenamefont {Cholia}, \citenamefont {Gunter},
  \citenamefont {Chevrier}, \citenamefont {Persson},\ and\ \citenamefont
  {Ceder}}]{Ong2013}%
  \BibitemOpen
  \bibfield  {author} {\bibinfo {author} {\bibfnamefont {S.~P.}\ \bibnamefont
  {Ong}}, \bibinfo {author} {\bibfnamefont {W.~D.}\ \bibnamefont {Richards}},
  \bibinfo {author} {\bibfnamefont {A.}~\bibnamefont {Jain}}, \bibinfo {author}
  {\bibfnamefont {G.}~\bibnamefont {Hautier}}, \bibinfo {author} {\bibfnamefont
  {M.}~\bibnamefont {Kocher}}, \bibinfo {author} {\bibfnamefont
  {S.}~\bibnamefont {Cholia}}, \bibinfo {author} {\bibfnamefont
  {D.}~\bibnamefont {Gunter}}, \bibinfo {author} {\bibfnamefont {V.~L.}\
  \bibnamefont {Chevrier}}, \bibinfo {author} {\bibfnamefont {K.~A.}\
  \bibnamefont {Persson}}, \ and\ \bibinfo {author} {\bibfnamefont
  {G.}~\bibnamefont {Ceder}},\ }\href {\doibase
  10.1016/j.commatsci.2012.10.028} {\bibfield  {journal} {\bibinfo  {journal}
  {Comput. Mater. Sci.}\ }\textbf {\bibinfo {volume} {68}},\ \bibinfo {pages}
  {314} (\bibinfo {year} {2013})}\BibitemShut {NoStop}%
\bibitem [{\citenamefont {Brown}\ \emph {et~al.}(2006)\citenamefont {Brown},
  \citenamefont {Fox}, \citenamefont {Maslen}, \citenamefont {O'Keefe},\ and\
  \citenamefont {Willis}}]{Brown2006}%
  \BibitemOpen
  \bibfield  {author} {\bibinfo {author} {\bibfnamefont {P.}~\bibnamefont
  {Brown}}, \bibinfo {author} {\bibfnamefont {A.~G.}\ \bibnamefont {Fox}},
  \bibinfo {author} {\bibfnamefont {E.~N.}\ \bibnamefont {Maslen}}, \bibinfo
  {author} {\bibfnamefont {M.~A.}\ \bibnamefont {O'Keefe}}, \ and\ \bibinfo
  {author} {\bibfnamefont {B.~T.~M.}\ \bibnamefont {Willis}},\ }\href {\doibase
  doi:10.1107/97809553602060000600} {\emph {\bibinfo {title} {{International
  Tables for Crystallography (Vol. C)}}}}\ (\bibinfo {year} {2006})\ pp.\
  \bibinfo {pages} {554--595}\BibitemShut {NoStop}%
\bibitem [{\citenamefont {Chantler}(1995)}]{Chantler1995}%
  \BibitemOpen
  \bibfield  {author} {\bibinfo {author} {\bibfnamefont {C.~T.}\ \bibnamefont
  {Chantler}},\ }\href {\doibase 10.1063/1.555974} {\bibfield  {journal}
  {\bibinfo  {journal} {J. Phys. Chem. Ref. Data}\ }\textbf {\bibinfo {volume}
  {24}},\ \bibinfo {pages} {71} (\bibinfo {year} {1995})}\BibitemShut {NoStop}%
\bibitem [{\citenamefont {Zimmermann}\ \emph {et~al.}(2017)\citenamefont
  {Zimmermann}, \citenamefont {Horton}, \citenamefont {Jain},\ and\
  \citenamefont {Haranczyk}}]{Zimmermann2017}%
  \BibitemOpen
  \bibfield  {author} {\bibinfo {author} {\bibfnamefont {N.~E.}\ \bibnamefont
  {Zimmermann}}, \bibinfo {author} {\bibfnamefont {M.~K.}\ \bibnamefont
  {Horton}}, \bibinfo {author} {\bibfnamefont {A.}~\bibnamefont {Jain}}, \ and\
  \bibinfo {author} {\bibfnamefont {M.}~\bibnamefont {Haranczyk}},\ }\href
  {\doibase 10.3389/fmats.2017.00034} {\bibfield  {journal} {\bibinfo
  {journal} {Front. Mater.}\ }\textbf {\bibinfo {volume} {4}},\ \bibinfo
  {pages} {34} (\bibinfo {year} {2017})}\BibitemShut {NoStop}%
\bibitem [{\citenamefont {Ong}\ \emph {et~al.}(2015)\citenamefont {Ong},
  \citenamefont {Cholia}, \citenamefont {Jain}, \citenamefont {Brafman},
  \citenamefont {Gunter}, \citenamefont {Ceder},\ and\ \citenamefont
  {Persson}}]{Ong2015}%
  \BibitemOpen
  \bibfield  {author} {\bibinfo {author} {\bibfnamefont {S.~P.}\ \bibnamefont
  {Ong}}, \bibinfo {author} {\bibfnamefont {S.}~\bibnamefont {Cholia}},
  \bibinfo {author} {\bibfnamefont {A.}~\bibnamefont {Jain}}, \bibinfo {author}
  {\bibfnamefont {M.}~\bibnamefont {Brafman}}, \bibinfo {author} {\bibfnamefont
  {D.}~\bibnamefont {Gunter}}, \bibinfo {author} {\bibfnamefont
  {G.}~\bibnamefont {Ceder}}, \ and\ \bibinfo {author} {\bibfnamefont {K.~A.}\
  \bibnamefont {Persson}},\ }\href {\doibase 10.1016/j.commatsci.2014.10.037}
  {\bibfield  {journal} {\bibinfo  {journal} {Comput. Mater. Sci.}\ }\textbf
  {\bibinfo {volume} {97}},\ \bibinfo {pages} {209} (\bibinfo {year}
  {2015})}\BibitemShut {NoStop}%
\bibitem [{\citenamefont {Zheng}\ \emph {et~al.}(2018)\citenamefont {Zheng},
  \citenamefont {Mathew}, \citenamefont {Chen}, \citenamefont {Chen},
  \citenamefont {Tang}, \citenamefont {Dozier}, \citenamefont {Kas},
  \citenamefont {Vila}, \citenamefont {Rehr}, \citenamefont {Piper},
  \citenamefont {Persson},\ and\ \citenamefont {Ong}}]{Zheng2018}%
  \BibitemOpen
  \bibfield  {author} {\bibinfo {author} {\bibfnamefont {C.}~\bibnamefont
  {Zheng}}, \bibinfo {author} {\bibfnamefont {K.}~\bibnamefont {Mathew}},
  \bibinfo {author} {\bibfnamefont {C.}~\bibnamefont {Chen}}, \bibinfo {author}
  {\bibfnamefont {Y.}~\bibnamefont {Chen}}, \bibinfo {author} {\bibfnamefont
  {H.}~\bibnamefont {Tang}}, \bibinfo {author} {\bibfnamefont {A.}~\bibnamefont
  {Dozier}}, \bibinfo {author} {\bibfnamefont {J.~J.}\ \bibnamefont {Kas}},
  \bibinfo {author} {\bibfnamefont {F.~D.}\ \bibnamefont {Vila}}, \bibinfo
  {author} {\bibfnamefont {J.~J.}\ \bibnamefont {Rehr}}, \bibinfo {author}
  {\bibfnamefont {L.~F.}\ \bibnamefont {Piper}}, \bibinfo {author}
  {\bibfnamefont {K.~A.}\ \bibnamefont {Persson}}, \ and\ \bibinfo {author}
  {\bibfnamefont {S.~P.}\ \bibnamefont {Ong}},\ }\href {\doibase
  10.1038/s41524-018-0067-x} {\bibfield  {journal} {\bibinfo  {journal} {Npj
  Comput. Mater.}\ }\textbf {\bibinfo {volume} {4}},\ \bibinfo {pages} {1}
  (\bibinfo {year} {2018})}\BibitemShut {NoStop}%
\end{thebibliography}%

\end{document}